\pdfoutput=1 %arXiv admin added this line
\documentclass{article}%
\usepackage{amssymb}
\usepackage{amsfonts}
\usepackage{amsmath}
\usepackage{graphicx}%
\providecommand{\U}[1]{\protect\rule{.1in}{.1in}}
\pdfoutput=1
\begin{document}

\title{Characterization of All Possible Orbits in the Schwarzschild Metric Revisited}
\author{F.T. Hioe* and David Kuebel
\and Department of Physics, St. John Fisher College, Rochester, NY 14618,
\and and
\and Department of Physics \& Astronomy, University of Rochester, Rochester, NY 14627}
\maketitle

\begin{abstract}
All possible orbital trajectories and their analytical expressions in the
Schwarzschild metric are presented in a single complete map characterized by
two dimensionless parameters. While three possible pairs of parameters with
different advantages are described, the parameter space that gives the most
convenient reduction to the Newtonian case is singled out and used which leads
to a new insight on Newtonian limits among other results. Numerous analytic
relations are presented. A comparison is made with the widely used formulation
and presentation given by S. Chandrasekhar.


\end{abstract}

\section{\bigskip Introduction}

As Einstein's theory of general relativity is approaching its centennial
celebration, the problem of understanding the orbital trajectories of a
particle around a very massive object still retains great interest. Although
current research has focused on the properties of spinning black holes and on
the effects of gravitational waves, the original analytic solutions for the
Schwarzschild metric are still used for the study of astronomical objects for
which the assumption of static spherical symmetry is reasonable. The orbital
solutions to the Schwarzschild metric also provide useful guides from which
various perturbation methods originate.

Although the orbital problem for the Schwarzschild metric has been solved in
terms of the equations for the orbits, a complete picture of the solution
space in terms of a single clear map has been lacking. Given, say, the
trajectory of a particle around some star or black hole, one would be
interested in the characterizing parameters of such a trajectory and how it
relates to other possible trajectories. Indeed, the question of constructing a
universal map and what physical parameters should be used for the coordinates
(the "latitude" and "longitude") of such a map did not seem to have been
seriously considered before our recent work [1-4]. It is well known [see e.g.
ref.5], for example, that the discriminant of a relevant cubic equation is
used to establish the criteria for distinguishing different types of possible
orbits in the Schwarzschild metric. However, because of the absence of a
universal map, clear boundary curves that separate different regions of the
parameter space for different types of orbits and trajectories were never
clearly presented before our work in ref.4.

The most widely used analytic solutions and analyses for orbits and
trajectories in the Schwarzschild metric were based on Chandrasekhar's work
[5] that originated with Darwin [6]. We shall make a comparison of our
analysis with theirs at the end of this paper. Two notable features of
Chandrasekhar's analysis are that (i) his analytic solutions for the orbits
are implicit, not explicit, expressions that include a parameter that is
related to the polar angle in a somewhat complicated implicit way, and (ii)
one of his two parameters for characterizing the orbital trajectories can be
real or purely imaginary, making the regions for different types of
trajectories difficult to be placed on a map. Among the earlier work, we
should mention the publications of Forsyth [7] and Whittaker [8] who gave some
special cases of the analytic solutions for the orbits that we present here
(and in refs. [1-4]) but in different forms. Hagihara [9] had a detailed
analysis of all possible orbits in terms of Weierstrassian elliptic functions
but his analysis involved some lengthy algebraic expressions and his division
of all possible orbits into seventeen cases also unnecessarily complicates the
picture. But he did present a map with two dimensionless parameters as
coordinates [9, Fig.13] even though he failed to clearly identify the physical
significance of the two parameters, and he classified the orbits largely
according to the roots of the relevant cubic equations.

Two important conclusions to be drawn from our derivation of analytic
solutions for the orbits and trajectories in terms of the Jacobian elliptic
functions are the following: (a) The consequence of having an analytic
solution for the orbits is not just that it is exact and convenient, as a
numerical solution of the differential equation could easily produce the same
precise trajectory. The more important consequence is that it cleverly picks
out the appropriate characterizing dimensionless parameters for the orbits and
these parameters are dimensionless in terms of some combinations of physical
quantities that could not have been easily guessed otherwise; and (b) The
well-studied Jacobian elliptic functions allow many simple relations among
different parameters for a number of special cases that are very useful for
checking numerical computations. It is with these considerations in mind that
we present the characterization of all possible orbits and trajectories on a
map with the special choice of two dimensionless physical parameters as the
coordinate axes.

In this paper, we summarize and extend the most essential features and
expressions of our previous analysis and present them in a complete picture.
We describe three pairs of dimensionless parameters that can be used for the
coordinates of the map of all possible orbits and trajectories, each of which
exhibits some special advantages. Among the useful outcomes are the following:
(1) All possible orbital trajectories of a particle in the Schwarzschild
metric are placed on a universal map; (2) The boundary between the two regions
that separate different types of trajectories is clearly shown on the map; (3)
One particular set of parameters serves to illuminate some interesting
properties with regard to the meaning and existence of Newtonian limiting
orbits. We also present an abundance of simple analytic results and relations
for some special cases in various parts of the map that are useful for many
purposes, not the least of which is to provide a way to check results from
numerical computations.

Here is a brief outline of this paper. In Section 2, we introduce the
fundamental differential equations in the Schwarzschild geometry and three
possible pairs of characterizing parameters. In Section 3, we present three
explicit analytic solutions for all possible orbital trajectories of a
particle in the Schwarzschild metric, define two regions and give examples of
orbital trajectories in the two regions. While these three fundamental
solutions for the orbits may have been given in different forms by other
authors, we have explicit analytic expressions for many related relevant
physical quantities, especially those for orbits of the hyperbolic type, that
were not given by other authors. One of these expressions (for the impact
parameter of a particle) corrects an erroneous expression given by
Chandrasekhar [5]. In Section 4 we describe the boundaries of the two regions
in the forms of curves expressed by simple equations. In Section 5, we present
analytic expressions for the orbital trajectories on these boundaries and
provide examples. In Section 6, we present a clear picture of how the orbital
trajectories change across the boundaries of the two regions. In Section 7, we
introduce a large collection of analytic relations that relate various
physical parameters at various special locations on our map. These have been
very useful for checking our numerical calculations and can be useful for
checking other numerical computations. However, this section could be skipped
on a first reading. In Section 8, we present the Newtonian and non-Newtonian
correspondences of our general relativistic results and show a sector of a
region on our map in which there exist elliptical orbits that have properties
which have not seemed to have been previously recognized. We also derive an
approximate formula for the general relativistic correction to a Newtonian
hyperbolic orbit. Our calculation gives a slightly different result from that
given by other authors for a proposed experiment involving a spaceship that
can be carried out in our solar system. In Section 9, we relate the parameters
that we use in our analysis with those used by Chandrasekhar [5]. This
comparison can be used for reference by those readers who are familiar with
Chandrasekhar's work. In Section 10, we summarize the principal results of
this paper.

\section{The Differential Equations and the Characterizing Parameters}

We consider the Schwarzschild geometry, i.e. the static spherically symmetric
gravitational field in the empty space surrounding some massive spherical
object such as a star or a black hole of mass $M$. The Schwarzschild metric
for the empty spacetime outside a spherical body in the spherical coordinates
$r,\theta,\phi$ is [10]%

\begin{equation}
dl^{2}=c^{2}\left(  1-\frac{\alpha}{r}\right)  dt^{2}-\left(  1-\frac{\alpha
}{r}\right)  ^{-1}dr^{2}-r^{2}d\theta^{2}-r^{2}\sin^{2}\theta d\phi^{2}%
\end{equation}

where%

\begin{equation}
\alpha=\frac{2GM}{c^{2}}%
\end{equation}

is the Schwarzschild radius, $G$ is the universal gravitation constant, and
$c$ is the speed of light. If $[x^{\mu}]=(t,r,\theta,\phi)$, then the
worldline $x^{\mu}(\tau)$, where $\tau$ is the proper time along the path, of
a particle moving in the equatorial plane $\theta=\pi/2$, satisfies the
equations [10]%

\begin{equation}
\left(  1-\frac{\alpha}{r}\right)  \overset{\cdot}{t}=\kappa,
\end{equation}

\begin{equation}
c^{2}\left(  1-\frac{\alpha}{r}\right)  \overset{\cdot}{t}^{2}-\left(
1-\frac{\alpha}{r}\right)  ^{-1}\overset{\cdot}{r}^{2}-r^{2}\overset{\cdot
}{\phi}^{2}=c^{2},
\end{equation}

\begin{equation}
r^{2}\overset{\cdot}{\phi}=h,
\end{equation}

where the derivative $\overset{\cdot}{}$ represents $d/d\tau$. The coordinates
$r$ and $\phi$ describe the position of the particle relative to the star or
black hole situated at the origin. The constant $h$ is identified as the
angular momentum per unit rest mass of the particle, and the constant $\kappa$
is identified to be the total energy per unit rest energy of the particle%

\begin{equation}
\kappa=\frac{E}{m_{0}c^{2}},
\end{equation}

where $E$ is the total energy of the particle in its orbit and $m_{0}$ is the
rest mass of the particle at $r=\infty$. Substituting eqs.(3) and (5) into (4)
gives the 'combined' energy equation%

\begin{equation}
\overset{\cdot}{r}^{2}+c^{2}+\frac{h^{2}}{r^{2}}\left(  1-\frac{\alpha}%
{r}\right)  -\frac{c^{2}\alpha}{r}=c^{2}\kappa^{2}.
\end{equation}

Substituting $dr/d\tau=(dr/d\phi)(d\phi/d\tau)=(h/r^{2})(dr/d\phi)$ into the
combined energy equation gives the differential equation for the orbit of the planet%

\begin{equation}
\left(  \frac{du}{d\phi}\right)  ^{2}=\alpha u^{3}-u^{2}+(2GM/h^{2}%
)u+c^{2}(\kappa^{2}-1)/h^{2}%
\end{equation}

where $u=1/r$. We define the dimensionless distance $q$ of the particle from
the star or black hole, measured in units of the Schwarzschild radius, by%

\begin{equation}
q\equiv\frac{r}{\alpha}\equiv\frac{1}{\alpha u}\equiv\frac{1}{v}.
\end{equation}

In terms of $v$, eq.(8) becomes%

\begin{equation}
\left(  \frac{dv}{d\phi}\right)  ^{2}=v^{3}-v^{2}+4s^{2}v+4s^{2}(\kappa
^{2}-1),
\end{equation}

where%

\begin{equation}
s^{2}=\left(  \frac{GM}{hc}\right)  ^{2}.
\end{equation}

By changing the variable from $v$ to another dimensionless quantity $U$
defined by%

\[
U\equiv\frac{1}{4}\left(  \frac{\alpha}{r}-\frac{1}{3}\right)  =\frac{1}%
{4}\left(  v-\frac{1}{3}\right)  =\frac{1}{4}\left(  \frac{1}{q}-\frac{1}%
{3}\right)  ,
\]

or%

\begin{equation}
v=\frac{1}{q}=\frac{1}{3}+4U,
\end{equation}

eq.(10) becomes%

\begin{equation}
\left(  \frac{dU}{d\phi}\right)  ^{2}=4U^{3}-g_{2}U-g_{3}%
\end{equation}

where%

\begin{align}
g_{2}  &  =\frac{1}{12}-s^{2}\nonumber\\
g_{3}  &  =\frac{1}{216}+\frac{1}{6}s^{2}-\frac{1}{4}\kappa^{2}s^{2}%
\equiv\frac{1}{216}-\frac{1}{12}s^{2}+\frac{1}{4}(1-e^{2})s^{4},
\end{align}

and where we have defined the dimensionless quantity $e^{2}$ given by%

\begin{equation}
e^{2}\equiv1+\frac{h^{2}c^{2}(\kappa^{2}-1)}{(GM)^{2}}\equiv1+\frac{\kappa
^{2}-1}{s^{2}}.
\end{equation}

It is clear that as soon as we use the dimensionless distance $q$ (and the
dimensionless angle $\phi$) to express the differential equation (8) in the
form of eq.(10) or (13), the solutions for the orbits should be conveniently
characterized by a pair of dimensionless parameters $(\kappa^{2},s^{2})$ or
$(g_{2},g_{3})$. Indeed, $g_{2}$ and $g_{3}$ are called the invariants of the
Weierstrassian elliptic functions [11] that would represent the solution for
$U$ for the differential equation (13). Notice that while the constant of
motion $\kappa$ defined by eq.(6), which is the total energy per unit rest
energy, can obviously be one of the dimensionless parameters, the other
constant of motion $h$ which is the angular momentum per unit rest mass is not
a dimensionless quantity and should not be used by itself as a parameter for
representing the orbits in a universal map. That $s$ defined by eq.(11) could
be used as the second characterizing parameter was recognized by one of us in
ref.[1]. In fact, the orbit solutions depend on $\kappa^{2}$ and $s^{2}$ as
shown by eqs.(10), (13) and (14). While we shall briefly discuss the parameter
spaces described by the coordinates $(\kappa^{2},s^{2})$ and $(g_{2},g_{3})$,
for most of this paper we shall use the parameter space characterized by the
coordinates $(e^{2},s^{2})$. The principal reason for this choice of
parameters is that the parameter $e$ conveniently reduces to the well known
parameter called eccentricity for the classical Newtonian case. In our earlier
work [1,2], there was an oversight on our part in which we unnecessarily
limited ourselves to the case $e^{2}\geq0$, but we remedied it in ref.4.

We call the two dimensionless parameters $e^{2}$ and $s^{2}$ defined by
eqs.(15) and (11) the energy eccentricity and gravitational field parameters,
or simply the energy and field parameters, respectively. They are the
principal parameters we use as coordinates for characterizing the orbital
trajectories of a particle in the gravitational field produced by the massive
object or black hole of mass $M$. We use the parameter space $(e^{2},s^{2})$,
and $(e^{2},s)$ as well, where $-\infty\leq e^{2}\leq\infty$, $s\geq0$ for
classifying and identifying all possible types of orbital trajectories. In
doing so, we shall put all possible trajectories in the Schwarzschild metric
on a two-dimensional universal map on which the two coordinate axes are real
and represent dimensionless physical quantities. For $e^{2}\geq0$, the three
cases of $0<e<1$, $e=1$ and $e>1$ can be used to classify the orbits as
elliptic-type, parabolic-type and hyperbolic-type, similar to those in the
Newtonian cases; For $e^{2}<0$ on the other hand, the orbits do not have any
Newtonian correspondences and this will be explained and discussed in Section 8.

\section{Three Analytic Solutions for the Orbital Trajectories in Two Regions}

We consider eq.(13) and derive its analytic solutions in this section.

The discriminant $\Delta$ of the cubic equation%

\begin{equation}
4U^{3}-g_{2}U-g_{3}=0
\end{equation}

is defined by%

\begin{equation}
\Delta=27g_{3}^{2}-g_{2}^{3}.
\end{equation}

For the case $\Delta\leq0$, the three roots of the cubic equation (16) are all
real. We call the three roots $e_{1},e_{2},e_{3}$ and arrange them so that
$e_{1}>e_{2}>e_{3}$; the special cases when two or three of the roots are
equal will be considered also. For the case $\Delta>0$, the cubic equation
(16) has one real root and two roots that are complex conjugates. The analytic
solutions of eq.(13) that we present below will give the distance $r$ of the
particle from the star or black hole in terms of the Jacobian elliptic
functions [11] that have the polar angle $\phi$ in their argument and that are
associated with a modulus $k$ that will be defined.

There are three relevant analytic solutions of eq.(13). The first two
solutions are for the case $\Delta\leq0$, and \ the third solution is for the
case $\Delta>0$. The region that contains all the coordinate points
$(e^{2},s^{2})$ for the first two solutions for the case $\Delta\leq0$ is
called Region I (see \ref{Fig.1}), and the region that contains all the
coordinate points $(e^{2},s^{2})$ for the third solution for the case
$\Delta>0$ is called Region II. There are two physically inadmissible regions:
the region below the horizontal axis $(s^{2}<0)$ and the region corresponding
to $\kappa^{2}<0$ which is the region (called Region II' in Fig.1) above the
curve $s^{2}>s_{2}^{2}$ where $s_{2}$ will be given in the next section. The
boundary between Regions I and II consists of two segments that we call
$s^{2}=s_{1}^{2}$ (the upper boundary of Region I) and $s^{2}=s_{1}^{\prime2}$
(the left boundary of Region I) that will be defined in the next section. The
three boundary curves $s^{2}=0$, $s^{2}=s_{1}^{\prime2}$ and $s^{2}=s_{1}^{2}$
all satisfy $\Delta=0$.

We now present the three relevant analytic solutions of eq.(13).

Solution (A1) For $\Delta\leq0$, $e_{1}>e_{2}\geq U>e_{3}$ (Applicable in
Region I).

Writing the right-hand side of eq.(13) as $4(e_{1}-U)(e_{2}-U)(U-e_{3})$, the
equation for the orbit is%

\begin{align}
\frac{1}{q}  &  =\frac{1}{3}+4e_{3}+4(e_{2}-e_{3})sn^{2}(\gamma\phi
,k)\nonumber\\
&  =\frac{1}{3}+4e_{3}+4(e_{2}-e_{3})\frac{1-cn(2\gamma\phi,k)}{1+dn(2\gamma
\phi,k)}.
\end{align}

For the case $e^{2}\leq1$, the initial point at $\phi=0$ gives $U=e_{3}$ (i.e.
$1/q=1/3+4e_{3}$) and thus $dU/d\phi=0$ from eq.(13) which means that the
particle at this point is in the direction perpendicular to the radius vector.
For the case $e^{2}>1$, the initial point is discussed in eq.(28). The
constant $\gamma$ appearing in the argument, and the modulus $k$, of the
Jacobian elliptic functions are given in terms of the three roots of the cubic
equation (16) by%

\begin{align}
\gamma &  =(e_{1}-e_{3})^{1/2},\\
k^{2}  &  =\frac{e_{2}-e_{3}}{e_{1}-e_{3}}.
\end{align}

where $e_{1},e_{2},e_{3}$ are given by%

\begin{align}
e_{1}  &  =2\left(  \frac{g_{2}}{12}\right)  ^{1/2}\cos\left(  \frac{\theta
}{3}\right)  ,\nonumber\\
e_{2}  &  =2\left(  \frac{g_{2}}{12}\right)  ^{1/2}\cos\left(  \frac{\theta
}{3}+\frac{4\pi}{3}\right)  ,\nonumber\\
e_{3}  &  =2\left(  \frac{g_{2}}{12}\right)  ^{1/2}\cos\left(  \frac{\theta
}{3}+\frac{2\pi}{3}\right)  ,
\end{align}

and where%

\begin{equation}
\cos\theta=g_{3}\left(  \frac{27}{g_{2}^{3}}\right)  ^{1/2}.
\end{equation}

The modulus $k$ of the elliptic functions has a range $0\leq k^{2}\leq1$. For
$k^{2}=0$, $sn(x,0)=\sin x,$ $cn(x,0)=\cos x,$ $dn(x,0)=1$. For $k^{2}=1$,
$sn(\gamma\phi,1)=\tanh(\gamma\phi)$, $cn(\gamma\phi,1)=dn(\gamma\phi,1)=\sec
h(\gamma\phi)$. The period of $cn(2\gamma\phi,k)$ is $4K(k) $, and the period
of $dn(2\gamma\phi,k)$ and of $sn^{2}(\gamma\phi,k)$ is $2K(k)$, where $K(k)$
is the complete elliptic integral of the first kind [11]. As $k^{2}$ increases
from $0$ to $1$, $K(k)$ increases from $\pi/2$ to $\infty$, and $k^{2}$ is a
useful parameter for the orbit.

The part showing Region I in \ref{Fig.1} is enlarged in \ref{Fig.2} in which
the left boundary of Region I is denoted in \ref{Fig.1} by the curve
$s_{1}^{\prime}$ for which $k^{2}=0$, and the upper boundary of Region I is
denoted in \ref{Fig.1} by the curve $s_{1}$ for which $k^{2}=1$. The
intersection point $V$ of $s_{1}^{\prime}$ and $s_{1}$, which we shall call
the vertex, is $(e^{2},s)=(-1/3,1/\sqrt{12})$ and corresponds to $e_{1}%
=e_{2}=e_{3}=0$. It is a special point of interest. Three other constant
$k^{2}$ curves for $k^{2}=0.01,$ $0.2$ and $0.5$ in Region I are also shown in
\ref{Fig.2}. The constant $k^{2}$ curves in Region II will be discussed later.

A typical orbit given by eq.(18) (not on any one of the three boundaries) in
Region I is a precessional elliptic-type orbit for $e^{2}<1$, a parabolic-type
orbit for $e^{2}=1$, and a hyperbolic-type orbit for $e^{2}>1$ [2,3]. For the
elliptic-type orbits ($e^{2}<1$), $e_{3}>-1/12$, the maximum distance
$r_{\max}$ (the aphelion) of the particle from the star or black hole and the
minimum distance $r_{\min}$ (the perihelion) of the particle from the star or
black hole, or their corresponding dimensionless forms $q_{\max}$ $(=r_{\max
}/\alpha)$ and $q_{\min}$ $(=r_{\min}/\alpha),$ are obtained from eq.(18) when
$\gamma\phi=0$ and when $\gamma\phi=K(k)$ respectively, and they are given by%

\begin{equation}
\frac{1}{q_{\max}}=\frac{1}{3}+4e_{3},
\end{equation}

and%

\begin{equation}
\frac{1}{q_{\min}}=\frac{1}{3}+4e_{2},
\end{equation}

where $e_{2}$ and $e_{3}$ are determined from eqs.(21), (22) and (14) in terms
of $e$ and $s$. The geometric eccentricity $\varepsilon$ of the orbit is
defined in the range $0\leq\varepsilon\leq1$ by%

\begin{equation}
\varepsilon\equiv\frac{r_{\max}-r_{\min}}{r_{\max}+r_{\min}}=\frac{q_{\max
}-q_{\min}}{q_{\max}+q_{\min}}.
\end{equation}

If we use $q_{\max}$ and $q_{\min}$ given by eqs.(23) and (24) and write%

\begin{equation}
\varepsilon=\frac{e_{2}-e_{3}}{1/6+e_{2}+e_{3}},
\end{equation}

then the definition of $\varepsilon$ can be extended to $\varepsilon>1$ but it
should be emphasized that any $\varepsilon$ outside the range between $0$ and
$1$ does not have the physical meaning given by eq.(25). We shall have more to
say about this point in Section 9.

The precessional angle $\Delta\phi$ is given by%

\begin{equation}
\Delta\phi=\frac{2K(k)}{\gamma}-2\pi.
\end{equation}

Whether the orbit is closed (and hence periodic) or non-closed (and hence
non-periodic) depends on whether $\Delta\phi/\pi$ is or is not a rational number.

An example of a precessional elliptical-type orbit for $e^{2}=0.25$ and
$s=0.2$ in Region I is shown in \ref{Fig.3}.

For the parabolic-type orbits ($e^{2}=1$), $e_{3}=-1/12$, and $q_{\max}%
=\infty$. The orbits are unbounded. The particle is initially at infinity at
$\phi=0$ and initially moves counterclockwise in a direction perpendicular to
the line joining it to the origin. The incoming trajectory coming from
infinity at $\phi=0$ goes around the origin counter-clockwise according to
eq.(18) as $\phi$ increases from $0$ and the outgoing trajectory makes an
angle $\phi=2K(k)/\gamma$ with the horizontal axis in going to infinity. A
particle coming from infinity can go around the star or black hole many times
as its polar angle $\phi$ increases from $0$ to $2K(k)/\gamma$ before going
off to infinity.

An example of a precessional parabolic-type orbit for $e^{2}=1$ and $s=0.4$ in
Region I is shown in \ref{Fig.4}.

For the hyperbolic-type orbits ($e^{2}>1$) [2,3], $e_{3}<-1/12$, the initial
point of the particle is not at $\phi=0$ but is at infinity at a polar angle
$\phi=\Psi_{1}$, where $\Psi_{1}$ is given by%

\begin{equation}
sn(\gamma\Psi_{1},k)=\sqrt{-\frac{\frac{1}{3}+4e_{3}}{4(e_{2}-e_{3})}},
\end{equation}

or%

\begin{equation}
\Psi_{1}=\gamma^{-1}sn^{-1}\left(  \sqrt{-\frac{\frac{1}{3}+4e_{3}}%
{4(e_{2}-e_{3})}},k\right)  ,
\end{equation}

and where $\gamma$ and $k$ are defined by eqs.(19) and (20). As $\phi$
increases from $\Psi_{1}$,%

\begin{equation}
\Psi_{2}\equiv\frac{2K(k)}{\gamma}-\Psi_{1}%
\end{equation}

is the next value of $\phi$ for $q$ to become infinite. Thus eq.(18) gives the
hyperbolic-type orbit in Region I for%

\begin{equation}
\Psi_{1}\leq\phi\leq\Psi_{2}.
\end{equation}

The orbit given by eq.(18) for $e^{2}>1$ describes a particle approaching the
star from infinity along an incoming asymptote at an angle $\Psi_{1}$ to the
horizontal axis, turning counter-clockwise about the star to its right on the
horizontal axis, and leaving along an outgoing asymptote at an angle $\Psi
_{2}$. A particle approaching along the incoming asymptote can go around the
star or black hole many times as its polar angle $\phi$ increases from
$\Psi_{1}$ to $\Psi_{2}$ before going off to infinity along the outgoing
asymptote. The $x$ coordinates of the points where the two asymptotes
intersect the horizontal axis relative to the origin are given by%

\begin{align}
x_{a1}  &  =-\frac{1}{2s^{2}\sqrt{e^{2}-1}}\frac{1}{\sin\Psi_{1}},\nonumber\\
x_{a2}  &  =\frac{1}{2s^{2}\sqrt{e^{2}-1}}\frac{1}{\sin\Psi_{2}}.
\end{align}

A negative (positive) sign for the $x_{ai}$ means that the corresponding
asymptote intersects the horizontal axis to the left (right) of the origin.
The minimum distance $q_{\min}$ of the particle from the star or black hole is
obtained from eq.(24) when the particle is at a polar angle $\phi=K(k)/\gamma
$, where $dr/d\phi=0$ at $q=q_{\min}$. A straight line through the origin $O$
(where the star or black hole is located) that makes an angle%

\begin{equation}
\chi=\frac{K(k)}{\gamma}%
\end{equation}

with the horizontal axis is the symmetry axis of the hyperbolic-type orbit.
Let $b$ be the impact parameter, i.e. the perpendicular distance from the
origin (where the star or black hole is located) to the asymptote. The
dimensionless impact parameter $b/\alpha$, where $\alpha$ is the Schwarzschild
radius, can be expressed as%

\begin{equation}
\frac{b}{\alpha}=\frac{1}{2s^{2}\sqrt{e^{2}-1}}.
\end{equation}

An example of a precessional hyperbolic-type orbit for $e^{2}=4$ and $s=0.2$
in Region I is shown in \ref{Fig.5}.

The orbital trajectories that occur on the left and upper boundaries of Region
I are special cases that deserve separate attention and will be discussed in
Section 5.

We now present the second solution.

Solution (A2) For $\Delta\leq0$, $U>e_{1}>e_{2}>e_{3}$ (Applicable in Region I).

We write the right-hand side of eq.(13) as $4(U-e_{1})(U-e_{2})(U-e_{3})$. The
equation for the orbital trajectory is%

\begin{equation}
\frac{1}{q}=\frac{1}{3}+4\frac{e_{1}-e_{2}sn^{2}(\gamma\phi,k)}{cn^{2}%
(\gamma\phi,k)},
\end{equation}

where $\gamma$, $k$, $e_{1}$, $e_{2}$ and $e_{3}$ are given by eqs.(19)-(21)
as in the Solution (A1). This solution gives a terminating orbit. The point at
$\phi=0$ has been chosen to be given by%

\begin{equation}
\frac{1}{q_{1}}=\frac{1}{3}+4e_{1}.
\end{equation}

The particle, starting from the polar angle $\phi=0$ at a distance $q_{1}$
from the black hole, plunges into the center of the black hole [2] when its
polar angle $\phi_{1}$ is given by $cn(\gamma\phi_{1},k)=0$, i.e. when%

\begin{equation}
\phi_{1}=\frac{K(k)}{\gamma}.
\end{equation}

Considering Solutions (A1) and (A2) together, each point $(e^{2},s)$ of the
parameter space in Region I allows two distinct orbits, one of which is
non-terminating (that consists of three types: elliptic, parabolic and
hyperbolic) and the other terminating. At the same coordinate point,
quantities that describe the two distinct orbits are related. For example, by
noting $e_{1}+e_{2}+e_{3}=0$ and from eqs.(23), (24) and (36), $q_{1}$ can be
expressed as%

\begin{equation}
\frac{1}{q_{1}}=1-\left(  \frac{1}{q_{\min}}+\frac{1}{q_{\max}}\right)  ,
\end{equation}

where $q_{\min}$ and $q_{\max}$ are the minimum and maximum distances for the
non-terminating orbit at the same coordinate point $(e^{2},s^{2})$. It will be
noted that $q_{1}$ is less than $q_{\min}$, i.e. for the terminating orbit the
particle is assumed initially to be closer to the black hole than the
$q_{\min}$ for the associated non-terminating orbit, except at $k^{2}=1$ where
$q_{1}=q_{\min}$ and the particle has a circular instead of a terminating
orbit that will be explained later. We note that $q_{1}$ is finite even for
$e^{2}\geq1$.

Three examples of these terminating orbits in Region I for the same coordinate
points of $s=0.2$ and $e^{2}=0.25,$ $1$, and $4$ are shown in \ref{Fig.6},
\ref{Fig.7} and \ref{Fig.8} (These were also the coordinate points used for
\ref{Fig.3}, \ref{Fig.4} and \ref{Fig.5}). Notice that the initial distances
of the particle from the black hole given by $q_{1}$ are all finite and are
less than $q_{\min}$ for the non-terminating orbits. This corrects an
erroneous statement made by us in ref. 3 after eq.(75).

We now present the third solution.

Solution (B) For $\Delta>0$ (Applicable in Region II).

Define%

\begin{align}
A  &  =\frac{1}{2}\left(  g_{3}+\sqrt{\frac{\Delta}{27}}\right)
^{1/3},\nonumber\\
B  &  =\frac{1}{2}\left(  g_{3}-\sqrt{\frac{\Delta}{27}}\right)  ^{1/3},
\end{align}

where $g_{3}$ and $\Delta$ are defined by eqs.(14) and (17). The real root of
the cubic equation (16) is given by%

\begin{equation}
a=A+B
\end{equation}

and the two complex conjugate roots $b$ and $\overline{b}$ are $-(A+B)/2\pm
(A-B)\sqrt{3}i/2$. We further define%

\begin{equation}
\gamma=[3(A^{2}+AB+B^{2})]^{1/4}%
\end{equation}

and%

\begin{equation}
k^{2}=\frac{1}{2}-\frac{3(A+B)}{4\sqrt{3(A^{2}+AB+B^{2})}}=\frac{1}{2}%
-\frac{3a}{4\gamma^{2}}.
\end{equation}

The curves of constant $k^{2}$, the squared modulus of the elliptic functions
that describe the orbital trajectories, radiate from the vertex point $V$
whose coordinates are $(e^{2},s^{2})=(-1/3,1/12)$ (see \ref{Fig.2}). In Region
I, $k^{2}$ increases counter-clockwise as one goes from the curve $k^{2}=0$
for $s=s_{1}^{\prime}$ to the curve $k^{2}=1$ for $s=s_{1}^{{}}$, and as one
keeps going counter-clockwise to Region II, $k^{2}$ decreases from $k^{2}=1$
\ and returns in value to $k^{2}=0$. The squared modulus $k^{2}$ is obtained
from eq.(20) for Region I and from eq.(42) for Region II. Two $k^{2}=const.$
curves in Region II are rather special: $k^{2}=1/2+\sqrt{3}/4=0.933012$ is a
horizontal line to the right of $V$ and $k^{2}=1/2-\sqrt{3}/4=0.0669872$ is a
horizontal line to the left of $V$. The $k^{2}=1/2$ curve in Regions I and II
has a simple analytic expression that will be given later.

Writing the right-hand side of eq.(13), with $U\geq a$, as
$4(U-a)(U-b)(U-\overline{b})$, the equation for the orbital trajectory is%

\begin{align}
\frac{1}{q}  &  =\frac{1}{3}+4a+4\gamma^{2}\frac{1-cn(2\gamma\phi
,k)}{1+cn(2\gamma\phi,k)}\nonumber\\
&  =\frac{1}{3}+4a+4\gamma^{2}tn^{2}(\gamma\phi,k)dn^{2}(\gamma\phi,k).
\end{align}

This is a terminating orbit and thus Region II allows only terminating orbits.

For $e^{2}<1$, $a$ is greater than $-1/12$, the initial distance $q_{2}$ of
the particle at $\phi=0$ (in a direction $dq/d\phi=0$) is given by%

\begin{equation}
\frac{1}{q_{2}}=\frac{1}{3}+4a.
\end{equation}

The particle plunges into the center of the black hole [2,3] when its polar
angle $\phi=\phi_{2}$ is given by%

\begin{equation}
\phi_{2}=\frac{K(k)}{\gamma},
\end{equation}

where $\gamma$ and $k$ are given by eqs.(41) and (42).

An example of a terminating orbit for $e^{2}=0.25$ and $s=0.3$ in Region II is
shown in \ref{Fig.9}.

For $e^{2}=1$, $a=-1/12$, and the initial distance of the particle at $\phi=0$
is at infinity (in a direction $dq/d\phi=0$). An example of a terminating
orbit for $e^{2}=1$ and $s=0.3$ in Region II is shown in \ref{Fig.10} in which
the initial position of the particle coming from infinity is at $\phi=0$.

For $e^{2}>1$, $a$ is less than $-1/12$, the particle comes from infinity at
the polar angle $\phi=\Phi_{2}$, where $\Phi_{2}$ is given by%

\begin{equation}
cn(2\gamma\Phi_{2},k)=\frac{1-D}{1+D},
\end{equation}

where%

\[
D=-\frac{1}{\gamma^{2}}\left(  \frac{1}{12}+a\right)  ,
\]

and where $\gamma$ and $k$ are given by eqs.(41) and (42). The orbit given by
eq.(43) describes a particle coming from infinity along an incoming asymptote
at an angle $\Phi_{2}$ to the horizontal axis, and as $\phi$ increases, the
orbit terminates at the center of the black hole [3] at a polar angle
$\phi_{2}$ given by eq.(45). There is thus only one asymptote, and the $x$
coordinate of the point where the asymptote intersects the horizontal axis
relative to the origin is given by%

\begin{equation}
x_{a}=-\frac{1}{2s^{2}\sqrt{e^{2}-1}}\frac{1}{\sin\Phi_{2}},
\end{equation}

and the dimensionless impact parameter is given by eq.(34).

An example of a terminating orbit for $e^{2}=4$ and $s=0.3$ in Region II is
shown in \ref{Fig.11} in which the particle comes from infinity along the
asymptote shown by the dotted line.

We have given three examples of these terminating orbits in Region II which
are shown in \ref{Fig.9}, \ref{Fig.10} and \ref{Fig.11} for the coordinate
points for $e^{2}=0.25,$ $1$, and $4$ \ and $s=0.3$. They will be discussed in
conjunction with the non-terminating orbits shown in \ref{Fig.3}, \ref{Fig.4}
and \ref{Fig.5} and with the terminating orbits shown in \ref{Fig.6},
\ref{Fig.7} and \ref{Fig.8} in Region I when we discuss how the orbital
trajectories change as we cross the upper boundary of Region I, $s_{1}$, to
Region II in Section 6. We shall also discuss in Section 6 how the orbital
trajectories change as we cross the left boundary of Region I, $s_{1}^{\prime
}$, to Region II, and for this purpose, we show an example of a terminating
orbit in Region II to the left of $s_{1}^{\prime}$ for $e^{2}=-0.25$ and
$s=0.2$ in \ref{Fig.12}.

Thus Region II, with orbits given by eq.(43), allows only terminating orbits
with a finite initial distance of the particle from the black hole given by
eq.(44) for $e^{2}<1$, and with an infinite initial distance for $e^{2}\geq1 $.

We now describe in greater details the boundaries for Regions I and II.

\section{The Boundaries of Regions I and II}

In this section, we present the boundary curves of Regions I and II, and it is
instructive to see these curves in three different parameter spaces:
$(e^{2},s^{2})$, $(\kappa^{2},s^{2})$ and $(-g_{3},-g_{2})$.

I. The parameter space $(e^{2},s^{2})$ or $(e^{2},s)$ (see \ref{Fig.1} and
\ref{Fig.2})

There are four relevant boundaries for our universal map in the parameter
space $(e^{2},s^{2})$ for $-\infty\leq e^{2}\leq+\infty$: (1) the bottom
boundary $s^{2}=0$ of Region I and of Region II ($s^{2}<0$ is physically
inadmissible), (2) the left boundary of Region I which we call $s=s_{1}%
^{\prime}$, (3) the top boundary of Region I which we call $s=s_{1}$, and (4)
the top boundary of Region II which we call $s=s_{2}$ (we note that $s>s_{2}$
corresponds to $\kappa^{2}<0$ and is physically inadmissible). A common
equation that characterizes boundaries (1)-(3) is $\Delta=0$.

(1) $\Delta=0$, $s^{2}=0$ and $k^{2}=0$.

It is easily verified that setting $s^{2}=0$ gives $g_{2}=1/12$, $g_{3}=1/216
$, $\theta=0$, $e_{1}=1/6$, $e_{2}=e_{3}=-1/12$ from eqs.(14), (22) and (21),
and thus $\Delta=0$ from eq.(17), $k^{2}=0$ and $\gamma=1/2$ from eqs.(20) and (19).

(2) $\Delta=0$, $k^{2}=0$, and $s^{2}=s_{1}^{\prime2}$ where $s_{1}^{\prime2}$
is given by%

\begin{equation}
s_{1}^{\prime2}=\frac{1-9e^{2}-\sqrt{(1+3e^{2})^{3}}}{27(1-e^{2})^{2}}%
\end{equation}

for $-1/3\leq e^{2}\leq0$, which results in $1/12\geq s_{1}^{\prime2}\geq0 $.
Eq.(48) can be inverted to give $e^{2}$ in terms of $s^{2}$ as%

\begin{equation}
e^{2}=\frac{1-18s^{2}+54s^{4}-\sqrt{(1-12s^{2})^{3}}}{54s^{4}},
\end{equation}

for $1/12\geq s^{2}\geq0$, which results in $-1/3\leq e^{2}\leq0$. We shall
refer to $s^{2}=s_{1}^{\prime2}$ as the left boundary of Region I. It can be
verified that along this boundary, $\theta=0$, $e_{1}=(1-12s_{1}^{\prime
2})^{1/2}/6$, $e_{2}=e_{3}=-(1-12s_{1}^{\prime2})^{1/2}/12$, $k^{2}=0 $ and
$\gamma=(1-12s_{1}^{\prime2})^{1/4}/2$.

(3) $\Delta=0$, $k^{2}=1$, and $s^{2}=s_{1}^{2}$ where $s_{1}^{2}$ is given by%

\begin{equation}
s_{1}^{2}=\frac{1-9e^{2}+\sqrt{(1+3e^{2})^{3}}}{27(1-e^{2})^{2}}%
\end{equation}

for $-1/3\leq e^{2}\leq\infty$, which results in $1/12\geq s_{1}^{2}\geq0$
(with a note that $s_{1}^{2}=1/16$ for $e=1$). Eq.(50) can be inverted to give%

\begin{equation}
e^{2}=\frac{1-18s^{2}+54s^{4}+\sqrt{(1-12s^{2})^{3}}}{54s^{4}},
\end{equation}

for $1/12\geq s^{2}\geq0$, which results in $-1/3\leq e^{2}\leq\infty$. We
shall refer to $s^{2}=s_{1}^{2}$ as the upper boundary of Region I. It can be
verified that along this boundary, $\theta=\pi$, $e_{1}=e_{2}=(1-12s_{1}%
^{2})^{1/2}/12$, $e_{2}=e_{3}=-(1-12s_{1}^{2})^{1/2}/6$, $k^{2}=1$ and
$\gamma=(1-12s_{1}^{2})^{1/4}/2$.

The intersection point $V$ of $s_{1}^{\prime}$ and $s_{1}$ is the vertex point
$(e^{2},s^{2})=(-1/3,1/12)$ where the innermost stable circular orbit occurs
and where all $k^{2}=const.$ curves meet.

(4) $s^{2}=s_{2}^{2}$, where $s_{2}^{2}$ is given by%

\begin{equation}
s_{2}^{2}=\frac{1}{1-e^{2}}.
\end{equation}

We shall refer to $s^{2}=s_{2}^{2}$ as the top boundary of Region II. The
physical requirement that $\kappa^{2}\geq0$,where $\kappa$ is given by eq.(6),
leads to the condition that $s^{2}\leq1/(1-e^{2})$. It can be shown that
eq.(52) determines that $q_{2}$, the initial position of the particle given by
eq.(44), is equal to $1$, and thus $s^{2}>s_{2}^{2}$ implies that the initial
position of the particle is inside the Schwarzschild horizon, i.e. the
physically inadmissible case of $q_{2}<1$. We call the region $s^{2}>s_{2}%
^{2}$ Region II'.

II. The Parameter Space $(\kappa^{2},s^{2})$ (see \ref{Fig.13})

This parameter space is equivalent to one used by Hagihara [9, Fig.13] but he
did not clearly identify the physical significance of the two parameters.

There are four relevant boundaries in the parameter space $(\kappa^{2},s^{2})$
for $\kappa^{2}\geq0$, $s^{2}\geq0$: (1) the bottom boundary $s^{2}=0$, (2)
the left boundary of Region I which we call $s=s_{1}^{\prime} $, (3) the top
boundary of Region I which we call $s=s_{1}$, and (4) the left boundary of
Region II defined by $\kappa^{2}=0$.

(1) $\Delta=0$, $s^{2}=0$ and $k^{2}=0$. This is the lower boundary for Region
I and for Region II.

(2) $\Delta=0$, $k^{2}=0$, and $s^{2}=s_{1}^{\prime2}$ where $s_{1}^{\prime2}$
is given by%

\[
s_{1}^{\prime2}=\frac{-(27\kappa^{4}-36\kappa^{2}+8)-\sqrt{\kappa^{2}%
(9\kappa^{2}-8)^{3}}}{32}%
\]

for $8/9\leq\kappa^{2}\leq1$, which results in $1/12\geq s_{1}^{\prime2}\geq
0$. It can be inverted to give $\kappa^{2}$ in terms of $s^{2}$ as%

\[
\kappa^{2}=\frac{1+36s^{2}-\sqrt{(1-12s^{2})^{3}}}{54s^{4}},
\]

for $1/12\geq s^{2}\geq0$, which results in $8/9\leq\kappa^{2}\leq1$. We shall
refer to $s^{2}=s_{1}^{\prime2}$ as the left boundary of Region I.

(3) $\Delta=0$, $k^{2}=1$, and $s^{2}=s_{1}^{2}$ where $s_{1}^{2}$ is given by%

\[
s_{1}^{2}=\frac{-(27\kappa^{4}-36\kappa^{2}+8)+\sqrt{\kappa^{2}(9\kappa
^{2}-8)^{3}}}{32}%
\]

for $8/9\leq\kappa^{2}\leq\infty$, which results in $1/12\geq s_{1}^{2}\geq0$.
It can be inverted to give%

\[
\kappa^{2}=\frac{1+36s^{2}+\sqrt{(1-12s^{2})^{3}}}{54s^{4}},
\]

for $1/12\geq s^{2}\geq0$, which results in $8/9\leq\kappa^{2}\leq\infty$. We
shall refer to $s^{2}=s_{1}^{2}$ as the upper boundary of Region I.

The intersection point $V$ of $s_{1}^{\prime}$ and $s_{1}$ is the vertex point
$(\kappa^{2},s^{2})=(8/9,1/12)$. Two special $k^{2}=const.$ curves,
$k^{2}=1/2\pm\sqrt{3}/4=0.933012$ and $0.0669872$, are shown as straight
horizontal lines to right and left of $V$ (see Section 7, 1d).

(4) $\kappa^{2}=0$.

This is the left boundary for Region II that is equivalent to the upper (and
left) boundary $s_{2}$ for Region II in the $(e^{2},s^{2})$ parameter space.

Fig.13 shows these four boundaries in the $(\kappa^{2},s^{2})$ parameter space
($s^{2}=0$ and $\kappa^{2}=0$ are the coordinate axes) that are similar to
those in the $(e^{2},s^{2})$ parameter space. However, from eq.(15), we have%

\[
s^{2}=\frac{1}{e^{2}-1}(\kappa^{2}-1),
\]

and thus the constant $e^{2}$ curves are straight lines with slopes
$1/(e^{2}-1)$ that originate from the same point $(\kappa^{2},s^{2})=(1,0)$,
compared to the parallel vertical lines in the $(e^{2},s^{2})$ parameter space.

III. The Parameter Space $(-g_{3},-g_{2})$ (see \ref{Fig.14}).

The reason that we use the coordinates $(-g_{3},-g_{2})$ instead of
$(g_{2},g_{3})$ is that they can be more easily identified with the parameter
spaces $(\kappa^{2},s^{2})$ and $(e^{2},s^{2})$ so that the horizontal axis
$-g_{3}$ increasing to the right corresponds to $\kappa^{2}$ (or $e^{2}$) axis
increasing to the right, and the vertical axis $-g_{2}$ increasing vertically
up corresponds to the $s^{2}$ axis increasing upward.

There are four relevant boundaries for our universal map in the parameter
space $(-g_{3},-g_{2})$ for $-\infty\leq-g_{3}\leq+\infty$, $-g_{2}\geq-1/12$:
(1) the bottom boundary $-g_{2}=-1/12$, (2) the left boundary of Region I
which we call $s=s_{1}^{\prime}$, (3) the right boundary of Region I which we
call $s=s_{1}$, and (4) the left boundary of Region II, $g_{3}=g_{3}^{(2)}$.

(1) $\Delta=0$, $-g_{2}=-1/12$ and $k^{2}=0$. This is the lower boundary for
Region I and for Region II.

(2) $\Delta=0$, $k^{2}=0$, and $-g_{3}=-g_{3}^{(0)}$ where $g_{3}^{(0)}$ is
given by%

\begin{equation}
g_{3}^{(0)}=\left(  \frac{g_{2}}{3}\right)  ^{3/2},
\end{equation}

for $-1/216\leq-g_{3}\leq0$, $-1/12\leq-g_{2}\leq0$. We shall refer to
$-g_{3}=-g_{3}^{(0)}$ as the left boundary of Region I, and is equivalent to
$s^{2}=s_{1}^{\prime2}$ for the $(\kappa^{2},s^{2})$ or $(e^{2},s^{2})$
parameter space.

(3) $\Delta=0$, $k^{2}=1$, and $-g_{3}=-g_{3}^{(1)}$ where $g_{3}^{(1)}$ is
given by%

\begin{equation}
g_{3}^{(1)}=-\left(  \frac{g_{2}}{3}\right)  ^{3/2},
\end{equation}

for $0\leq-g_{3}\leq1/216$, $-1/12\leq-g_{2}\leq0$. We shall refer to
$-g_{3}=-g_{3}^{(1)}$ as the right boundary of Region I, and is equivalent to
$s^{2}=s_{1}^{2}$ for the $(\kappa^{2},s^{2})$ or $(e^{2},s^{2})$ parameter space.

The origin $(-g_{3},-g_{2})=(0,0)$ is the vertex point $V$ where ISCO occurs
and where all $k^{2}=const.$ curves meet. The vertical axis $g_{3}=0$
coincides with $k^{2}=0.5$; the horizontal axis to the right of the origin
$(-g_{3}>0)$ coincides with $k^{2}=0.933012$, and the horizontal axis to the
left of the origin $(-g_{3}<0)$ coincides with $k^{2}=0.0669872$ (see Section
7, 1d).

(4) $g_{3}=g_{3}^{(2)}$ where $g_{3}^{(2)}$ is given by%

\begin{equation}
g_{3}^{(2)}=\frac{1}{2\cdot3^{3}}-\frac{1}{2\cdot3}g_{2}.
\end{equation}

This is the left boundary for Region II that is equivalent to $\kappa^{2}=0$
or $s^{2}=s_{2}^{2}=1/(1-e^{2})$ for the $(\kappa^{2},s^{2})$ or $(e^{2}%
,s^{2})$ parameter space.

The triangular region bounded by the three curves (1), (2) and (3) given above
is Region I. The rest of the region above the straight lines given by (1) and
(4) is Region II.

While the boundaries in the parameter space appear very symmetrical in the
parameter space $(-g_{3},-g_{2})$ compared to those in the parameter space
$(e^{2},s^{2})$, the $e^{2}=const.$ curves are parabolas represented by the equation%

\begin{equation}
g_{3}=-\frac{1+3e^{2}}{2^{6}\cdot3^{3}}+\frac{1+e^{2}}{2^{3}\cdot3}g_{2}%
+\frac{1-e^{2}}{2^{2}}g_{2}^{2}%
\end{equation}

with the exception that $e^{2}=1$ is a straight line. These curves all
originate from the point $(-g_{3},-g_{2})=(-1/216,-1/12)$.

The results which are described in the $(e^{2},s^{2})$ [or $(e^{2},s)$]
parameter space can be easily translated into the $(\kappa^{2},s^{2})$ and
$(-g_{3},-g_{2})$ parameter spaces using eqs.(11), (14), (15) and the
correspondences given in this section.

\section{Orbital Trajectories on the Boundaries}

The orbital trajectories on the boundaries of Regions I and II are important
special cases. Consider the orbital trajectories given by the three solutions
given by eqs.(18), (35) and (43) separately.

Solution (A1)

The left boundary $k^{2}=0$ given by $s_{1}^{\prime2}$ corresponds to
$\theta=0$ in eq.(22) and $\varepsilon=0$ in eq.(26), and we have
$e_{1}=(1-12s_{1}^{\prime2})^{1/2}/6=-2e_{2}=-2e_{3}$. The curve for $k^{2}=0
$ and thus $s_{1}^{\prime2}$ extends from $e^{2}=-1/3,$ $s^{2}=1/12$ to
$e^{2}=0,$ $s^{2}=0$. The orbital trajectories given by eq.(18) on this
boundary curve are circular orbits the radii $q_{c}$ of which are given by%

\begin{equation}
\frac{1}{q_{c}}=\frac{1}{3}\left\{  1-\left(  1-12s_{1}^{\prime2}\right)
^{1/2}\right\}  =\frac{2\left\{  1-3e^{2}-(1+3e^{2})^{1/2}\right\}
}{9(1-e^{2})},
\end{equation}

or%

\begin{equation}
q_{c}=\frac{1}{4s_{1}^{\prime2}}\left\{  1+\left(  1-12s_{1}^{\prime2}\right)
^{1/2}\right\}  =\frac{1}{2e^{2}}\left\{  -1+3e^{2}-(1+3e^{2})^{1/2}\right\}
.
\end{equation}

For the values of $s_{1}^{\prime2}$ that range from $0$ to $1/12$ (and
$-1/3\leq e^{2}\leq0$), the range of radii for these stable circular orbits is%

\begin{equation}
3\leq q_{c}\leq\infty,
\end{equation}

or%

\[
\frac{6GM}{c^{2}}\leq r_{c}\leq\infty.
\]

The circular orbit with the radius $q_{c}=3$ (or $r_{c}=6GM/c^{2}$) at the
vertex $V$ represented by the coordinates $(e^{2},s^{2})=(-1/3,1/12)$ or
$(\kappa^{2},s^{2})=(8/9,1/12)$ is called the innermost stable circular orbit (ISCO).

Because every orbital trajectory given by Solution (A1) precesses according to
eq.(27), these circular orbits also precess with a precession angle%

\begin{equation}
\Delta\phi=2\pi\left\{  (1-12s_{1}^{\prime2})^{-1/4}-1\right\}
\end{equation}

for $0<s_{1}^{\prime2}\leq1/12$ even though this precession angle is not
observable unless the orbit is slightly perturbed from its circular path [12].
In terms of the radius $r_{c}$ of the circular orbit, the precession angle is
given by%

\begin{equation}
\Delta\phi=2\pi\left\{  \left(  1-\frac{3\alpha}{r_{c}}\right)  ^{-1/2}%
-1\right\}  .
\end{equation}

Thus excluding the case of zero gravitational field $s_{1}^{\prime}=0$ for
which the circular orbits have an infinite radius, all (perfectly) circular
orbits have a non-zero precession angle. The precession angle for the
innermost stable circular orbit (ISCO), which occurs at the vertex point $V$
is infinite.

The upper boundary $k^{2}=1$ given by $s_{1}^{2}$ corresponds to $\theta=\pi$
in eq.(22), and we have $e_{1}=e_{2}=\left(  1-12s_{1}^{2}\right)
^{1/2}/12=-e_{3}/2$. Substituting these into eqs.(23), (24) and (26) gives a
useful relationship $\left(  1-12s_{1}^{2}\right)  ^{1/2}=2\varepsilon
/(3+\varepsilon)$. The curve for $k^{2}=1$ given by $s_{1}^{2}$ extends from
$e^{2}=-1/3,$ $s^{2}=1/12$ to $e^{2}=\infty,$ $s^{2}=0$. Equation (18) on this
boundary becomes%

\begin{equation}
\frac{1}{q}=\frac{1}{3}+\frac{1}{3}\sqrt{1-12s_{1}^{2}}\left(  \frac{1-5\sec
h(2\gamma\phi)}{1+\sec h(2\gamma\phi)}\right)  ,
\end{equation}

where%

\begin{equation}
\gamma=\frac{1}{2}\left(  1-12s_{1}^{2}\right)  ^{1/4}.
\end{equation}

For $-1/3\leq e^{2}<1$ ($1/12\geq s_{1}^{2}>1/16$), the particle starts from
an initial position $q_{\max}$ at $\phi=0$ given by%

\begin{equation}
\frac{1}{q_{\max}}=\frac{1}{3}-\frac{2}{3}\left(  1-12s_{1}^{2}\right)
^{1/2}=\frac{1-\varepsilon}{3+\varepsilon}%
\end{equation}

and ends up at $\phi=\infty$ circling the star or black hole with a radius
that asymptotically approaches $q_{\min}$ given by%

\begin{equation}
\frac{1}{q_{\min}}=\frac{1}{3}+\frac{1}{3}\left(  1-12s_{1}^{2}\right)
^{1/2}=\frac{1+\varepsilon}{3+\varepsilon}.
\end{equation}

For $e^{2}=1$, the particle starts at infinity at the polar angle $\phi=0$,
and for $e^{2}>1$ ($s_{1}^{2}<1/16$), the particle starts from infinity at the
polar angle $\phi=\Psi_{1}$ given from eq.(32) by%

\begin{equation}
\tanh^{2}(\gamma\Psi_{1})=\frac{2}{3}-\frac{1}{3}\left(  1-12s_{1}^{2}\right)
^{-1/2},
\end{equation}

and ends up at $\phi=\infty$ circling the star or black hole with a radius
that asymptotically approaches $q_{\min}$ given by eq.(63). The angle
$\Psi_{1}$ can range from $0$ for $e^{2}=1$ (parabolic-type orbit) to
$2\tanh^{-1}(1/\sqrt{3})=1.31696$ $=75.456^{\circ}$ for very large $e^{2}$.

We refer to all these orbits given by eq.(62) as the asymptotic orbits.

Examples of asymptotic orbits for $e^{2}=0.25$, $1$ and $4$ are shown in
\ref{Fig.15}, \ref{Fig.16} and \ref{Fig.17}.

Solution (A2)

Substituting $k^{2}=0$ and $s^{2}=s_{1}^{\prime2}$ into eq.(35) gives the
equation for the orbital trajectories for Solution (A2) on the left boundary
of Region I as%

\begin{equation}
\frac{1}{q}=\frac{1}{3}\left[  1-\left(  1-12s_{1}^{\prime2}\right)
^{1/2}\right]  +\frac{\left(  1-12s_{1}^{\prime2}\right)  ^{1/2}}{\cos
^{2}(\gamma\phi)},
\end{equation}

where%

\begin{equation}
\gamma=\frac{1}{2}\left(  1-12s_{1}^{\prime2}\right)  ^{1/4}.
\end{equation}

Thus on the left boundary of Region I, the terminating orbit (inside Region I)
given by Solution (A2) remains a terminating orbit. The particle, starting
from the polar angle $\phi=0$ at a distance $q_{1}$ given by $1/q_{1}%
=1/3+2\left(  1-12s_{1}^{\prime2}\right)  ^{1/2}/3$ plunges into the center of
the black hole when $\phi=\phi_{1}=\pi/\left(  1-12s_{1}^{\prime2}\right)
^{1/4}$.

On the other hand, substituting $k^{2}=1$ and $s^{2}=s_{1}^{2}$ into eq.(35)
leads to $q=$ $const.$ independent of $\phi$ (because $e_{1}=e_{2}$ and the
dependence on $\phi$ cancels out). Thus on the upper boundary of Region I the
terminating orbit (inside Region I) given by solution (A2) becomes an unstable
circular orbit of radius $q_{u}$\ given by%

\begin{equation}
\frac{1}{q_{u}}=\frac{1}{3}+\frac{1}{3}\left(  1-12s_{1}^{2}\right)
^{1/2}=\frac{2\left\{  1-3e^{2}+(1+3e^{2})^{1/2}\right\}  }{9(1-e^{2})}%
=\frac{1+\varepsilon}{3+\varepsilon},
\end{equation}

or%

\begin{equation}
q_{u}=\frac{1}{4s_{1}^{2}}\left\{  1-\left(  1-12s_{1}^{2}\right)
^{1/2}\right\}  =\frac{1}{2e^{2}}\left\{  -1+3e^{2}+(1+3e^{2})^{1/2}\right\}
,
\end{equation}

and%

\begin{equation}
\varepsilon=\frac{3-q_{u}}{q_{u}-1}.
\end{equation}

It is an unstable circle because its radius $q_{u}$ is identical to the radius
$q_{\min}$ of the asymptotic circle given by eq.(65). For the values of
$s^{2}$ that have a range $0\leq s_{1}^{2}\leq1/12$ $($ and $\infty\geq
e^{2}\geq-1/3)$, we have the range of values for the radii of these unstable
circular orbits along the upper boundary $s^{2}=s_{1}^{2}$ of Region I given by%

\begin{equation}
1.5\leq q_{u}<3,
\end{equation}

or%

\[
\frac{3GM}{c^{2}}\leq r_{u}<\frac{6GM}{c^{2}}.
\]

Notice the similarity between the expressions given by eqs.(69) and (70) for
the radii of the unstable circles on $s_{1}$ and those for the radii of the
stable circles on $s_{1}^{\prime}$ given by eqs.(57) and (58). Unlike the
stable circles for which $\varepsilon=0$, these unstable circular orbits do
not have $\varepsilon=0$ except when $q_{u}\rightarrow3$ which is the radius
of the innermost stable circular orbit (ISCO) at the vertex $V$.

Solution (B)

Substituting $k^{2}=0$ and $s^{2}=s_{1}^{\prime2}$ into eq.(43) gives the
equation for the terminating orbits for Solution (B) on the left boundary of
Region I. This equation coincides with eq.(67) which is the expression that
describes the terminating orbits for Solution (A2) on the same boundary, i.e.
with $q_{2}$ of eq.(44) becoming $q_{1}$ of eq.(36) and with $a=e_{1}%
=(1-12s_{1}^{\prime2})/6$ on the boundary.

Substituting $k^{2}=1$ and $s^{2}=s_{1}^{2}$ into eq.(43), the terminating
orbits for Solution (B) on the lower boundary of Region II (which is the upper
boundary of Region I) become the same asymptotic orbits given by eq.(62) which
describes the non-terminating orbits of Solution (A1) on the same boundary.
That is, on the boundary $s^{2}=s_{1}^{2}$, $q_{2}$ of eq.(44) becomes
$q_{\max}$ of eq.(64) and with $a=e_{3}=-(1-12s_{1}^{2})/6$. The value of
$q_{\min}$ is given by eq.(65) for $-1/3\leq e^{2}<1$. When $e^{2}\geq1$ the
particle starts from infinity ($q_{2}=\infty$) at a polar angle $\phi=\Psi
_{1}$ that is given by eq.(66).

Thus for the orbits on the boundaries given by Solutions (A1), (A2) and (B),
we have (1) on $s_{1}^{\prime}$, the stable circular orbits of radii $q_{c}$
given by eq.(58) and the terminating orbits given by eq.(67), and (2) on
$s_{1}$, the unstable circular orbits of radii $q_{u}$ given by eq.(70) and
the asymptotic orbits given by eq.(62). Even though these results reproduce
the conclusions obtained from the effective potential calculation, they place
the orbits and their corresponding analytic expressions clearly on the
boundary curves $s_{1}^{\prime}$ and $s_{1}$ shown in \ref{Fig.2}.

\section{Trajectory Changes Across the Boundaries}

The left hand side of eq.(7) excluding the $\overset{\cdot}{r}^{2}$term can be
considered to be an effective potential. In a similar manner we define, from
eq.(13), an "effective potential" $W$ to be%

\[
W=-(4U^{3}-g_{2}U-g_{3}).
\]

Although only the condition that $W\leq0$ is physically realizable, it is
useful to see the entire curve $W$ above and below the horizontal $U$ axis and
we shall examine it and its variation in this section.

We want to show the changes in the orbital trajectory in conjunction with the
changes of this effective potential as we consider various coordinate points
on a counterclockwise path in parameter space around the vertex point $V$ in
\ref{Fig.2}. We have chosen a specific path that generally represents the
changes in the trajectories and in the effective potential as boundaries are
crossed. The points where the $W$ curve intersects the $U$ axis are the real
roots of eq.(16). Recall that Solution (A1) for Region I requires $e_{1}\geq
e_{2}\geq U\geq e_{3}$, Solution (A2) for Region I requires $U\geq e_{1}\geq
e_{2}\geq e_{3}$, and Solution (B) for Region II requires $U\geq a$. For $U$
defined by eq.(12), $U\rightarrow+\infty$ implies $r\rightarrow0$.

We start by considering a coordinate point inside Region I for which $e^{2}<1$
that we choose to be $(e^{2},s)=(0.25,0.2)$ (see \ref{Fig.2}). We know from
Section 3 that it allows a precessional elliptic-type orbit given by eq.(18)
shown in \ref{Fig.3}, and a terminating orbit given by eq.(35) shown in Fig.6.
Note that $q_{1}$ is less than $q_{\min}$ and the two possible orbits describe
two separate initial conditions. The corresponding $W$ curve is shown in
\ref{Fig.18}. Noting that $e_{3}>-1/12$ for this example, the bowl-like
"potential" below the $U$ axis between $U=e_{3}$ and $U=e_{2}$ allows an
elliptic-type orbit for a particle between $q=q_{\max}$ and $q=q_{\min}$ given
by eqs. (23) and (24). On the other hand, the down-hill "potential" below the
$U$ axis for $U\geq e_{1}$ allows a terminating orbit for a particle that
starts at $U=e_{1}$ and ends at $U=+\infty$ or from $q=q_{1}$ given by eq.(36)
to $q=0$. Thus the possible orbits shown in \ref{Fig.3} and \ref{Fig.6} are
appropriately matched with the $W$ curve.

Going counterclockwise around the vertex $V$ (see \ref{Fig.2}), we choose the
next coordinate point $(0.25,0.26481)$ which is straight above the previous
point but is on the boundary $s_{1}$ characterized by $k^{2}=1$. This point
allows an asymptotic orbit given by eq.(62) and shown in \ref{Fig.15} with
$q_{\max}$ and $q_{\min}$ given by eqs. (64) and (65), and it also allows a
circular orbit with a radius $q_{u}$ given by eq.(69) with $q_{u}=q_{\min}$.
The corresponding $W$ curve is shown in \ref{Fig.19} which indicates that
$e_{1}=e_{2}$. The bowl-like potential between $U=e_{3}$ and $U=e_{1}=e_{2}$
allows a particle starting from $U=e_{3}$ or $q=q_{\max}$ to circle the black
hole with a radius that asymptotically approaches $U=e_{1}=e_{2}$ or
$q=q_{\min}$. The down-hill potential for $U\geq e_{1}=e_{2}$ allows the
particle to remain (unstably) at $U=e_{1}=e_{2}$ or for it to have a circular
orbit of radius $q=q_{u}$. This circular orbit is clearly unstable because a
small perturbation would make it into a terminating or an asymptotic orbit.

We next choose a point $(0.25,0.3)$ in Region II which is again straight above
the previous point. It allows only a terminating orbit given by eq.(43) and
shown in \ref{Fig.9}. The corresponding $W$ curve is shown in \ref{Fig.20}
with one real root at $U=a$. The transition from Region I to Region II across
$s_{1}$ clearly shows that $e_{3}$ becomes $a$. Noting that $a>-1/12$, the
down-hill potential for $U\geq a$ allows a particle that starts at a finite
distance from the black hole at $U=a$ or $q=q_{2}$ given by eq.(44) and ends
at $U=+\infty$ or $q=0$. The asymptotic orbit shown in \ref{Fig.15} becomes a
terminating orbit shown in \ref{Fig.9} as we go from the boundary $s_{1}$ to
Region II. The unstable circular orbit on $s_{1}$ also becomes a terminating
orbit (as the initial distance becomes less than $q_{u}$) which is part of the
same terminating orbit in Region II arising from the asymptotic orbit on the
boundary (see the transition of the $W$ curve below the horizontal axis from
\ref{Fig.19} to \ref{Fig.20}). These examples show how Solutions (A1) and (A2)
merge into Solution (B) across the boundary $s_{1}$.

We choose the next coordinate point $(-0.5,0.3)$ in Region II that is
horizontally left to the previous point (see \ref{Fig.2}). The $W$ curve is
shown in \ref{Fig.21} and indicates the one real root $U=a$ has increased in
the positive direction, i.e. the initial distance $q_{2}$ of the particle from
the black hole has decreased compared to the previous point. The terminating
orbit is similar to \ref{Fig.9} except that the initial distance is smaller.
If the chosen coordinate point were to be further left in \ref{Fig.2} so that
it fell on the left upper boundary $s_{2}$ of Region II (see \ref{Fig.1}), $a$
would be equal to $1/6$ and the initial position of the particle would be on
the Schwarzschild horizon corresponding to $q_{2}=1$.

We next select the coordinate point $(-0.25,0.2)$ in Region II below the
previous point but close to the left boundary $s_{1}^{\prime}$ of Region I
(see \ref{Fig.2}). The terminating orbit is shown in Fig.12. The $W$ curve is
shown in Fig.22 where the real root has become more positive while the left
part of the curve is slightly above and almost touches the $U$ axis. If the
chosen coordinate point were to be further down in Fig.2 so that it fell on
the lower boundary $s=0$ of Region II (see Figs. 1 and 2), $a$ would be equal
to $1/6$ and the initial position of the particle would be on the
Schwarzschild horizon corresponding to $q_{2}=1$.

We now go to the right and choose the next coordinate point $(e^{2}%
,s)=(-0.099275,0.2)$ exactly on the boundary $s_{1}^{\prime}$ characterized by
$k^{2}=0$ (see \ref{Fig.2}). The $W$ curve is shown in \ref{Fig.23}. It
touches the $U$ axis at $U=e_{2}=e_{3}$ and allows a circular orbit with a
radius $q_{c}$ given by eq.(57). The down-hill potential given by $U\geq
e_{1}$ gives a terminating orbit with $q_{1}$ given by eq.(36). The circular
orbit is a stable one as a small perturbation that could change \ref{Fig.23}
into one similar to \ref{Fig.18} clearly confines the perturbation to be
bounded, and the possible terminating orbit solution is well separated from it
as $e_{1}$ is greater than $e_{2}=e_{3}$, or $q_{1}$ is less than $q_{c}$. The
terminating orbit for this point is similar to the terminating orbit of the
previous point except for a small difference in $q_{1}$. The transition of the
terminating orbit from Region II to Region I is a smooth and gradual one (with
$a$ becoming $e_{1}$). However, when we cross the boundary $s_{1}^{\prime}$, a
stable circular orbit becomes possible on the boundary and it then becomes an
elliptical orbit as the coordinate point moves into the interior of Region I.
As we noted previously [see eqs. (60) and (61)], the circular orbits, like the
elliptical orbits, precess.

We now return to our starting point $(0.25,0.2)$ for which the $W$ curve is
shown in \ref{Fig.18}, and the possible trajectories are shown by \ref{Fig.3}
and \ref{Fig.6}.

In the plots for the "potential" curve $W$, it is useful to mark the $U$ axis
where $U\leq-1/12$ by a thick line for the following identification. As we
have already noted, Solution (A1) gives orbits that are elliptic-type,
parabolic-type and hyperbolic-type for $e_{3}>,=,$ and $<-1/12$ respectively,
and Solution (B) gives terminating orbits for a particle that is initially at
a distance $q_{2}$ from the black hole that is finite, infinite at $\phi=0$,
and infinite at $\Psi_{2}>0$ for $a>,=$ and $<-1/12$ respectively. The
examples which illustrate the crossing of the $s_{1}$ boundary that are shown
by \ref{Fig.3}, \ref{Fig.15} and \ref{Fig.9} for the orbital trajectories, and
by \ref{Fig.18}, \ref{Fig.19} and \ref{Fig.20} for the $W$ curves, are for the
cases $e_{3}$ and $a$ greater than $-1/12$ that are representative of
$e^{2}<1$. If we chose to cross the $s_{1}$ boundary along a path on $e^{2}%
=1$, the $W$ curve would have $e_{3}=-1/12$ or $a=-1/12$, and the
corresponding representative trajectories would be like those shown in Figs.
4, 16 and 10; and if we chose to cross the $s_{1}$ boundary along the path on
$e^{2}>1$, the $W$ curve would have $e_{3}<-1/12$ or $a<-1/12$ (i.e. the left
side of the $W$ curve would cut the thick line representing $U\leq-1/12$) and
the corresponding representative trajectories would be like those shown in
\ref{Fig.5}, \ref{Fig.17} and \ref{Fig.11}. As for the terminating orbits in
Region I, they are not affected much by the values of $e^{2}$, as are shown by
\ref{Fig.6}, \ref{Fig.7} and \ref{Fig.8}, and when we cross the boundary
$s_{1}$, they all become unstable circular orbits on the boundary and then
become parts of the terminating orbits in Region II.

We see that the shape of the $W$ curve and whether or not it crosses the thick
line that represents $U\leq-1/12$ can be used to foresee the possible type of
orbits and trajectories and to relate them to the approximate location of the
coordinate point on the map.

The $W$ curve at the vertex $V$ located at $(e^{2},s)=(-1/3,1/\sqrt{12})$ is
given by $W=-4U^{3}$ and it has three real roots coinciding at $e_{1}%
=e_{2}=e_{3}=0$.

\section{Analytic Expressions for Special Cases}

The exact analytic orbit equations (18), (35) and (43), and their
corresponding equations on the boundaries, eqs.(57), (62), (67) and (69), give
us a clear picture of what these orbits look like and what their coordinates
are on a map characterized by $(e^{2},s^{2})$ and with boundaries of Regions I
and II characterized by eqs.(48), (50), and (52). While generally we are
required to start the numerical computation of the orbits with eq.(21) or
(39), there are a number of special values of squared modulus $k^{2}$ and
squared energy parameter $e^{2}$ that allow various exact relations that are
very interesting and useful. We have exploited the theory of Jacobian elliptic
functions to come up with many of these relationships. A few of these analytic
relations have been presented in a previous paper [2], but it is convenient to
collect all of them in one place and we present and summarize them in the following.

(I) Curves of $k^{2}=const.$ and the Orbit Equations on These Curves.

(1a) For $k^{2}=0$, Sections 4 and 5 give many special results for this
special $k^{2}$ value. In particular, we have eq.(48) [or (49)] that gives the
coordinates $(e^{2},s_{1}^{\prime2})$ of this curve. The radii of the stable
circular orbits from Solution (A1) are given by eqs.(57)-(59) and the
precession angle is given by eq.(60) or (61). The terminating orbit equation
from Solution (A2) is given by eq.(67).

(1b) For $k^{2}=1/2$, we have a simple analytic expression that relates the
values of $s^{2}$ and $e^{2}$ on it:%

\begin{equation}
s^{2}=\frac{1}{6(1-e^{2})}\left(  1-\sqrt{\frac{1+2e^{2}}{3}}\right)
\end{equation}

for $e^{2}\geq-1/2$, $0\leq s^{2}\leq1/9$ (in Region I and part of Region II)
and for which $s^{2}=1/18$ for $e^{2}=1$; and a simple expression%

\begin{equation}
s^{2}=\frac{1}{6(1-e^{2})}\left(  1+\sqrt{\frac{1+2e^{2}}{3}}\right)  ,
\end{equation}

for $1\geq e^{2}\geq-1/2$, $s^{2}\geq1/9$ (in Region II). Equations (73) and
(74) can be inverted to give%

\begin{equation}
e^{2}=1-\frac{18s^{2}-1}{54s^{4}}.
\end{equation}

Equations (73) and (74) or eq.(75) are shown as a single continuous curve
marked by $k^{2}=0.5$ in Fig.2. In the parameter space $(\kappa^{2},s^{2})$,
the $k^{2}=0.5$ curve is given by%

\[
s^{2}=\frac{1}{18(3\kappa^{2}-2)}.
\]

In the parameter space $(-g_{3},-g_{2})$, the $k^{2}=0.5$ curve is simply the
vertical axis $g_{3}=0$.

By noting that in Region I, $e_{1}=-e_{3}=\sqrt{g_{2}}/2$, $e_{2}=0$,
$\gamma=\sqrt[4]{g_{2}}$, the orbit equation (18) becomes%

\begin{equation}
\frac{1}{q}=\frac{1}{3}-2\sqrt{g_{2}}cn^{2}(\gamma\phi,1/\sqrt{2}),
\end{equation}

where the values of $s^{2}$ in the range $1/12\geq s^{2}\geq0$ for $g_{2}$ are
given by eq.(73), and $q_{\min}=3$ independent of $e^{2}$. For the
elliptic-type orbits corresponding to $-1/3\leq e^{2}<1$, the initial distance
$q_{\max}$ at $\phi=0$ of the particle from the star or black hole is given by%

\begin{equation}
\frac{1}{q_{\max}}=\frac{1}{3}-2\sqrt{g_{2}}.
\end{equation}

For the parabolic-type orbit corresponding to $e^{2}=1$, $q_{\max}=\infty$ at
$\phi=0$. For the hyperbolic-type orbit corresponding to $e^{2}>1$, the
particle approaches from infinity along an incoming asymptote at an angle
$\Psi_{1}$ given by eq.(29) to the horizontal axis, goes around the star or
black hole, and leaves along an outgoing asymptote at an angle $\Psi_{2}$, where%

\begin{equation}
\Psi_{1}=\gamma^{-1}sn^{-1}\left(  \sqrt{1-\frac{1}{6\sqrt{g_{2}}}},\frac
{1}{\sqrt{2}}\right)  ,
\end{equation}

and%

\begin{equation}
\Psi_{2}=\frac{2K(1/\sqrt{2})}{\sqrt[4]{g_{2}}}-\Psi_{1},
\end{equation}

where $K(1/\sqrt{2})=1.85407$.

The terminating orbit equation (35) in Region I given by Solution (A2) becomes%

\begin{equation}
\frac{1}{q}=\frac{1}{3}+\frac{2\sqrt{g_{2}}}{cn^{2}(\gamma\phi,1/\sqrt{2})}.
\end{equation}

The particle, starting from the initial distance $q_{1}$ given by%

\begin{equation}
\frac{1}{q_{1}}=\frac{1}{3}+2\sqrt{g_{2}}%
\end{equation}

plunges into the center of the black hole when its polar angle $\phi_{1}$ is
given by%

\begin{equation}
\phi_{1}=\frac{K(1/\sqrt{2})}{\sqrt[4]{g_{2}}}.
\end{equation}

For the terminating equation (43) in Region II given by Solution (B), the
values of $s^{2}$ are given by eq.(73) for the range $-1/3\leq e^{2}\leq-1/2
$, $1/12\leq s^{2}\leq1/9$, and are given by eq.(74) for the range $e^{2}%
\geq-1/2$, $s^{2}\geq1/9$. By noting from eqs.(39)-(42) that $A=-B=\sqrt
{12s^{2}-1}/12$, $a=0$, $\gamma=[(12s^{2}-1)/3]^{1/4}/2$, the orbit equation becomes%

\begin{equation}
\frac{1}{q}=\frac{1}{3}+4\gamma^{2}\left(  \frac{1-cn(2\gamma\phi,1/\sqrt{2}%
)}{1-cn(2\gamma\phi,1/\sqrt{2})}\right)  .
\end{equation}

The $k^{2}=1/2$ curve in Region II is confined to $e^{2}<1$ as $s^{2}$
increases from $1/12$ to $\infty$, and thus the initial distance $q_{2}$ of
the particle from the black hole is always bounded. Indeed $q_{2}=3$
independent of $e^{2}$. The particle, starting from this distance from the
black hole, plunges into the center of the black hole when its polar angle
$\phi_{2}$ is given by%

\begin{equation}
\phi_{2}=\frac{K(1/\sqrt{2})}{\gamma}.
\end{equation}

The above results supplement the results given in Appendix C of ref.[2] that
have a gap in the range $-1/3\leq e^{2}<0$.

(1c) For $k^{2}=1$, Sections 4 and 5 give many special results for this
special $k^{2}$ value. In particular, we have eq.(50) [or (51)] that gives the
coordinates $(e^{2},s_{1}^{2})$ of this curve. The \ non-terminating orbit
equation given by Solution (A1) becomes an asymptotic orbit eq.(62). The
relevant quantities for $-1/3\leq e^{2}\leq1$ are given by eqs.(64) and (65),
and for $e^{2}>1$ are given by eq.(66) and the description after it. The
terminating orbit given by Solution (A2) becomes an unstable circular orbit
with the radius given by eq.(69) or (70) that has a range given by eq.(72).
The terminating orbit in Region II given by Solution (B) becomes an asymptotic
orbit given by eq.(62).

(1d) $k^{2}=1/2\pm\sqrt{3}/4$ are special values because they correspond to
$s^{2}=1/12$ or $g_{2}=0$. Substituting $s^{2}=1/12$ into $g_{3}$ and $k^{2}$
given by eqs.(14) and (42) show the following: $k^{2}=1/2+\sqrt{3}/4=0.933012$
for all values of $g_{3}>0$ which holds when $\kappa^{2}<8/9$ or when
$e^{2}<-1/3$; and $k^{2}=1/2-\sqrt{3}/4=0.0669872$ for all values of $g_{3}<0$
which holds when $\kappa^{2}>8/9$ or when $e^{2}>-1/3$. Thus the $k^{2}%
=1/2\pm\sqrt{3}/4$ curves are represented by a single horizontal straight line
($s^{2}=1/12$ or $g_{2}=0$) and passing through the vertex point $V$ in all
our three parameter spaces shown in \ref{Fig.2}, \ref{Fig.13} and \ref{Fig.14}.

(II) Geometric Eccentricity $\varepsilon$ Defined by Eq.(25) for $-1/3\leq$
$e^{2}\leq1$ in Region I.

(2a) $\varepsilon\rightarrow e$ as $s\rightarrow0$ for $0\leq\varepsilon<1;$
and $\varepsilon=1$ when $e=1$, i.e. $\varepsilon=e=1$ for all values of $s$
(see ref.[2]).

(2b) $\varepsilon=0$ for $k^{2}=0$.

(2c) For $k^{2}=1/2$, the following equations relate the values of $s^{2}$ and
$e^{2}$ on the curve given by eq.(73) to the value of $\varepsilon$ defined by
eq.(25) for non-terminating orbits in Region I in the range $-1/3\leq
e^{2}\leq1$.%

\begin{equation}
e^{2}=1-\frac{12(1+\varepsilon)^{2}(1-\varepsilon)(1+3\varepsilon
)}{(3+6\varepsilon-\varepsilon^{2})^{2}},
\end{equation}

or%

\begin{equation}
\varepsilon=\frac{-3(1+e^{2})+2\sqrt{3(1+2e^{2})}+2\sqrt{1-e^{2}}%
\sqrt{-3(1+e^{2})+2\sqrt{3(1+2e^{2})}}}{7-e^{2}-2\sqrt{3(1+2e^{2})}},
\end{equation}

with $\varepsilon=1$ when $e=1$; and%

\begin{equation}
s^{2}=\frac{3+6\varepsilon-\varepsilon^{2}}{36(1+\varepsilon)^{2}},
\end{equation}

or%

\begin{equation}
\varepsilon=\frac{3(1-12s^{2})+2\sqrt{3(1-12s^{2})}}{1+36s^{2}}.
\end{equation}

(2d) For $k^{2}=1$, the following equations relate the values of $s^{2}$ and
$e^{2}$ on the curve given by eq.(50) or (51) to the value of $\varepsilon$
defined by eq.(25) for non-terminating orbits in Region I in the range
$-1/3\leq e^{2}\leq1$.%

\begin{equation}
e^{2}=\frac{(1+\varepsilon)(-3+5\varepsilon)}{(3-\varepsilon)^{2}},
\end{equation}

or%

\begin{equation}
\varepsilon=\frac{-(1+3e^{2})+4\sqrt{1+3e^{2}}}{5-e^{2}},
\end{equation}

and%

\begin{equation}
s^{2}=\frac{(3-\varepsilon)(1+\varepsilon)}{4(3+\varepsilon)^{2}},
\end{equation}

or%

\begin{equation}
\varepsilon=\frac{1-12s^{2}+2\sqrt{1-12s^{2}}}{1+4s^{2}}.
\end{equation}

(III) The Special Case of $e^{2}=1$.

(3a) For $0\leq s^{2}\leq1/16$ in Region I. The $k^{2}$ and $\gamma$ given by
eqs.(20) and (19) for the non-terminating and terminating orbit equations (18)
and (35) can be expressed simply in terms of $s^{2}$ by the following equations:%

\begin{equation}
s^{2}=\frac{k^{2}}{4(1+k^{2})^{2}},
\end{equation}

or%

\begin{equation}
k^{2}=\frac{1-8s^{2}-\sqrt{1-16s^{2}}}{8s^{2}},
\end{equation}

\begin{equation}
\gamma^{2}=\frac{1}{4(1+k^{2})}=\frac{2s^{2}}{1-\sqrt{1-16s^{2}}}.
\end{equation}

For the non-terminating orbits, $q_{\max}=\infty$, and%

\begin{equation}
q_{\min}=\frac{1+k^{2}}{k^{2}}.
\end{equation}

For the terminating orbits, the initial distance $q_{1}$ of the particle from
the black hole is%

\begin{equation}
q_{1}=1+k^{2},
\end{equation}

and the particle plunges into the center of the black hole, for $0\leq
k^{2}<1$, when its polar angle $\phi$ is equal to $\phi_{1}$ given by eq.(37).
For the particular case of $k^{2}=1$, the terminating orbit becomes a circle
of radius $q_{c}=q_{1}=2$.

(3b) For $s^{2}>1/16$ in Region II. The $k^{2}$ and $\gamma$ given by eqs.(42)
and (41) have the following simple expressions:%

\begin{equation}
k^{2}=\frac{1}{2}+\frac{1}{8s},
\end{equation}

\begin{equation}
\gamma^{2}=\frac{s}{2}.
\end{equation}

The particle, coming from infinity at $\phi=0$, plunges into the center of the
black hole at a polar angle $\phi=\phi_{2}$ given by eq.(45).

(IV) The Special Case of $e^{2}=-1/3.$ The $\gamma$, $k^{2}$ and $q_{2}$ given
by eqs.(41), (42) and (44) for the terminating orbits in Region II have the
following expressions:%

\begin{equation}
\gamma^{2}=\frac{1}{4\sqrt{3}}\left\{  (1-12s^{2})^{2/3}[1+(1-12s^{2}%
)^{1/3}+(1-12s^{2})^{2/3}]\right\}  ^{1/2},
\end{equation}

\begin{equation}
k^{2}=\frac{1}{2}-\frac{\sqrt{3}}{4}\frac{(1-12s^{2})^{1/3}}{\left\vert
(1-12s^{2})^{1/3}\right\vert }\frac{1+(1-12s^{2})^{1/3}}{[1+(1-12s^{2}%
)^{1/3}+(1-12s^{2})^{2/3}]^{1/2}},
\end{equation}

and%

\begin{equation}
\frac{1}{q_{2}}=\frac{1}{3}\left\{  1+(1-12s^{2})^{1/3}+(1-12s^{2}%
)^{2/3}\right\}  .
\end{equation}

These expressions for $\gamma^{2}$, $k^{2}$ and $q_{2}$ follow by substituting
into eqs. (39)-(42) and (44) $\Delta=s^{4}(1-12s^{2})^{2}/(2^{4}\cdot3)$ for
$e^{2}=-1/3$. The particle, initially at a distance $q_{2}$ from the black
hole, terminates its orbit at the center of the black hole at a polar angle
$\phi=\phi_{2}$ given by eq.(45).

(V) The Maximum Value of $q_{2}$ \ In Region II For A Given Value of $e^{2}<0
$.

For any given value of $e^{2}\leq-1/3$, as we increase $s^{2}$ from $0$ to
$\infty$, $q_{2}$ in Region II given by eq.(44) increases from $1$ at
$s^{2}=0$ to a maximum value given by%

\begin{equation}
q_{2\max}=\frac{1-e^{2}}{-e^{2}}%
\end{equation}

which occurs at a value of $s^{2}$ given by%

\begin{equation}
s^{2}=\frac{-e^{2}}{2(1-e^{2})^{2}},
\end{equation}

and as $s^{2}$ increases beyond this value, $q_{2}$ decreases and becomes
equal to $1$ when $s^{2}=s_{2}^{2}$ given by eq.(52), and $q_{2}$ continues to
decrease to $0$ as $s^{2}$ increases to $\infty$. The same expressions for
$q_{2\max}$ and the value of $s^{2}$ where $q_{2\max}$ occurs apply for
$-1/3<e^{2}<0$ except that we need to exclude the values of $s_{1}^{\prime
2}\leq s^{2}\leq s_{1}^{2}$ in Region I for which $q_{2}$ is not defined. On
the boundary of Region I and of II, the value of $q_{2}$ is given by
$1/q_{2}=1/3+4a$, where $a=(1-12s_{1}^{\prime2})/6$ on $s_{1}^{\prime2}$ and
$a=-(1-12s_{1}^{2})/6$ on $s_{1}^{2}$.

Equations (103) and (104) can be proved as follows. We note that $q_{2}$ is
given by eq.(44) and that $a$ satisfies eq.(16), i.e. $4a^{3}-g_{2}a-g_{3}=0$
from which setting $da/d(s^{2})=0$ gives $a=-1/12+(1-e^{2})s^{2}/2$.
Substituting this value of $a$ into the cubic equation satisfied by $a$ gives
the expression for $s^{2}$ given by eq.(104) which in turn gives a simple
expression for $a$ and the simple expression for $q_{2\max}$ given by eq.(103).

The $\gamma^{2}$ and $k^{2}$ for the $s^{2}$ given by eq.(104) where
$q_{2}=q_{2\max}$ can be obtained from eqs.(41) and (42) and have the
following expressions:%

\begin{equation}
\gamma^{2}=\frac{-e^{2}}{4(1-e^{2})},
\end{equation}

and%

\begin{equation}
k^{2}=\frac{1}{2}+\frac{1+2e^{2}}{4(-e^{2})},
\end{equation}

from which the polar angle $\phi_{2}$ with which the particle plunges into the
center of the black hole can be calculated from eq.(45).

While the descriptions of the orbital trajectories in Regions I and II already
provide a clear picture of what a particle's orbit would look like in
different parts of the map (see \ref{Fig.1} and \ref{Fig.2}), the various
analytic formulas given in this section can be used to provide quantitative
checks of the parameters associated with the orbits at various specific
coordinate points of the map. We mention three examples: (i) the values of
$k^{2}$ given by eq.(94) along $e^{2}=1$ in Region I (from $s^{2}=0$ to
$1/16$), and those given by eq.(98) in Region II (from $s^{2}=1/16$ to
$\infty$) provide useful checks on how the various $k^{2}=const.$ curves
radiating from the special vertex point $V$ cross the $e^{2}=1$ line, (ii) as
can be checked from eqs.(103) and (104), the values of $q_{2}$ of eq.(44)
increase from $1$ to $4$ along $e^{2}=-1/3$ as $s^{2}$ increases from $0$ to
$3/32$, and decrease from $4$ to $1$ as $s^{2}$ increases from $3/32$ to $3/4$
(the upper boundary of Regions II), and (iii) the geometrical eccentricity
$\varepsilon$ given by eqs.(85)-(91) for the precessing elliptic-type orbits
corresponding to the specific $k^{2}=1/2$ can be easily tracked as it varies
from $\varepsilon=1$ for $e^{2}=1,$ $s^{2}=1/18$ to $\varepsilon
=(2/\sqrt{-3+2\sqrt{3}}-1)^{-1}=0.516588$ for $e^{2}=0,$ $s^{2}=(3-\sqrt
{3})/18=0.0704416$.

\section{Newtonian and Non-Newtonian Correspondences}

We have used $e^{2}$ defined by eq.(15) as one of the two principal parameters
for representing the orbit. In this section, we discuss the relation of this
parameter to the corresponding Newtonian parameter. It will be seen that some
relativistic bounded orbits do not have corresponding Newtonian orbits when
appropriate limits are taken.

From eqs.(15) and (7), the squared energy parameter $e^{2}$ can be expressed as%

\begin{equation}
e^{2}=\left(  \frac{r^{2}\overset{\cdot}{\phi}}{GM}\right)  ^{2}\left\{
\overset{\cdot}{r}^{2}+\left(  r\overset{\cdot}{\phi}-\frac{GM}{r^{2}%
\overset{\cdot}{\phi}}\right)  ^{2}-\frac{2GM}{c^{2}}r\overset{\cdot}{\phi
}^{2}\right\}  .
\end{equation}

We first show how in the Newtonian limit the non-terminating orbits in Region
I given by eq.(18) become the Newtonian elliptical, parabolic or hyperbolic
orbit and how the energy parameter $e$ becomes the "eccentricity" of the
Newtonian orbit.

In Newtonian mechanics, the energy $E_{0}$ of a particle of mass $m$ in a
gravitational field produced by a massive body of mass $M$ is the sum of its
kinetic and potential energies and is given by%

\begin{equation}
E_{0}=\frac{1}{2}m\left(  \overset{\cdot}{r}^{2}+r^{2}\overset{\cdot}{\phi
}^{2}\right)  -\frac{GmM}{r},
\end{equation}

where the derivative $\overset{\cdot}{}$ represents $d/dt$, $t$ being the
ordinary time. If we define the Newtonian eccentricity $e_{N}$ to be%

\begin{equation}
e_{N}=\left\{  1+\frac{2E_{0}h^{2}}{m(GM)^{2}}\right\}  ^{1/2},
\end{equation}

the Newtonian equation of motion $\overset{\cdot\cdot}{r}-r\overset{\cdot
}{\phi}^{2}=-GM/r^{2}$ obtained from eq.(108) can be integrated to give the
equation for the orbit given by%

\begin{equation}
\frac{1}{r}=\frac{GM}{h^{2}}(1-e_{N}\cos\phi).
\end{equation}

Equation (110) is used with the usual identification that $0\leq e_{N}<1$
gives an elliptic orbit, $e_{N}=1$ gives a parabolic orbit, and $e_{N}>1$
gives a hyperbolic orbit. Note that $e_{N}$ is defined for the range of all
positive values $0\leq e_{N}\leq\infty$.

Before we discuss how eqs.(107) and (109) are related, and how eq.(110) is an
approximation of eq.(18), it is useful to review briefly the curves of conic
sections purely from the mathematical point of view. It is known that the
equation in the polar coordinates $(r,\phi)$%

\begin{equation}
\frac{1}{r}=\frac{1}{p}(1-\varepsilon_{C}\cos\phi)
\end{equation}

represents a curve traced by a point, say $P$, about a focus, say $F$, on the
$x$-axis, such that the ratio of the distance $r\equiv PF$ to the
perpendicular distance $PD$ to a directrix which is a line parallel to the $y
$-axis, is a constant equal to $\varepsilon_{C}$, i.e. $\varepsilon_{C}\equiv
PF/PD$, and $\varepsilon_{C}$ is defined for all positive values and thus in
the range $0\leq\varepsilon_{C}\leq\infty$. The parameters $\varepsilon_{C}$
and $p$ characterize three types of curves. The curve is an ellipse for
$0\leq\varepsilon_{C}<1$ and $p=a(1-\varepsilon_{C}^{2})$, a parabola for
$\varepsilon_{C}=1$ and $p=2r_{\min}$, and a hyperbola for $\varepsilon_{C}>1$
and $p=a(\varepsilon_{C}^{2}-1)$. For the elliptic curve, the parameters $a$
and $b=a\sqrt{1-\varepsilon_{C}^{2}}$ are associated with the semi-major and
semi-minor axes, and $\varepsilon_{C}$ can be shown to be related to $r_{\min
}$ and $r_{\max}$ by $\varepsilon_{C}=(r_{\max}-r_{\min})/(r_{\max}+r_{\min
}).$ This expression for $\varepsilon_{C}$ in terms of $r_{\max}$ and
$r_{\min}$ is not valid for the hyperbolic curve. For the hyperbolic curve,
the parameters $a$ and $b=a\sqrt{\varepsilon_{C}^{2}-1}$ are used to
characterize the angles with respect to the $x$-axis of the two asymptotes.

By identifying the equation of curves of the conic sections given by eq.(111)
with the Newtonian orbit equation (110), we say that the Newtonian orbit is an
elliptic, parabolic or hyperbolic curve depending on the value of $e_{N}$, and
we identify $e_{N}$ with $\varepsilon_{C}$ for the entire range of positive
values from $0$ to $\infty$, and by equating $GM/h^{2}=1/p $.

We now want to see what the Newtonian limit implies from our general
relativistic orbit equation (18). First, from eqs.(108) and (109), the
Newtonian eccentricity $e_{N}$ can be expressed as%

\begin{equation}
e_{N}^{2}=\left(  \frac{r^{2}\overset{\cdot}{\phi}}{GM}\right)  ^{2}\left\{
\overset{\cdot}{r}^{2}+\left(  r\overset{\cdot}{\phi}-\frac{GM}{r^{2}%
\overset{\cdot}{\phi}}\right)  ^{2}\right\}  .
\end{equation}

Comparing eqs.(107) and (112), it is seen that our energy parameter
$e\rightarrow e_{N}$ in the Newtonian limit implies the approximation
$\tau\rightarrow t$ and $c\rightarrow\infty$. However, since setting
$c=\infty$ is not consistent with reality, we will proceed by stating that
$e\rightarrow e_{N}$ if we take the approximation $\tau\rightarrow t$ and%

\begin{equation}
\overset{\cdot}{r}^{2}+\left(  r\overset{\cdot}{\phi}-\frac{GM}{r^{2}%
\overset{\cdot}{\phi}}\right)  ^{2}>>\frac{2GM}{c^{2}}r\overset{\cdot}{\phi
}^{2}.
\end{equation}

Next, we proceed from eq.(18) to eq.(110) in steps which, by noting eqs.(107)
and (113), must be taken with care around the point $e^{2}=0$. For small
$s^{2}$, expanding the parameters and functions in eqs.(18)-(22) in power
series in $s$, we find [3], for $e^{2}>0$,%

\begin{align}
e_{1}  &  =\frac{1}{6}-s^{2}-(e^{2}+3)s^{4}+..\nonumber\\
e_{2}  &  =-\frac{1}{12}+\frac{1}{2}(e+1)s^{2}+\frac{(e+1)^{3}}{2e}%
s^{4}+..\nonumber\\
e_{3}  &  =-\frac{1}{12}-\frac{1}{2}(e-1)s^{2}+\frac{(e-1)^{3}}{2e}s^{4}+...
\end{align}

\begin{align}
k^{2}  &  =4es^{2}\left(  1+\frac{1+9e^{2}-2e^{3}}{e^{2}}s^{2}+..\right)  ,\\
\gamma &  =\frac{1}{2}[1-(3-e)s^{2}+..],
\end{align}

\begin{equation}
K(k)=\frac{\pi}{2}(1+es^{2}+..),
\end{equation}

and%

\[
sn(\gamma\phi,k)=\sin\frac{\phi}{2}+\left[  \frac{1}{4}e\left(  \sin\frac
{\phi}{2}+\sin\frac{3\phi}{2}\right)  -\frac{3\phi}{2}\cos\frac{\phi}%
{2}\right]  s^{2}+...
\]

Substituting the above approximations into eqs.(18), (26) and (27) gives, to
the order of $s^{2}$,%

\[
\frac{1}{q}=2s^{2}\{1-\varepsilon\cos[(1-\delta)\phi]\},
\]

which, in terms of $r$, is the approximate orbit equation%

\begin{equation}
\frac{1}{r}=\frac{GM}{h^{2}}\{1-\varepsilon\cos[(1-\delta)\phi]\},
\end{equation}

where $\varepsilon$, to the order of $s^{2}$, is given by%

\begin{equation}
\varepsilon\simeq e+(e^{-1}-e^{3})s^{2},
\end{equation}

and where $\delta$, to the order of $s^{2}$, is given by%
\begin{equation}
\delta\simeq3s^{2}.
\end{equation}

$\delta$ is related to the precessional angle $\Delta\phi$ given in eq.(27) by
$\Delta\phi\simeq2\pi\delta\simeq6\pi s^{2}=6\pi\lbrack GM/(hc)]^{2}$ and it
is independent of $e$ (to the order $s^{2}$) for $0<e\leq\infty$. Thus if we
ignore terms of order $s^{2}$, we recover the Newtonian orbit equation given by%

\begin{equation}
\frac{1}{r}=\frac{GM}{h^{2}}(1-e\cos\phi).
\end{equation}

The question that can be posed at this point is whether we can define the
Newtonian limit by stating that it is the general relativistic result for
small $s$ if we ignore terms of order $s^{2}$ and higher.

To proceed, we note that the case $e=0$ is excluded from the expansions given
by eqs.(114)-(117) and it is also clear that $e^{2}=0$ does not satisfy the
condition given by eq.(113) and must be treated separately.

For $e^{2}=0$, instead of eqs.(114)-(117), we find%

\begin{align}
e_{1}  &  =\frac{1}{6}-s^{2}-3s^{4}-20s^{6}+...\nonumber\\
e_{2}  &  =-\frac{1}{12}+\frac{1}{2}s^{2}+\frac{\sqrt{2}}{2}s^{3}+\frac{3}%
{2}s^{4}+\frac{21\sqrt{2}}{8}s^{5}+10s^{6}+...\nonumber\\
e_{3}  &  =-\frac{1}{12}+\frac{1}{2}s^{2}-\frac{\sqrt{2}}{2}s^{3}+\frac{3}%
{2}s^{4}-\frac{21\sqrt{2}}{8}s^{5}+10s^{6}+....
\end{align}

\begin{align}
k^{2}  &  =4\sqrt{2}s^{3}\left(  1+\frac{45}{4}s^{2}+...\right)  ,\\
\gamma &  =\frac{1}{2}-\frac{3}{2}s^{2}+\frac{\sqrt{2}}{2}s^{3}-\frac{27}%
{4}s^{4}+\frac{33\sqrt{2}}{8}s^{5}+...
\end{align}

\begin{equation}
K(k)=\frac{\pi}{2}\left(  1+\sqrt{2}s^{3}+...\right)  ,
\end{equation}

and%

\[
sn(\gamma\phi,k)=\sin\frac{\phi}{2}-\frac{3\phi}{2}s^{2}\cos\frac{\phi}%
{2}+\frac{\sqrt{2}}{4}\left(  \sin\frac{\phi}{2}+\sin\frac{3\phi}{2}\right)
s^{3}+...
\]

Substituting the above approximations into eqs.(18), (26) and (27) gives the
orbit equation for $e^{2}=0$ in the form given by eq.(118) with the same
$\delta$ given by eq.(120) but, instead of eq.(119), $\varepsilon$ is now
given by%

\begin{equation}
\varepsilon=\sqrt{2}s\left(  1+\frac{9}{4}s^{2}\right)
\end{equation}

to the order of $s^{2}$. Thus we have an elliptic orbit that precesses with
the same angle $\delta$ given by eq.(120) but with an eccentricity equal to
$\sqrt{2}s+9\sqrt{2}s^{3}/4$ to the order $s^{3}$. If we ignore terms of order
$s^{2}$, the orbit equation becomes%

\begin{equation}
\frac{1}{r}=\frac{GM}{h^{2}}(1-\sqrt{2}s\cos\phi),
\end{equation}

which is a (non-precessing) elliptical orbit with a non-Newtonian eccentricity
that is dependent on the speed of light. To the best of our knowledge, this
general relativistic elliptical orbit with an eccentricity of the order $s$
has not been noted by other authors. As an example, an elliptic orbit with
eccentricity $\varepsilon=0.079005$ could have resulted from the general
relativistic orbit with $e=0$ [remembering that $e$ is defined by eq.(107) and
not eq.(112)] and $s=0.0554787$ for which the approximation formula (126)
gives $\varepsilon\simeq0.079002$ and the first term alone gives $\sqrt
{2}s=0.078459$, and eq.(120) gives $\delta\simeq0.009234$. The orbit for $e=0$
from general relativity becomes a Newtonian circular orbit if we ignore the
first order term in $s$, i.e. ignoring the second order term $s^{2}$ and
higher order terms is not sufficient to get a Newtonian limit for this case.

It is clear that the entire region characterized by $e^{2}\leq0$ or%

\begin{equation}
\overset{\cdot}{r}^{2}+\left(  r\overset{\cdot}{\phi}-\frac{GM}{r^{2}%
\overset{\cdot}{\phi}}\right)  ^{2}\leq\frac{2GM}{c^{2}}r\overset{\cdot}{\phi
}^{2}%
\end{equation}

is non-Newtonian in character. This includes all bound orbits in Region I in
the range $-1/3\leq e^{2}\leq0$, and all circular orbits that occur on the
curve $s_{1}^{\prime2}$ from the vertex $V$ to the origin $O$ for which
$k^{2}=0$ and $\varepsilon=0$. All circular orbits precess even though the
precession angle is not observable [12], and the precession angle is given by
eq.(61) which is non-zero unless the radius of the circle is infinite which
occurs on $s_{1}^{\prime}=0$ for zero gravitational field. For small $s$, the
precession angle is given approximately by%

\[
\Delta\phi\simeq6\pi s^{2}\simeq6\pi\frac{GM}{c^{2}r_{c}},
\]

where $r_{c}\simeq h^{2}/(GM)$ is the radius of the circular orbit. For small
$s_{1}^{\prime2}$, the values of $e^{2}$ along the $s_{1}^{\prime}$ curve
where the circular orbits occur are given by%

\[
e^{2}\simeq-2s_{1}^{\prime2}.
\]

For small $s_{1}^{\prime}$ and for $s$ just above $s_{1}^{\prime}$ inside
Region I given by $s^{2}=s_{1}^{\prime2}+(\Delta s)^{2}$, it can be shown that
to the order $\Delta s$ we have an elliptic orbit similar to eq.(127) given by%

\begin{equation}
\frac{1}{r}=\frac{GM}{h^{2}}[1-\sqrt{2}(\Delta s)\cos\phi].
\end{equation}

To summarize the above results, for small values of $s$, the general
relativistic correction to the Newtonian elliptic orbit is second order in $s
$ for $e^{2}>0$ for which the correction appears in the angle of precession,
but is first order in $s$ or $\Delta s$ for $e^{2}\leq0$ for which the
correction appears in the eccentricity of the orbit. This division gives a new
meaning and significance to the parameter $e^{2}$. The new feature of this
result is that if we ignore terms of order $s^{2}$ and higher but not terms of
order $s$, then there exist non-Newtonian elliptic orbits of eccentricity
$\sqrt{2}s$ given by eq.(127).

For the hyperbolic-type orbit ($e^{2}>1$), in the Newtonian limit for small
$s^{2}$, eq.(28) becomes%

\begin{equation}
\sin^{2}\frac{\Psi_{1}}{2}\simeq\frac{1}{2}-\frac{1}{2e},
\end{equation}

from which we find%

\begin{equation}
\Psi_{1}\simeq\cos^{-1}(\frac{1}{e})\equiv\phi_{0},
\end{equation}

and the complementary angle $\Psi_{1}^{\prime}\equiv2\pi-\Psi_{2}$, where
$\Psi_{2}$ is given by eq.(30), becomes $\Psi_{1}^{\prime}\simeq\phi_{0}$
also. The Newtonian hyperbolic equation is thus given by eq.(110) or (121)
with $e_{N}=$ $e>1$ where $\cos^{-1}(1/e)\leq\phi\leq2\pi-\cos^{-1}(1/e)$. If
we define%

\begin{equation}
\Theta_{GR}\equiv\Psi_{1}+\Psi_{1}^{\prime}%
\end{equation}

and%

\begin{equation}
\Theta_{Newton}=2\phi_{0},
\end{equation}

the difference $\Delta\phi\equiv\Theta_{Newton}-\Theta_{GR}$ can be taken to
be an analog of the precession angle given by eq.(27) for a hyperbolic orbit,
and for small $s^{2}$, it was shown [3] to be given by%

\begin{equation}
\Delta\phi\simeq\left[  6\pi-6\phi_{0}+2(2+e^{-2})\sqrt{e^{2}-1}\right]
s^{2}.
\end{equation}

This result is different from the approximation used by Longuski et al. [13].
As discussed in ref.13, an experimental test can be carried out to check this result.

We see that our parameter $e^{2}$ defined by eq.(15) or (107) not only has the
advantage of being directly associated with the total energy but is also
directly related with $e_{N}^{2}$ defined by eq.(109) or (112) in the
Newtonian limit for the range $0<e^{2}\leq\infty$. The region characterized by
$e^{2}\leq0$ alerted us that there is a region other than the high
gravitational field ($s^{2}>>0$) region in which the orbits are non-Newtonian
in character; these include all those orbits for which $e^{2}\leq0$, and all
stable circular orbits along $s_{1}^{\prime}$ for all values of $s$. In
particular, for small $s$ and to the order $s$, a non-Newtonian elliptic orbit
given by eq.(127) or (129) for $e^{2}\leq0$ does not have a corresponding
counterpart for $e^{2}>0$ [see eq.(121)].

Thus Newtonian mechanics is recovered from general relativity if we ignore
terms of order $s^{2}$ and higher for $e^{2}>0$, and if we ignore terms of
order $s$ \ and higher for $e^{2}\leq0$, and not from setting $s=0$. If we set
$s=0$, then eq.(18) from Solution (A1) [for $0\leq e^{2}\leq\infty$] gives
$q=\infty$ that can be interpreted as an orbit of infinite radius, and eq.(35)
from Solution (A2) [for $0\leq e^{2}\leq\infty$] and eq.(43) from Solution (B)
[for $-\infty\leq e^{2}\leq0$] both give the same terminating orbit
$q=\cos(\phi/2)$ for which the particle starts from the Schwarzschild horizon
$q=1$ and terminates at the center of the black hole. While these orbits may
not mean much by themselves, they are the proper limits of the corresponding
orbits as $s^{2}\rightarrow0$. The small $s$ region for the terminating orbits
given by Solutions (A2) and (B) is non-Newtonian because the initial position
of the particle is very close to the black hole.

One can raise the question of whether, instead of using the energy parameter
$e$ and eqs.(18), (35) and (43) for the orbital trajectories, it is more
convenient to start with an orbit equation analogous to the Newtonian orbit
equation with an eccentricity $\varepsilon$ appropriately defined. This is the
approach used by Darwin and extended by Chandrasekhar that we shall discuss
and compare with our analysis in the next section.

\section{The Analyses of Darwin and Chandrasekhar}

Chandrasekhar [5] extended Darwin's work [6] and classified the orbits from
the solutions of eq.(8) into three types: (a) The radial geodesics (b) The
bound orbits $(\kappa^{2}<1)$ and (c) The unbound orbits $(\kappa^{2}\geq1)$.
The bound and unbound orbits are divided according to three categories (i),
(ii) and (iii) which he called orbits of the first and second kind, and orbits
with imaginary eccentricities, respectively. Then there are various special
cases within each category. Chandrasekhar used various relations and
inequalities for characterizing each type of orbit. While all possible orbits
(albeit expressed in completely different expressions from ours) are in indeed
in place, they are not clearly located on a "map" with coordinates that can be
easily identified with the commonly used physical parameters. By briefly
describing Chandrasekhar's work and analysis of these orbits and comparing
them with ours, we intend to clarify this point in this section.

We do not discuss the radial geodesics (of zero angular momentum) which is a
simple special case. For the purpose of comparing Chandrasekhar's work with
ours, it is easier to group the orbits according to the three categories (i),
(ii) and (iii) which can be identified to those corresponding to what we call
Solutions (A1), (A2) and (B).

We first describe the orbits of the first kind which, as we shall see,
correspond to the non-terminating orbits in Region I (solution A1) given by
eq.(18) in our case. For convenience of comparison, we use our notations to
describe Chandrasekhar's solution and parameters. To make it resemble the
equation for the Newtonian orbit, his solution of eq.(8) is expressed as%

\begin{equation}
u=\frac{1}{\ell}(1+\varepsilon\cos\chi),
\end{equation}

where $\ell$ is the (generalized) "latus rectum" and $\varepsilon$ is the
(generalized) "eccentricity" that are defined from the three real roots
$u_{1},u_{2},u_{3}$ of the cubic equation under the condition $\Delta\leq0$
with $u_{1}\leq u_{2}\leq u_{3}$ as%

\begin{align}
u_{1}  &  =\frac{1}{\ell}(1-\varepsilon),\nonumber\\
u_{2}  &  =\frac{1}{\ell}(1+\varepsilon),\nonumber\\
u_{3}  &  =\frac{c^{2}}{2GM}-\frac{2}{\ell}.
\end{align}

The "polar angle" $\chi$ in eq.(135) is related to the polar angle $\phi$ of
the spherical coordinate used in eqs.(3)-(5) by%

\begin{equation}
\phi=\frac{2}{(1-6\mu+2\mu\varepsilon)^{1/2}}F\left(  \frac{\pi}{2}-\frac
{\chi}{2},k\right)  ,
\end{equation}

where%

\begin{equation}
\mu=\frac{GM}{c^{2}\ell},
\end{equation}

and $F(\theta,k)$ is the incomplete elliptic integral of the first kind with
modulus $k$ given by%

\begin{equation}
k^{2}=\frac{4\mu\varepsilon}{1-6\mu+2\mu\varepsilon}.
\end{equation}

The three roots of the cubic equation%

\begin{equation}
\alpha u^{3}-u^{2}+(2GM/h^{2})u+c^{2}(\kappa^{2}-1)/h^{2}=0
\end{equation}

given by eq.(136) are related to our three roots of the cubic equation (16) by%

\begin{align}
\alpha u_{3}  &  =1-4\mu=\frac{1}{3}+4e_{1},\nonumber\\
\alpha u_{2}  &  =2\mu(1+\varepsilon)=\frac{1}{3}+4e_{2},\nonumber\\
\alpha u_{1}  &  =2\mu(1-\varepsilon)=\frac{1}{3}+4e_{3},
\end{align}

where the dimensionless parameter $\mu$ is related to largest root $e_{1}$
given in eq.(21) in our formulation by%

\begin{equation}
\mu=\frac{1}{6}-e_{1},
\end{equation}

and where the geometrical eccentricity $\varepsilon$ is related to $e_{2}$ and
$e_{3}$ in our formulation by eq.(26). However, Chandrasekhar not only
extended the range of values of $\varepsilon$ to beyond $1$ but also used it
for the orbits of the second kind. It should be noted that for orbits of the
first kind, $\varepsilon$ has a simple geometrical meaning only when
$0\leq\varepsilon\leq1$. For orbits of the second kind, $\varepsilon$ has no
geometrical meaning for any value of $\varepsilon$.

Indeed characterizing the orbits of the first and second kinds according to
$(\mu,\varepsilon)$ is equivalent to characterizing these orbits according to
$(e_{1},e_{2})$ since $e_{3}=-(e_{1}+e_{2})$. The relationships of the
parameters $\mu,\varepsilon$ and our parameters $e^{2},s^{2}$ can be shown to
be given by the following equations:%

\begin{equation}
s^{2}=\mu\lbrack1-\mu(3+\varepsilon^{2})],
\end{equation}

and%

\begin{equation}
1-e^{2}=\frac{(1-4\mu)(1-\varepsilon^{2})}{[1-\mu(3+\varepsilon^{2})]^{2}}.
\end{equation}

It is clear from these two equations that rather complicated combinations of
the parameters $\mu$ and $\varepsilon$ are required to express the parameters
$s^{2}$ and $e^{2}$ which are related simply to the angular momentum and total
energy of a particle in a gravitational field. These dimensionless energy and
field parameters $e^{2}$ and $s^{2}$ arise naturally in the analytic solutions
of eq.(13). A large part of Chandrasekhar's work includes expressing the
constraints on $\mu$ and $\varepsilon$ that are required for different types
of orbits. An inequality which is repeatedly stated that must hold for the
orbits of the first and second kinds, bound or unbound, is $1-\mu
(3+\varepsilon^{2})\geq0$, which, as can be seen from eq.(143), is simply a
requirement that $s^{2}\geq0$. Another required inequality for the orbits of
the first and second kinds, bound or unbound, is $1-6\mu-2\mu\varepsilon\geq
0$, which can be shown to be simply $e_{1}\geq e_{2}$. Other required
inequalities such as $\mu<1/4$, $0\leq\varepsilon^{2}<1$ for the bound orbits,
and $\mu<1/4$, $\varepsilon^{2}\geq1$ for the unbound orbits, for the first
and second kinds, are clearly associated with the identifications $0\leq
e^{2}<1$ and $e^{2}\geq1$ respectively. We again mention that while
$\varepsilon$ has a simple physical meaning of geometrical eccentricity for
the orbits of the first kind in the range $0\leq\varepsilon<1$, it does not
have any simple physical meaning for the other two kinds of orbits for any
value of $\varepsilon$; whereas our parameter $e^{2}$ defined by eq.(15) not
only has a simple physical meaning for all values in the range $-\infty\leq
e^{2}\leq+\infty$ and for all orbits, it also has a well-defined Newtonian
limit as $e\rightarrow e_{N\text{ }}$ given by eq.(109) in the limit of very
small $s$ values for the range $0<e\leq\infty$. Other relationships such as
$\varepsilon=0$ implies $k^{2}=0$, and $2\mu(3+\varepsilon)=1$ implies
$k^{2}=1$ can be seen easily from eq.(139).

For the case of stable circular orbits arising from the orbits of the first
kind for which $\varepsilon=0$, we have $\mu=(2q_{c})^{-1}$, where $3\leq
q_{c}\leq\infty$ is given by eq.(58), and%

\begin{equation}
\kappa^{2}=\frac{\left(  1-\frac{1}{q_{c}}\right)  ^{2}}{1-\frac{3}{2q_{c}}}.
\end{equation}

In terms of our parameter $s_{1}^{\prime2}$, we have%

\begin{equation}
s_{1}^{\prime2}=\frac{1-\frac{3}{2q_{c}}}{2q_{c}},
\end{equation}

and the corresponding value of $e^{2}$ is%

\begin{equation}
e^{2}=\frac{\frac{9}{4}-q_{c}}{\left(  \frac{3}{2}-q_{c}\right)  ^{2}}.
\end{equation}

The parameter $\mu$ can be expressed in terms of $s_{1}^{\prime2}$ and the
corresponding $e^{2}$ as%

\begin{equation}
\mu=\frac{1}{6}\left\{  1-\left(  1-12s_{1}^{\prime2}\right)  ^{1/2}\right\}
=\frac{1-3e^{2}-(1+3e^{2})^{1/2}}{9(1-e^{2})}.
\end{equation}

For ISCO (the innermost stable circular orbit), $q_{c}=3$, $\mu=1/6$,
$\kappa^{2}=8/9$, $e^{2}=-1/3$ and $s^{2}=1/12$.

For the orbits of the second kind which correspond to the terminating orbits
in Region I (Solution A2) in our case, the roots of the cubic equation (140)
are still given by the same eq.(141), and the parameters $\mu$ and
$\varepsilon$ are still defined or related to our formulation by eqs.(142) and
(26), but the orbit equation is given by%

\begin{equation}
u=\left(  \frac{c^{2}}{2GM}-\frac{2}{\ell}\right)  +\left(  \frac{c^{2}}%
{2GM}-\frac{3+\varepsilon}{\ell}\right)  \tan^{2}\frac{\xi}{2},
\end{equation}

where $\xi$ is related to the polar angle $\phi$ by%

\begin{equation}
\phi=\frac{2}{(1-6\mu+2\mu\varepsilon)^{1/2}}F\left(  \frac{\xi}{2},k\right)
,
\end{equation}

where the modulus of the incomplete elliptic integral of the first kind
$F(\theta,k)$ is given by the same expression (139). Equations (143), (144),
and the statements made following eq.(144) all hold for the orbits of the
second kind. Remember that the orbit equation (149) for the special case
$\varepsilon=0$ and hence $k^{2}=0$ still gives, not a circular orbit, but a
terminating orbit given by eq.(67). On the other hand, the orbit equation
(149) for the special case $2\mu(3+\varepsilon)=1$ and hence $k^{2}=1$ gives
an unstable circular orbit with a radius given by eq.(69), and $\mu$ can be
expressed in terms of the radius of this circle by $\mu=(q_{u}-1)/(4q_{u})$ or
$1/4-1/(4q_{u})$, where $1.5\leq q_{u}<3$. The expressions for $\kappa^{2}$,
$s_{1}^{2}$ and $e_{1}^{2}$ in terms of $q_{u} $ are remarkably similar to
those in eqs.(145)-(147), and they are given by%

\begin{equation}
\kappa^{2}=\frac{\left(  1-\frac{1}{q_{u}}\right)  ^{2}}{1-\frac{3}{2q_{u}}},
\end{equation}

\begin{equation}
s_{1}^{2}=\frac{1-\frac{3}{2q_{u}}}{2q_{u}},
\end{equation}

and%

\begin{equation}
e_{1}^{2}=\frac{\frac{9}{4}-q_{u}}{\left(  \frac{3}{2}-q_{u}\right)  ^{2}},
\end{equation}

while the expression for $\mu$ in terms of $s_{1}^{2}$ and the corresponding
$e^{2}$ is%

\begin{equation}
\mu=\frac{1}{4}-\frac{1}{12}\left\{  1+\left(  1-12s_{1}^{2}\right)
^{1/2}\right\}  =\frac{1}{4}-\frac{1-3e^{2}+(1+3e^{2})^{1/2}}{18(1-e^{2})}.
\end{equation}

We also have the following expression for $\kappa^{2}$ in terms of
$\varepsilon$:%

\begin{equation}
1-\kappa^{2}=\frac{1-\varepsilon^{2}}{9-\varepsilon^{2}},
\end{equation}

and the expressions for $s_{1}^{2}$ and the corresponding $e^{2}$ in terms of
$\varepsilon$ are given in eqs.(91) and (89).

As we pointed out in eqs.(64) and (65) in our discussion of Solution (A1) for
$k^{2}=1$, the geometrical eccentricity $\varepsilon$ has a physical meaning
expressed by eq.(25) for $q_{\max}$ and $q_{\min}$ of the orbit of the first
kind (Solution A1) for $0\leq\varepsilon\leq1$ or $-1/3\leq e^{2}\leq1$. We
have used the same $\varepsilon$ defined by eq.(26) for the orbit of Solution
(A2) to express $q_{u}$ which is also the radius of the asymptotic circle
given by eq.(65) or (69). In Section 6, we described the relationship between
$q_{u}$ given by eq.(69) and $q_{\min}$ given by eq.(65) and the continuations
of the orbits across the boundary from Region I to II.

For both orbits of the first and second kinds for the special case of
$2\mu(3+\varepsilon)=1$, $e^{2}$ and $s^{2}$ are related to $\varepsilon$
defined by eq.(26) by eqs.(89) and (91) which show that $e^{2}\rightarrow
\infty$ and $s^{2}\rightarrow0$ as $\varepsilon\rightarrow3$ and thus the
values of $\varepsilon$ are restricted to $<3$ for the special case
$2\mu(3+\varepsilon)=1$. The inequality given by eq.(72) implies
$0\leq\varepsilon\leq3$ from eq.(71).

For the orbits with imaginary eccentricities which correspond to the
terminating orbits in Region II in our case, the roots of the cubic equation
(140) consist of one real root and two complex conjugate roots and are related
to the roots in our formulation by%

\begin{align}
\alpha u_{1}  &  =1-4\mu=\frac{1}{3}+4a,\nonumber\\
\alpha u_{2}  &  =2\mu(1+i\varepsilon)=\frac{1}{3}+4b,\nonumber\\
\alpha u_{2}  &  =2\mu(1-i\varepsilon)=\frac{1}{3}+4\overline{b},
\end{align}

where $a$, $b$ and $\overline{b}$ are defined by eq.(40) and the line below
it. The parameters $\mu$ and $\varepsilon$ are related to $a$, $b$ and
$\overline{b}$ and to $A$ and $B$ of eqs.(39) and (40) by%

\begin{equation}
\mu=\frac{1}{6}-a=\frac{1}{6}-(A+B),
\end{equation}

and%

\begin{equation}
\varepsilon=-i\frac{6(b-\overline{b})}{1-6a}=\frac{6\sqrt{3}(A-B)}{1-6(A+B)}.
\end{equation}

As in the case $\varepsilon>1$ for Solution (A1) and for all values of
$\varepsilon$ defined by eq.(26) for Solution (A2), the $\varepsilon$ here
defined by eq.(158) for Solution B has no simple physical meaning at all. The
orbit equation is expressed in the form%

\begin{equation}
u=\frac{1}{\ell}(1+\varepsilon\tan\frac{\xi}{2}),
\end{equation}

where $\xi$ is related to the polar angle $\phi$ by an elliptic integral, and
the squared modulus $k^{2}$ of the elliptic integral is related to
$\varepsilon$ and $\mu$ in a somewhat more complicated way through several
equations (see [5] p.112). The relationships of the parameters $\mu
,\varepsilon$ and our parameters $e^{2},s^{2}$ are given by the following equations:%

\begin{equation}
s^{2}=\mu\lbrack1-\mu(3-\varepsilon^{2})],
\end{equation}

and%

\begin{equation}
1-e^{2}=\frac{(1-4\mu)(1+\varepsilon^{2})}{[1-\mu(3-\varepsilon^{2})]^{2}}.
\end{equation}

The bound orbits for orbits with imaginary eccentricities are specified by
$\mu<1/4$ and the unbound orbits by $\mu\geq1/4$ and $1-\mu(3-\varepsilon
^{2})>0$, and there is no upper limit for $\varepsilon^{2}$ for either bound
or unbound orbits. The parameter $\varepsilon$ is related to the three roots
$a$, $b$, and $\overline{b}$ of the cubic equation (16) by eq.(158). The bound
and unbound orbits are specified simply in our case by $e^{2}<1$ (which
ensures that $a>-1/12$ for eq.(43)) and $e^{2}\geq1$ (which ensures that
$a\leq-1/12$).

The motivation of Darwin's and Chandrasekhar's analyses is to express the
orbit equation (135) in a way which resembles the Newtonian orbit equation
with the introduction of the "latus rectum" $\ell$ (and thus $\mu$) and
"eccentricity" $\varepsilon$. However, this is done at the expense that not
only the "polar angle" $\chi$ or $\xi$ is not easily related to the polar
angle $\phi$ of the spherical coordinate, but also that $\ell$ and
$\varepsilon$ (that can become imaginary) do not have simple physical meanings
for most of these orbits. The two parameters $\ell$ and $\varepsilon$ in
essence represent roots of the cubic equation (140), and the entirety of all
possible orbits cannot be conveniently placed in a parameter space that is
directly related to the commonly used physical quantities.

Incidentally, for the unbound orbits of the first kind (Solution A1 for
$e^{2}>1$) and the unbound terminating orbits in Region II (solution B for
$e^{2}>1$), an impact parameter $b$ can be defined. Chandrasekhar gave this
impact parameter incorrectly as (see [5], p.114)%

\begin{equation}
b=\frac{h\kappa/c}{\sqrt{\kappa^{2}-1}}.
\end{equation}

The correct impact parameter is [3]%

\begin{equation}
b=\frac{h/c}{\sqrt{\kappa^{2}-1}},
\end{equation}

where $h$ and $\kappa$ are defined by eqs.(5) and (6). In terms of $e^{2}$ and
$s^{2}$, the dimensionless impact parameter is given by eq.(34). The possible
cause of error in eq.(162) is the following oversight: When a particle of rest
mass $m_{0}$ initially very far away from the star or black hole has an
initial velocity $v_{\infty}$ and an impact parameter $b$ from the star or
black hole, the angular momentum $h$ per unit rest mass of the particle is not
$h=v_{\infty}b$, but should be$\ h=mv_{\infty}b/m_{0}$, where $m=m_{0}%
/\sqrt{1-v_{\infty}^{2}/c^{2}}$.

\section{Summary and Commentary}

\begin{figure}
[ptb]
\begin{center}
\includegraphics[
natheight=11.000400in,
natwidth=8.500200in,
height=5.2641in,
width=5.303in
]%
{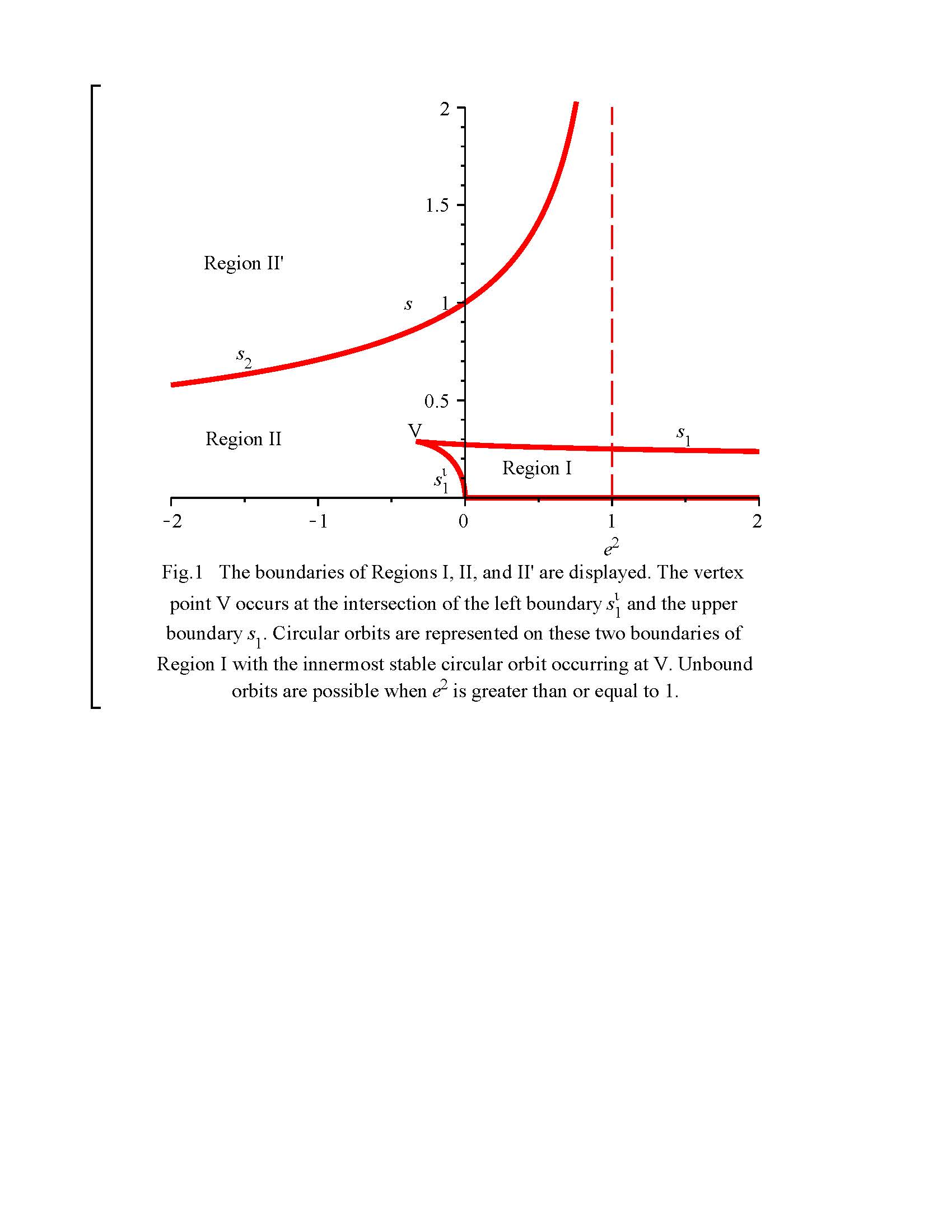}%
\caption{The boundaries of Regions I, II and II' are displayed. The vertex
point $V$ occurs at the intersection of the left boundary $s_{1}^{\prime}$ and
the upper boundary $s_{1}$.}%
\label{Fig.1}%
\end{center}
\end{figure}

\begin{figure}
[ptb]
\begin{center}
\includegraphics[
natheight=11.000400in,
natwidth=8.500200in,
height=5.1188in,
width=4.7193in
]%
{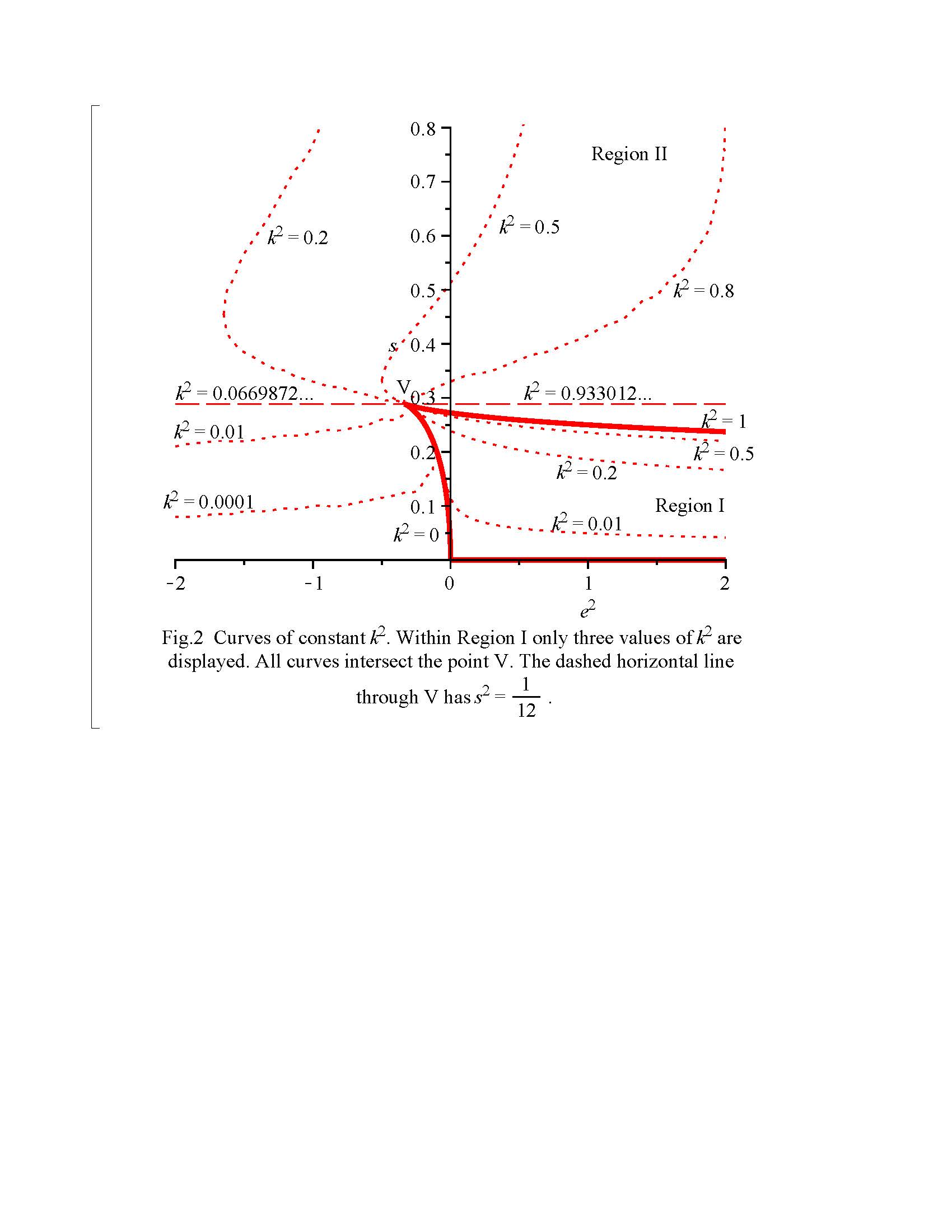}%
\caption{Curves of constant $k^{2}$. Within Region I only three values of
$k^{2}$ are displayed. All curves intersect the point $V$. The dashed
horizontal line through $V$ has $s^{2}=1/12$.}%
\label{Fig.2}%
\end{center}
\end{figure}

We have presented three possible analytic solutions for the orbit of a
particle in the Schwarzschild geometry in explicitly simple forms by eqs.(18),
(35) and (43), and we have presented a universal map with coordinates
$(e^{2},s)$ [as in \ref{Fig.1} and \ref{Fig.2}] on which all possible orbits
arising from these three solutions can be placed, where the two dimensionless
parameters $e^{2}$ and $s^{2}$ [defined by eqs.(15) and (11)] which we call
the energy and field parameters respectively, are in the ranges $-\infty\leq
e^{2}\leq\infty$ and $0\leq s\leq\infty$. We have presented but not used two
other maps with coordinates $(\kappa^{2},s^{2})$ (\ref{Fig.13}) and
$(-g_{3},-g_{2})$ (\ref{Fig.14}) that could be interesting alternatives where
$\kappa$, $g_{2}$ and $g_{3}$ are defined by eqs.(6) and (14). The
non-terminating orbit solution (18) and the terminating orbit solution (35)
occur in Region I ($\Delta\leq0$), and the other terminating orbit solution
(43) occurs in Region II ($\Delta>0$), where $\Delta$ is defined by eq.(17).
In contrast to the analyses of other authors that used various mathematical
inequalities for separating and discussing the different regions, Region I is
explicitly shown (in \ref{Fig.1}) to be bounded on the left by $s=s_{1}%
^{\prime}$ given by eq.(48) that is obtained from setting $\Delta=0$ and
$k^{2}=0$ [see eqs.(20) and (42)], and it is bounded on the top by $s=s_{1}$
given by eq.(50) that is obtained from setting $\Delta=0$ and $k^{2}=1$, and
Region II is bounded on the top for $e^{2}\leq1$ by $s=s_{2}$ given by eq.(52).

\begin{figure}
[ptb]
\begin{center}
\includegraphics[
natheight=11.000400in,
natwidth=8.500200in,
height=5.3074in,
width=4.7297in
]%
{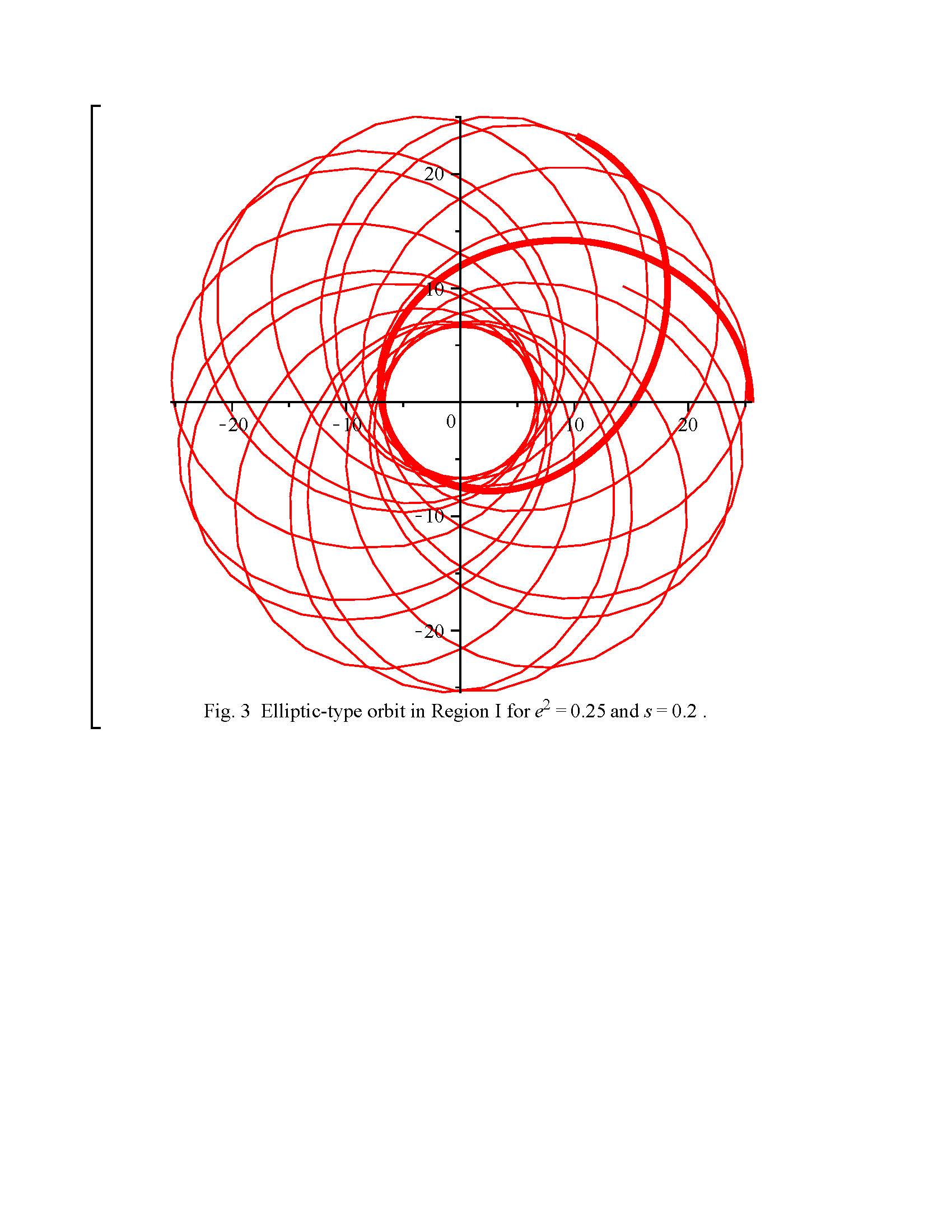}%
\caption{Elliptic-type orbit in Region I for $e^{2}=0.25$ and $s=0.2$.}%
\label{Fig.3}%
\end{center}
\end{figure}

\begin{figure}
[ptb]
\begin{center}
\includegraphics[
natheight=11.000400in,
natwidth=8.500200in,
height=4.785in,
width=4.3033in
]%
{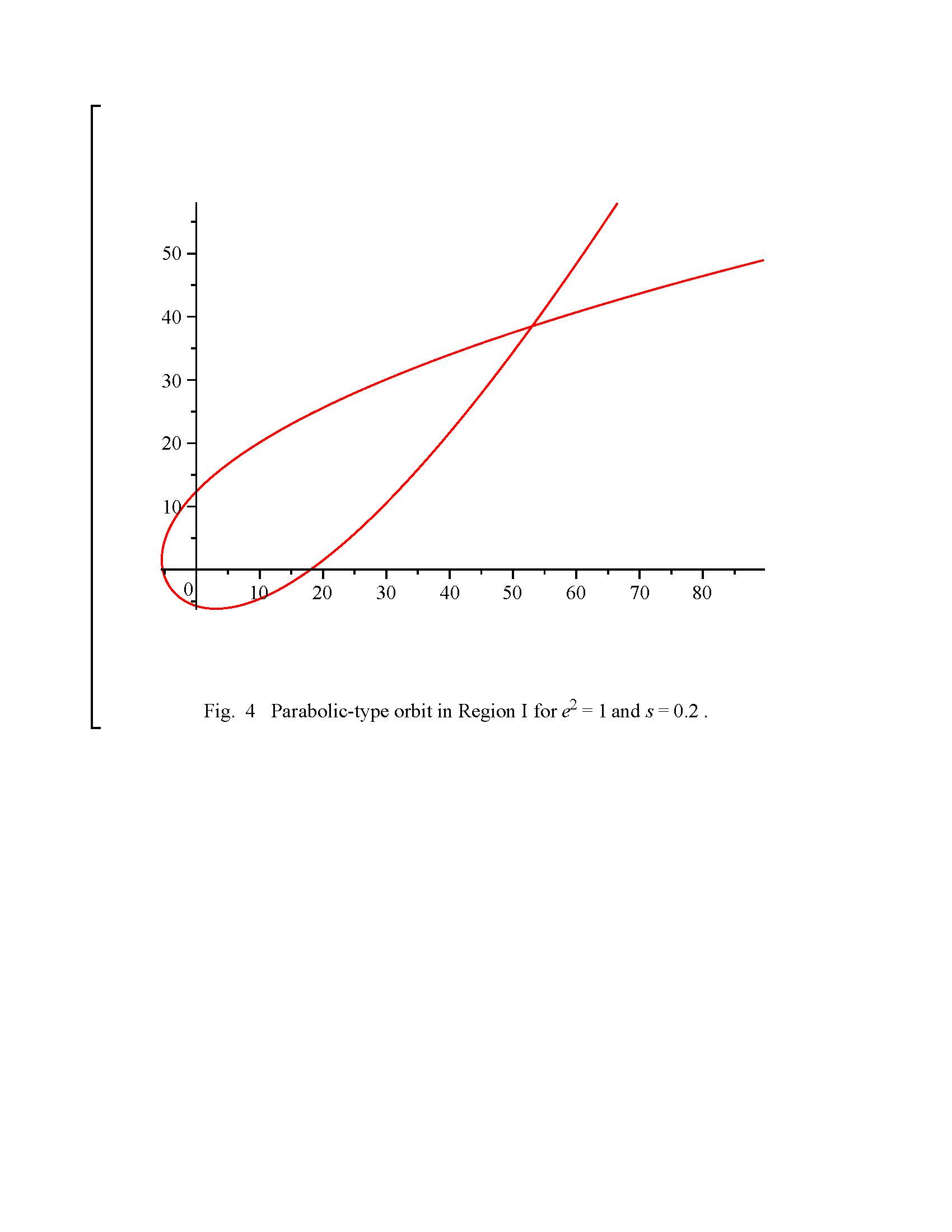}%
\caption{Parabolic-type orbit in Region I for $e^{2}=1$ and $s=0.2$.}%
\label{Fig.4}%
\end{center}
\end{figure}

\begin{figure}
[ptb]
\begin{center}
\includegraphics[
natheight=11.000400in,
natwidth=8.500200in,
height=6.0494in,
width=5.1984in
]%
{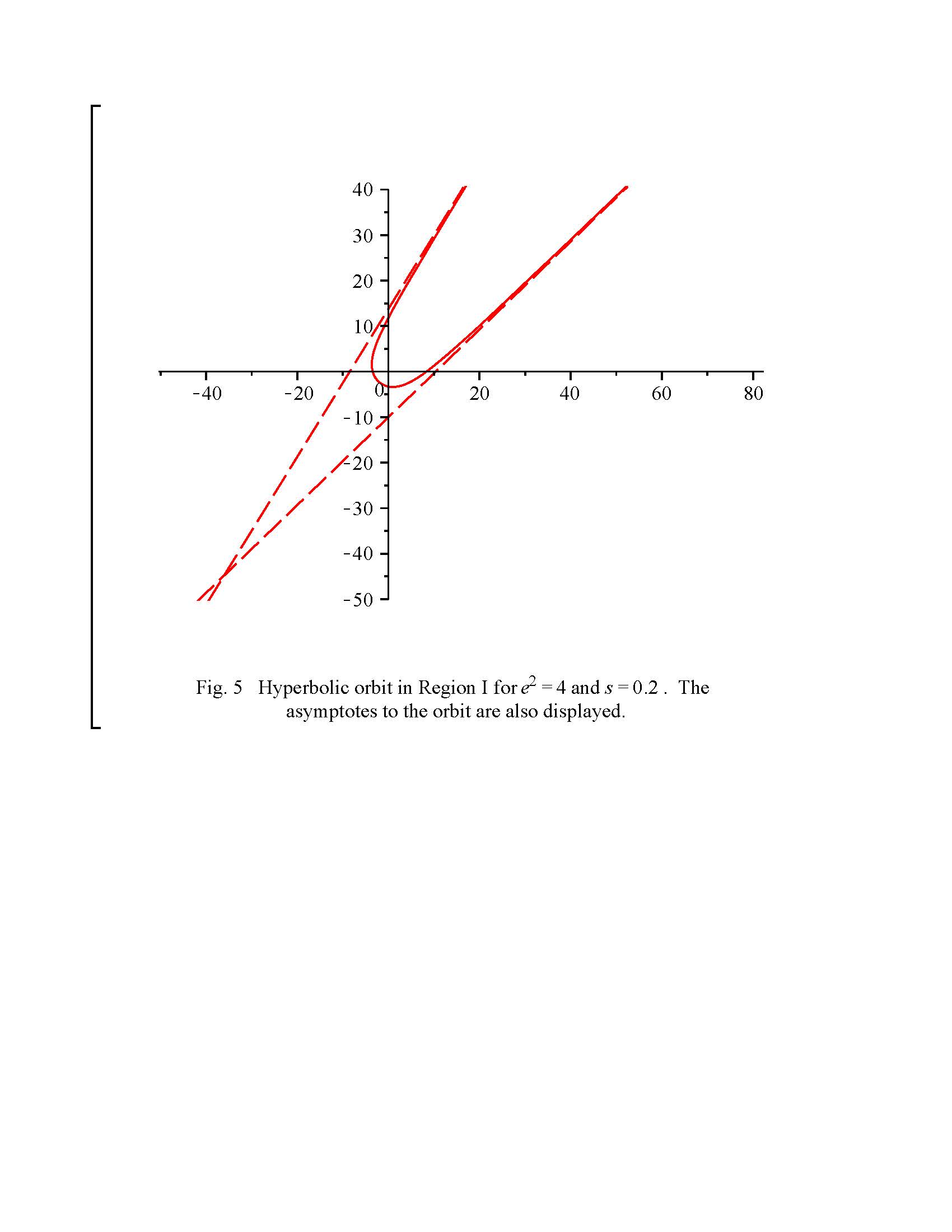}%
\caption{Hyperbolic-type orbit in Region I for $e^{2}=4$ and $s=0.2$. The
asymptotes to the orbit are also displayed.}%
\label{Fig.5}%
\end{center}
\end{figure}

\begin{figure}
[ptb]
\begin{center}
\includegraphics[
natheight=11.000400in,
natwidth=8.500200in,
height=6.1471in,
width=5.3964in
]%
{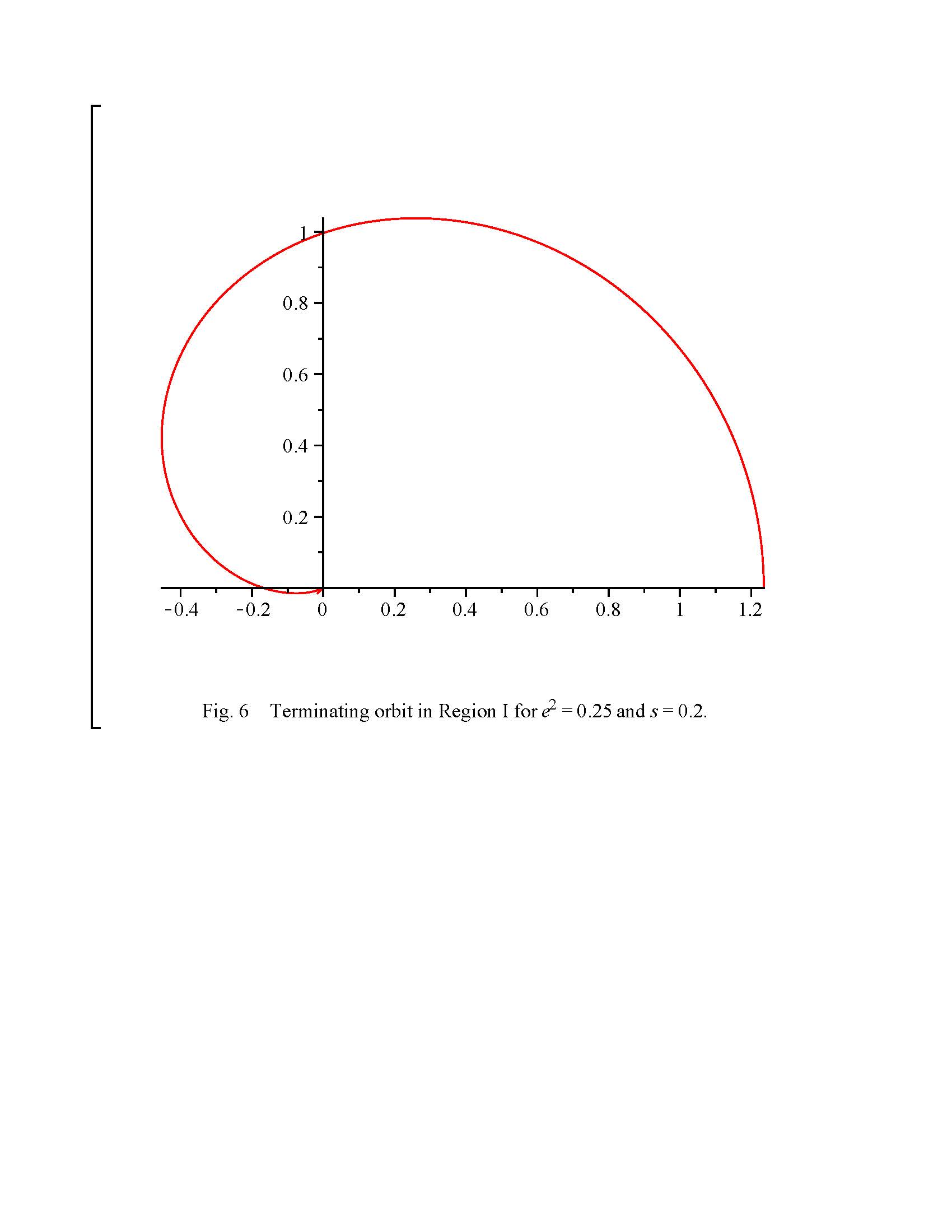}%
\caption{Terminating orbit in Region I for $e^{2}=0.25$ and $s=0.2$.}%
\label{Fig.6}%
\end{center}
\end{figure}

\begin{figure}
[ptb]
\begin{center}
\includegraphics[
natheight=11.000400in,
natwidth=8.500200in,
height=5.9785in,
width=5.2823in
]%
{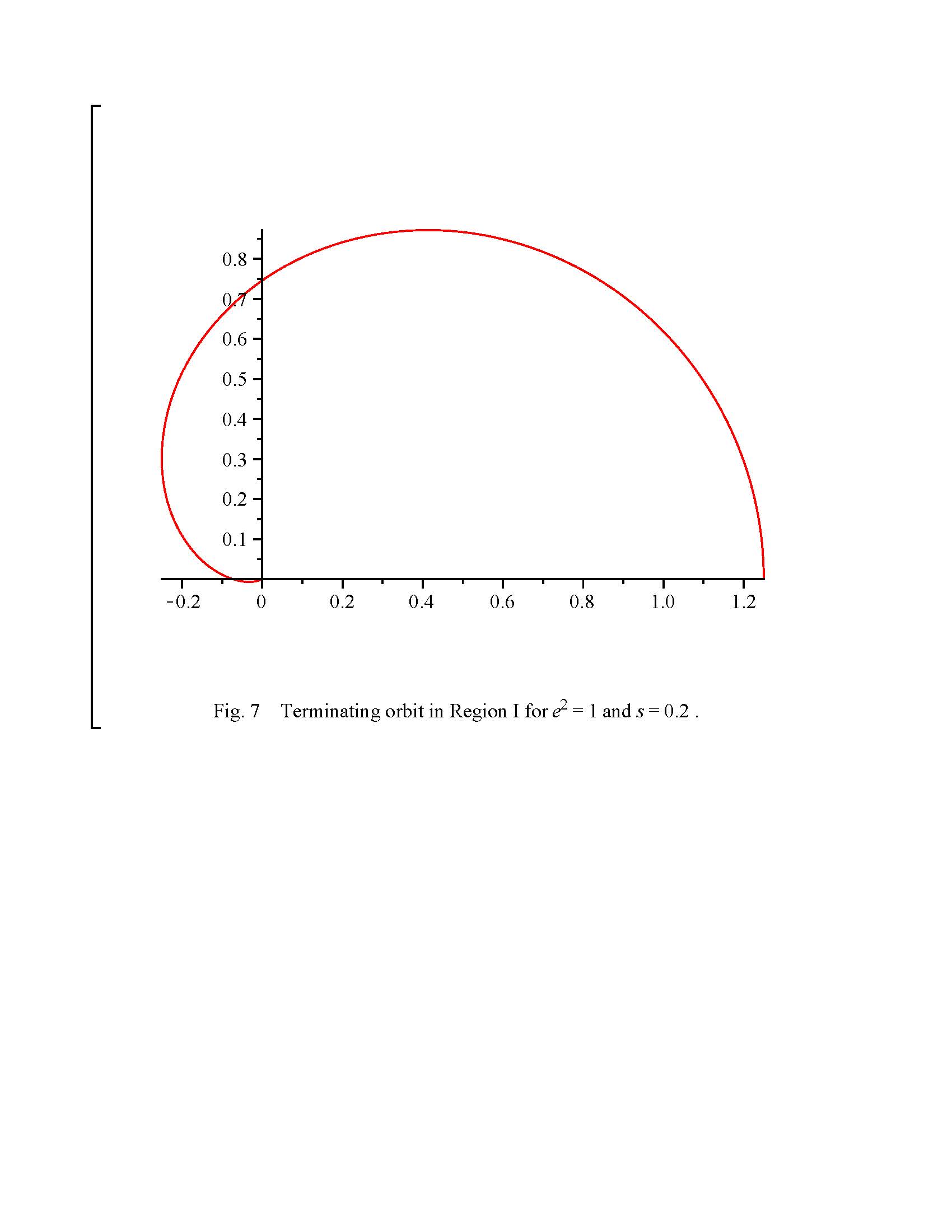}%
\caption{Terminating orbit in Region I for $e^{2}=1$ and $s=0.2$.}%
\label{Fig.7}%
\end{center}
\end{figure}

\begin{figure}
[ptb]
\begin{center}
\includegraphics[
natheight=11.000400in,
natwidth=8.500200in,
height=6.026in,
width=5.3342in
]%
{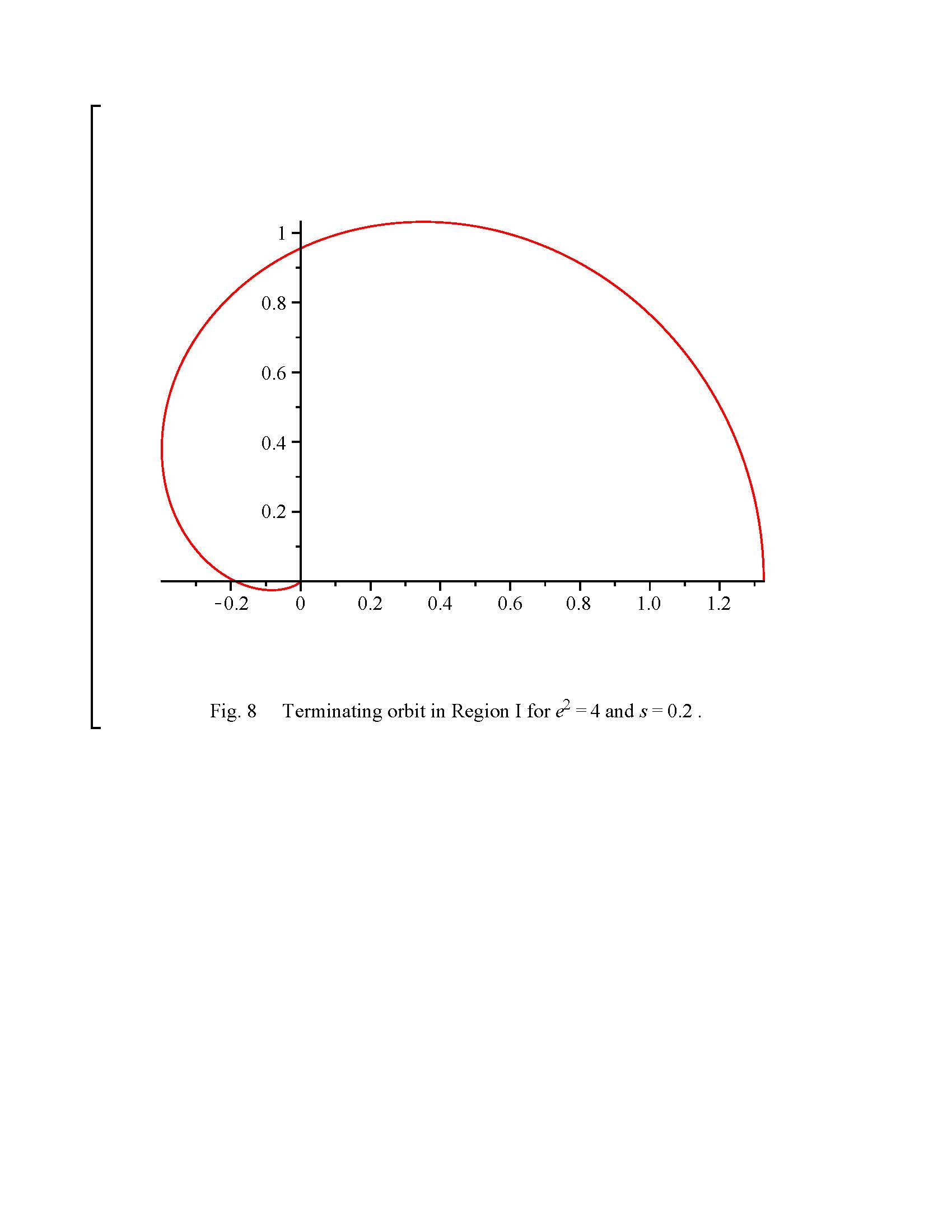}%
\caption{Terminating orbit in Region I for $e^{2}=4$ and $s=0.2$.}%
\label{Fig.8}%
\end{center}
\end{figure}

\begin{figure}
[ptb]
\begin{center}
\includegraphics[
natheight=11.000400in,
natwidth=8.500200in,
height=5.8574in,
width=5.188in
]%
{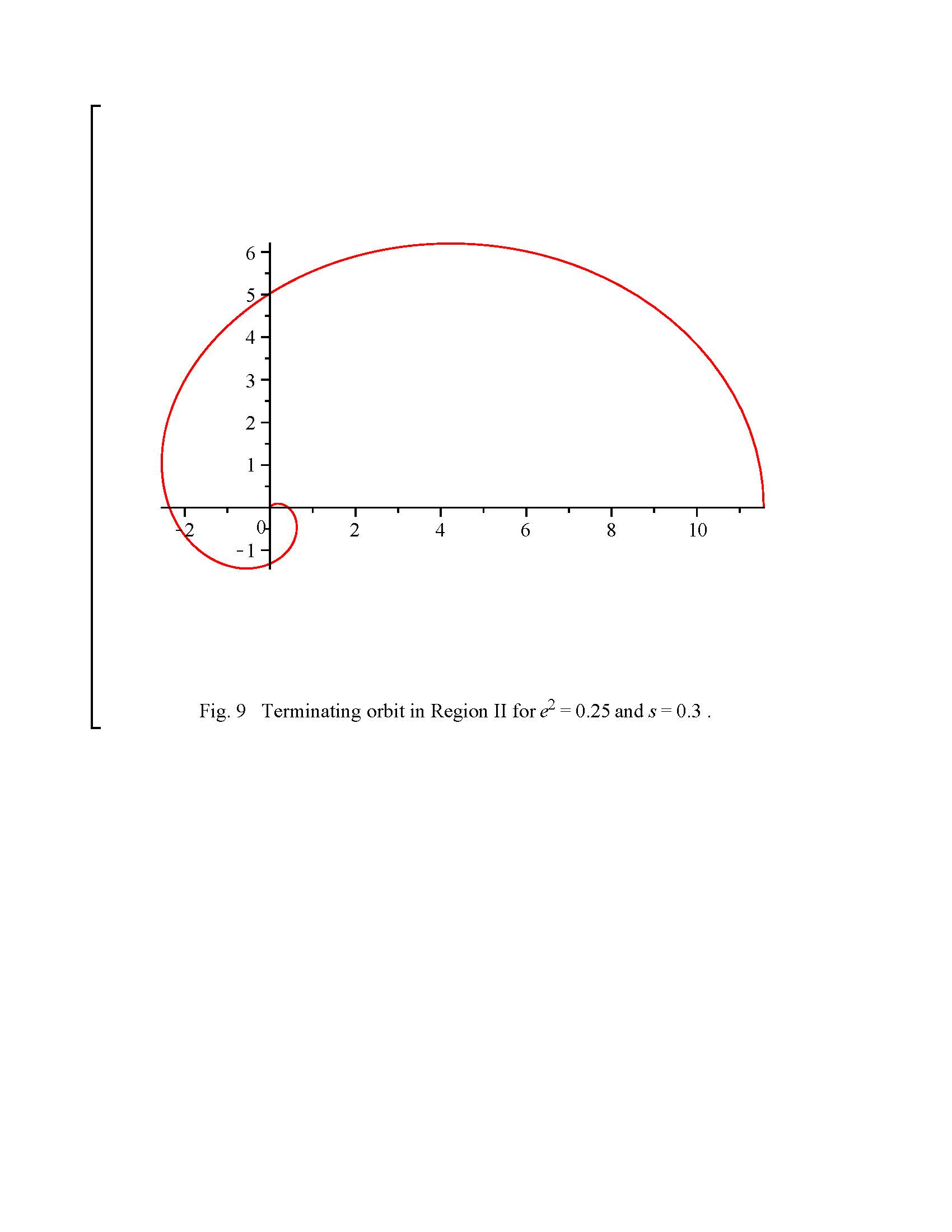}%
\caption{Terminating orbit in Region II for $e^{2}=0.25$ and $s=0.3$.}%
\label{Fig.9}%
\end{center}
\end{figure}

\begin{figure}
[ptb]
\begin{center}
\includegraphics[
natheight=11.000400in,
natwidth=8.500200in,
height=5.6654in,
width=5.0427in
]%
{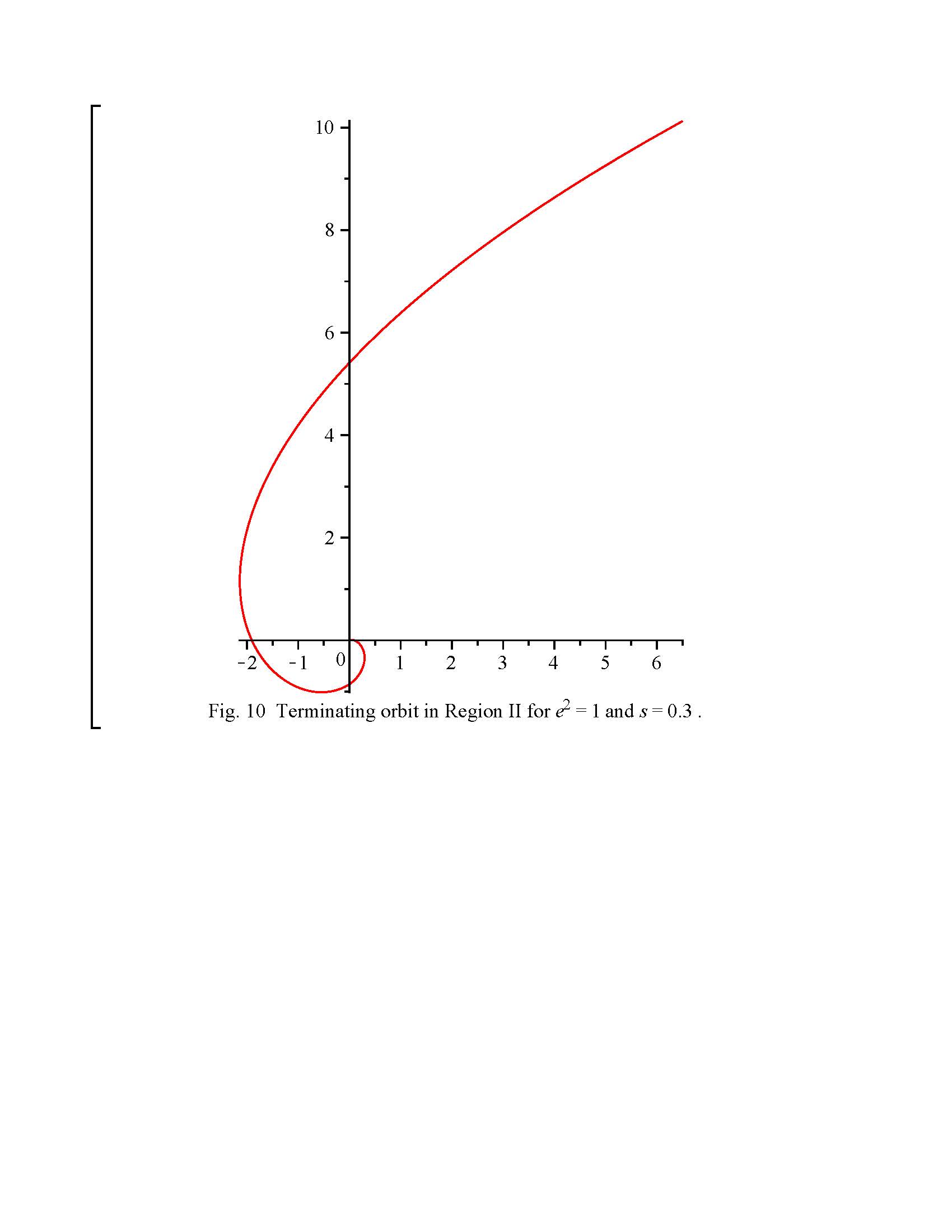}%
\caption{Terminating orbit in Region II for $e^{2}=1$ and $s=0.3$.}%
\label{Fig.10}%
\end{center}
\end{figure}

\begin{figure}
[ptb]
\begin{center}
\includegraphics[
natheight=11.000400in,
natwidth=8.500200in,
height=5.732in,
width=5.1984in
]%
{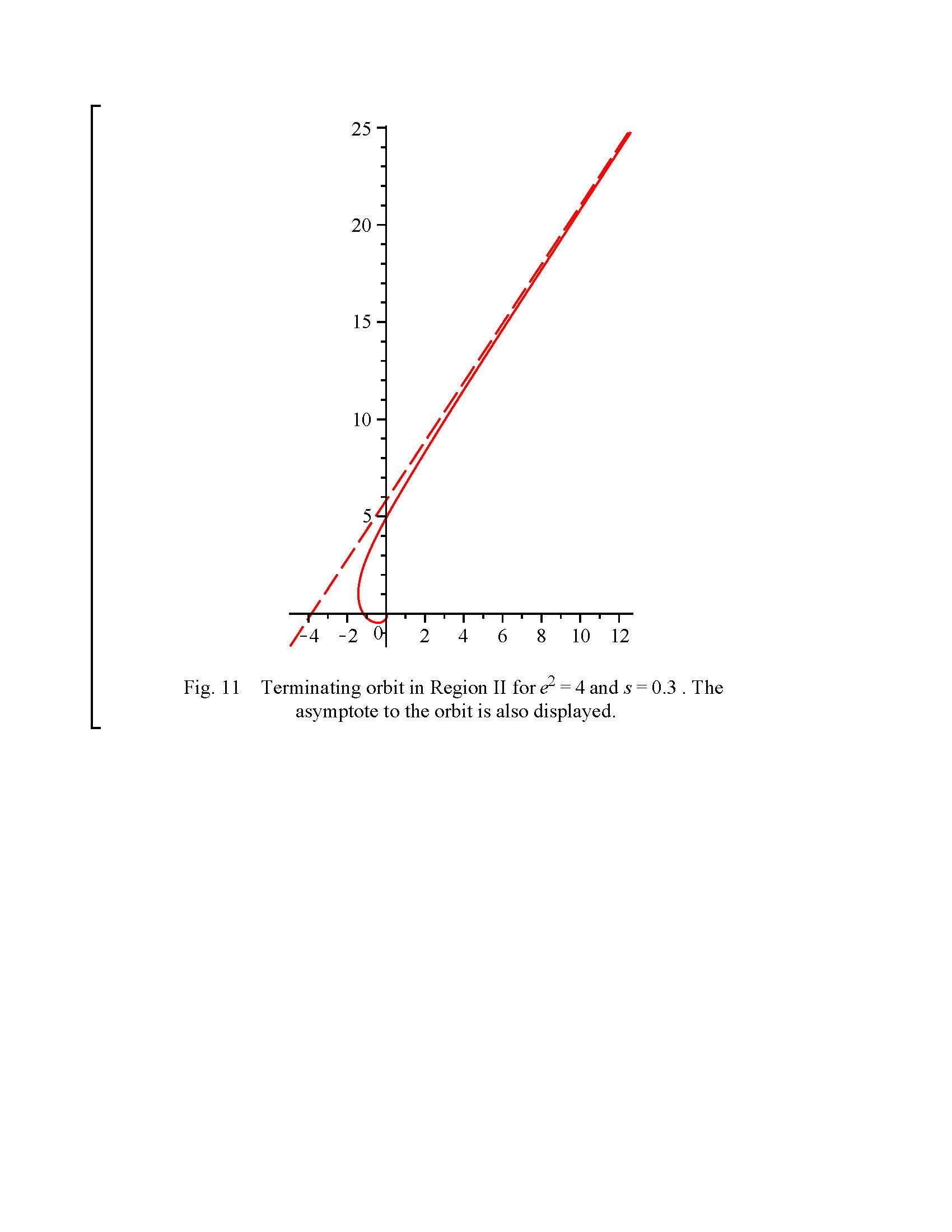}%
\caption{Terminating orbit in Region II for $e^{2}=4$ and $s=0.3$. The
asymptote to the orbit is also displayed.}%
\label{Fig.11}%
\end{center}
\end{figure}

\begin{figure}
[ptb]
\begin{center}
\includegraphics[
natheight=11.000400in,
natwidth=8.500200in,
height=5.9975in,
width=5.3238in
]%
{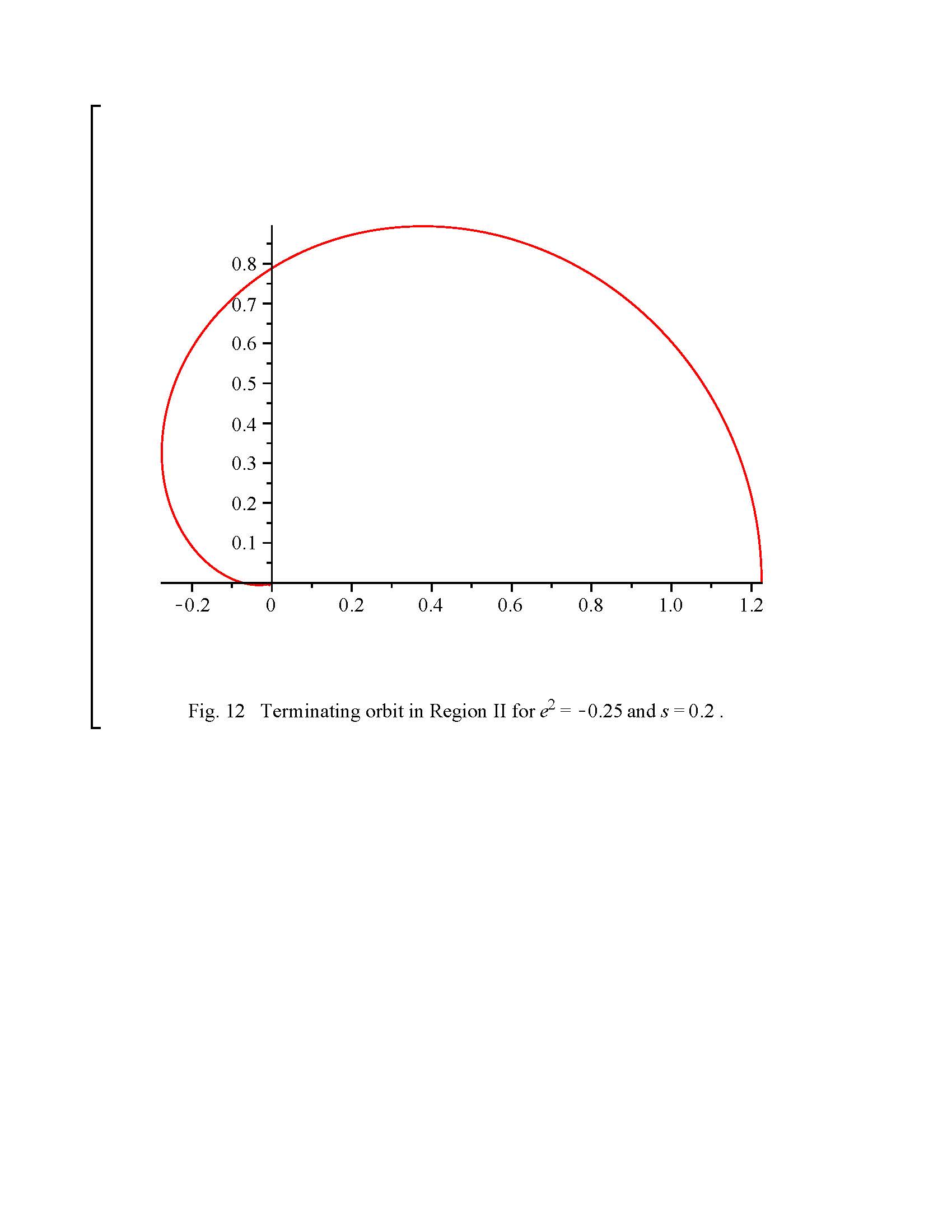}%
\caption{Terminating orbit in Region II for $e^{2}=-0.25$ and $s=0.2$.}%
\label{Fig.12}%
\end{center}
\end{figure}

\begin{figure}
[ptb]
\begin{center}
\includegraphics[
natheight=11.000400in,
natwidth=8.500200in,
height=6.0347in,
width=5.4596in
]%
{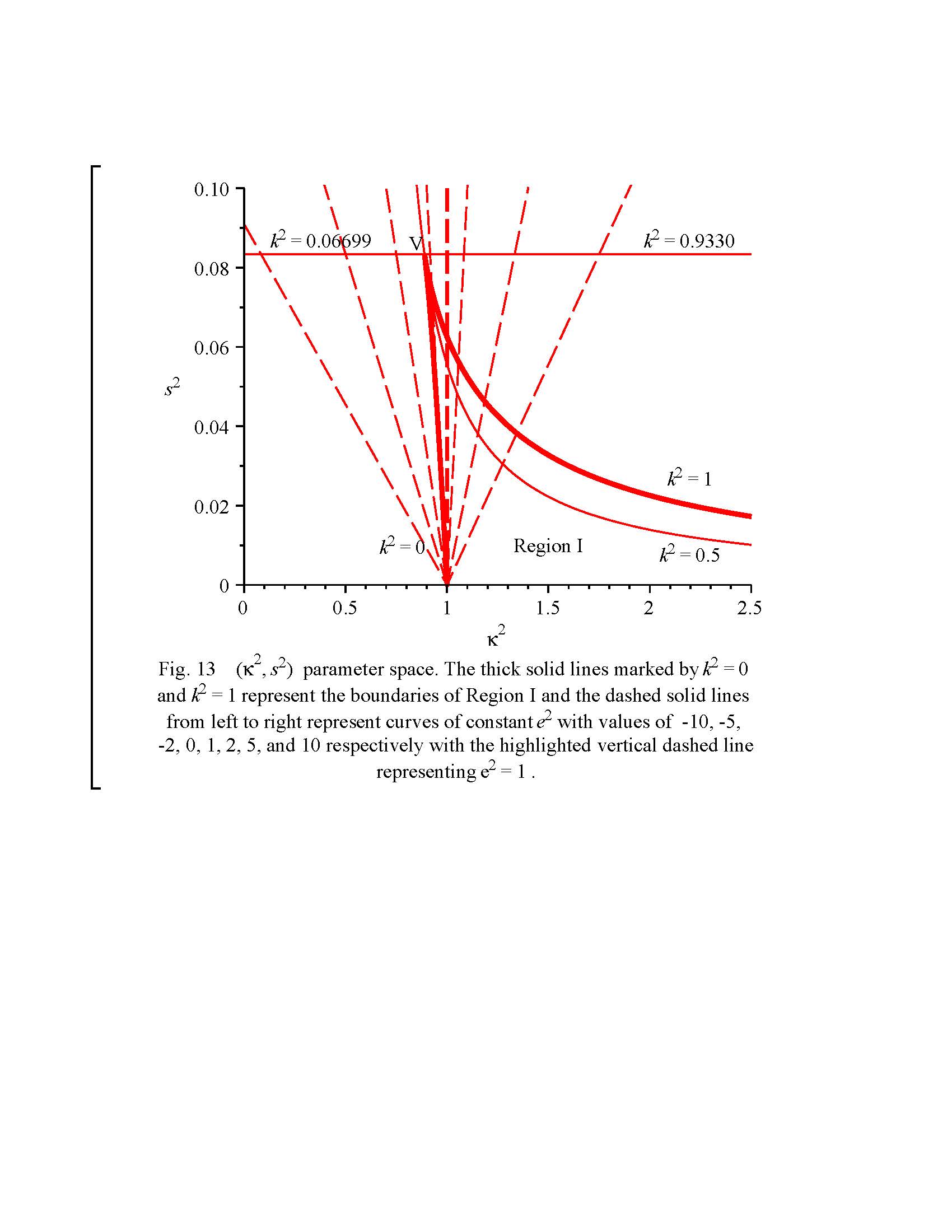}%
\caption{$(\kappa^{2},s^{2})$ parameter space. The thick lines marked by
$k^{2}=0$ and $k^{2}=1$ represent the boundaries of Region I and the dashed
solid lines represent curves of constant $e^{2}$ with values of, from left to
right, $-10,-5,-2,0,1,2,5,$ and $10$ respectively with the highlighted
vertical dashed line representing $e^{2}=1$.}%
\label{Fig.13}%
\end{center}
\end{figure}

\begin{figure}
[ptb]
\begin{center}
\includegraphics[
natheight=11.000400in,
natwidth=8.500200in,
height=6.2958in,
width=5.7303in
]%
{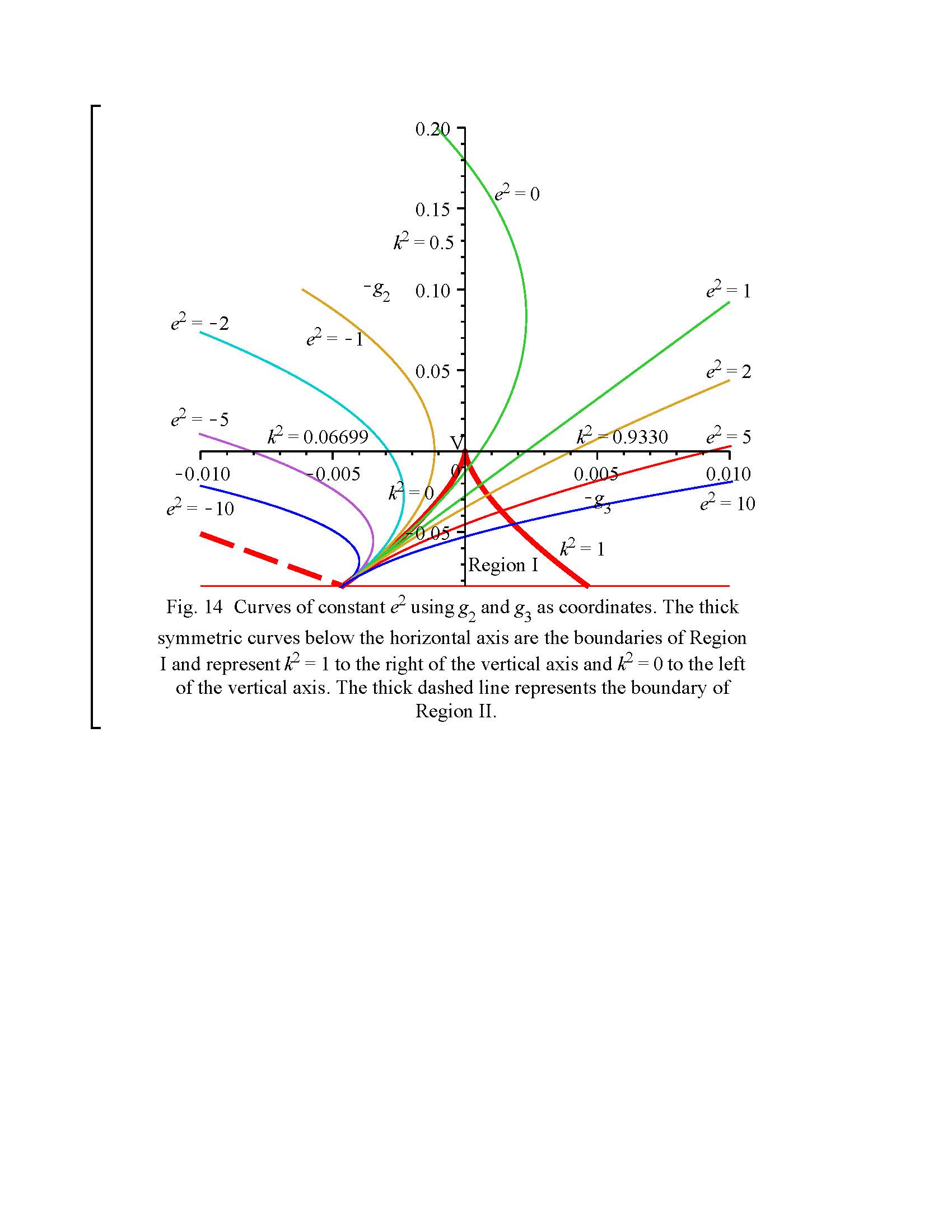}%
\caption{Curves of constant $e^{2}$ using $g_{2}$ and $g_{3}$ as coordinates.
The thick symmetric curves below the horizontal axis are the boundaries of
Region I and represent $k^{2}=1$ to the right of the vertical axis and
$k^{2}=0$ to the left of the vertical axis. The thick dashed line represents
the boundary of Region II.}%
\label{Fig.14}%
\end{center}
\end{figure}

The use of $e^{2}$ and $s^{2}$ for characterizing an orbit is seen to arise
naturally from eq.(10) or (13), and the advantage of this characterization is
enhanced by the fact that the non-terminating orbits inside Region I
(excluding its boundaries) can be conveniently classified into three types:
the elliptic-type for $-1/3\leq e^{2}<1$ (\ref{Fig.3}), the parabolic-type for
$e^{2}=1$ (\ref{Fig.4}), and the hyperbolic-type for $e^{2}>1$ (\ref{Fig.5}).
With the star or black hole at the origin, the elliptic-type orbit is a bound
precessing elliptic orbit with the precessional angle $\Delta\phi$ given by
eq.(27) and with $q_{\max}$ and $q_{\min}$ given by eqs.(23) and (24). The
parabolic-type orbit is an unbound orbit in which a particle approaches from
infinity at $\phi=0$ and can, in general, go around the star or black hole a
number of times as its polar angle increases from $0$ to $2K(k)/\gamma$ before
going off to infinity. The hyperbolic-type orbit is also an unbound orbit in
which a particle approaches from infinity along an incoming asymptote at an
angle $\Psi_{1}$ given by eq.(29) to the horizontal axis, turning
counter-clockwise about the star to its right on the horizontal axis, and
leaving along an outgoing asymptote at an angle $\Psi_{2}$ given by eq.(30).
The $x$ coordinates of the points where the two asymptotes intersect the
horizontal axis relative to the origin are given by eq.(32), a straight line
through the origin that makes an angle $\chi$ with the horizontal axis given
by eq.(33) is the symmetry axis of the hyperbolic-type orbit, and the impact
parameter or the perpendicular distance from the origin to the asymptote is
given by eq.(34). As for the terminating orbits inside Region I, a particle
initially at a distance $q_{1}$ given by eq.(36) from the black hole at
$\phi=0$ plunges into the center of the black hole when the polar angle
$\phi=\phi_{1}$ given by eq.(37), and $q_{1}$ is finite for all values of
$e^{2}$ (\ref{Fig.6}, \ref{Fig.7} and \ref{Fig.8}). Region II allows only
terminating orbits. The initial distance $q_{2}$ of the particle from the
black hole is finite and is given by eq.(44) for $-\infty\leq e^{2}<1$
(\ref{Fig.9} and \ref{Fig.12}), it is infinite at $\phi=0$ for $e^{2}=1$
(\ref{Fig.10}), and it is infinite at $\phi=\Psi_{2}$ given by eq.(46) for
$e^{2}>1$ (\ref{Fig.11}). The particle plunges into the center of the black
hole when its polar angle $\phi=\phi_{2}$ given by eq.(45).

\begin{figure}
[ptb]
\begin{center}
\includegraphics[
natheight=11.000400in,
natwidth=8.500200in,
height=6.4515in,
width=5.7821in
]%
{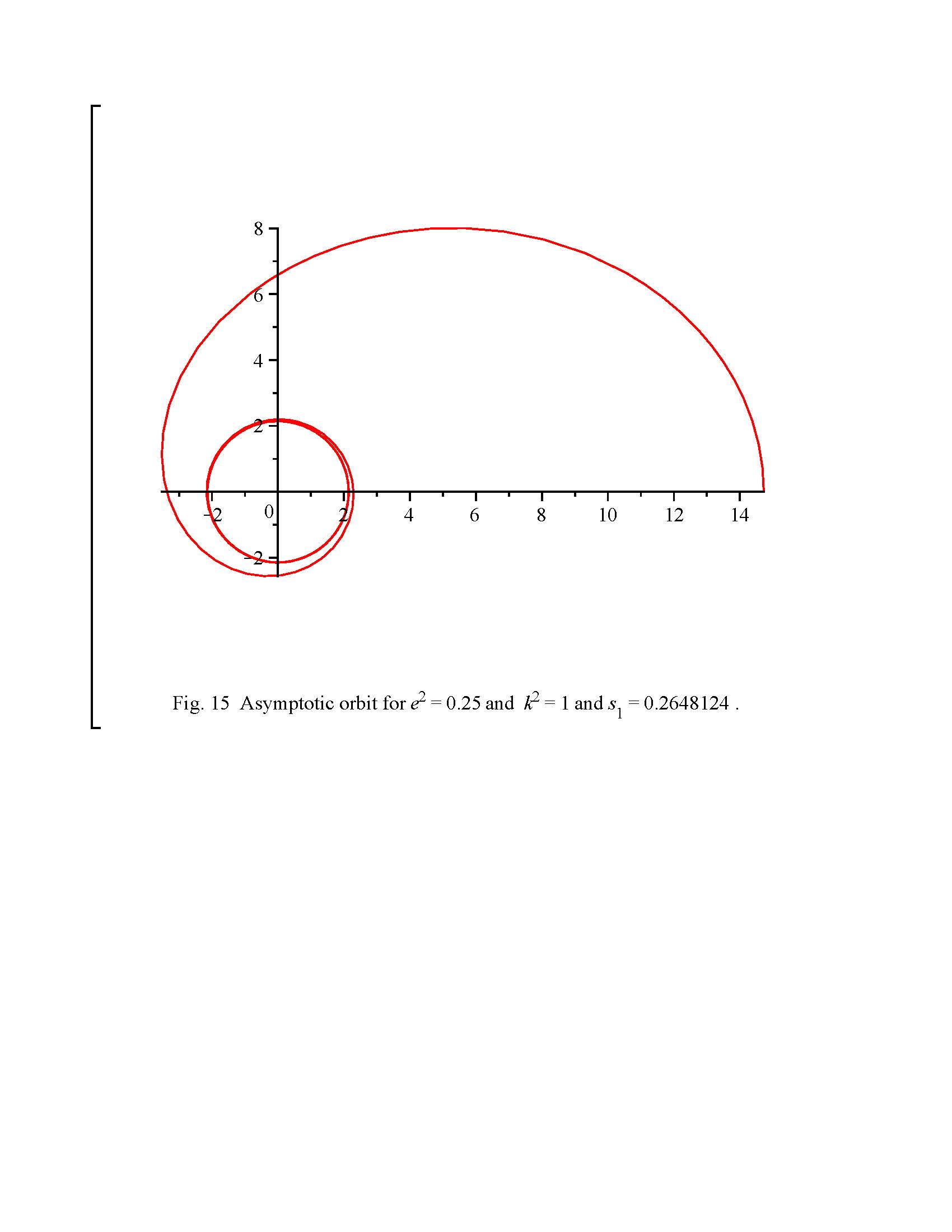}%
\caption{Asymptotic orbit for $e^{2}=0.25$ and $k^{2}=1$ and $s_{1}%
=0.2648124$.}%
\label{Fig.15}%
\end{center}
\end{figure}

\begin{figure}
[ptb]
\begin{center}
\includegraphics[
natheight=11.000400in,
natwidth=8.500200in,
height=6.6936in,
width=5.9387in
]%
{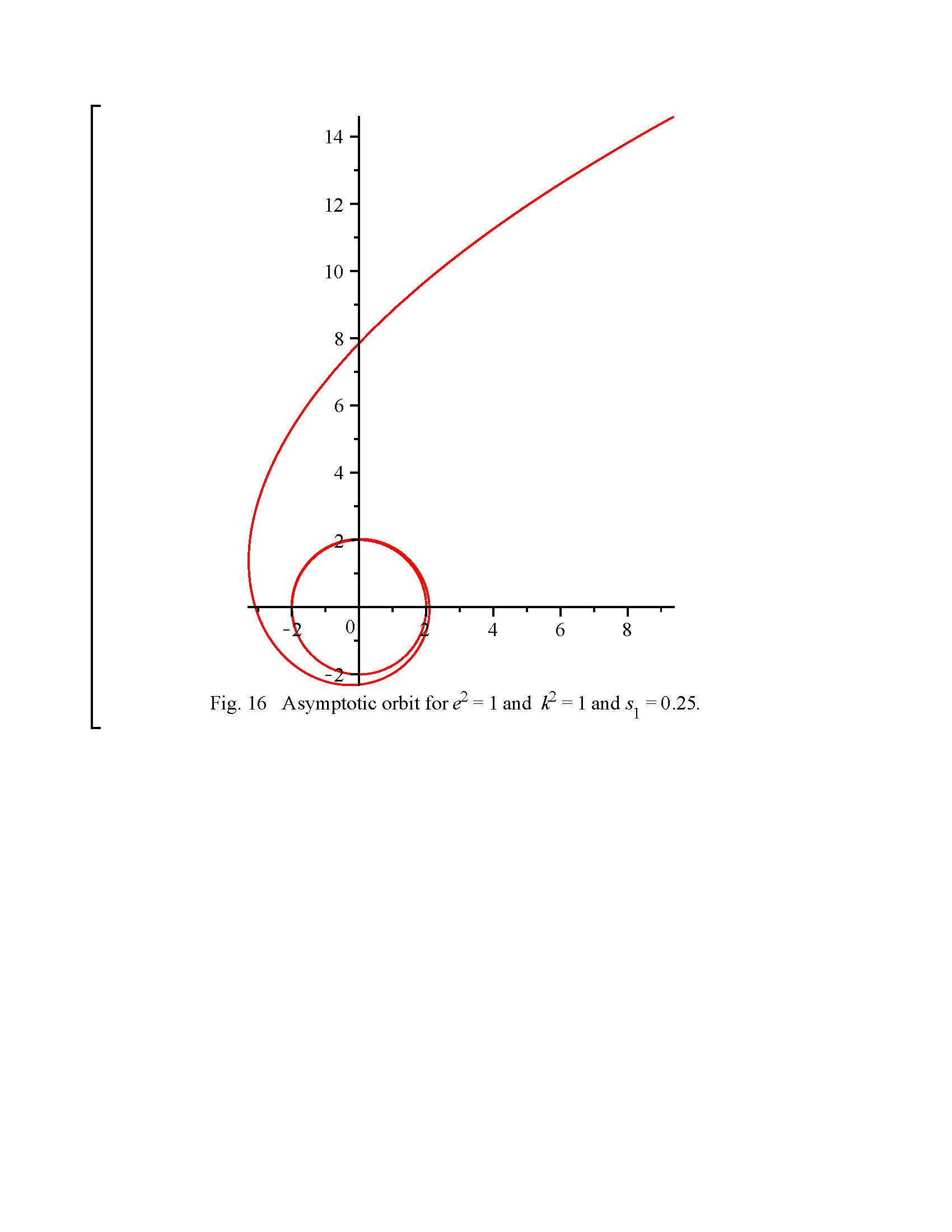}%
\caption{Asymptotic orbit for $e^{2}=1$ and $k^{2}=1$ and $s_{1}=0.25$.}%
\label{Fig.16}%
\end{center}
\end{figure}

\begin{figure}
[ptb]
\begin{center}
\includegraphics[
natheight=11.000400in,
natwidth=8.500200in,
height=6.5267in,
width=5.8237in
]%
{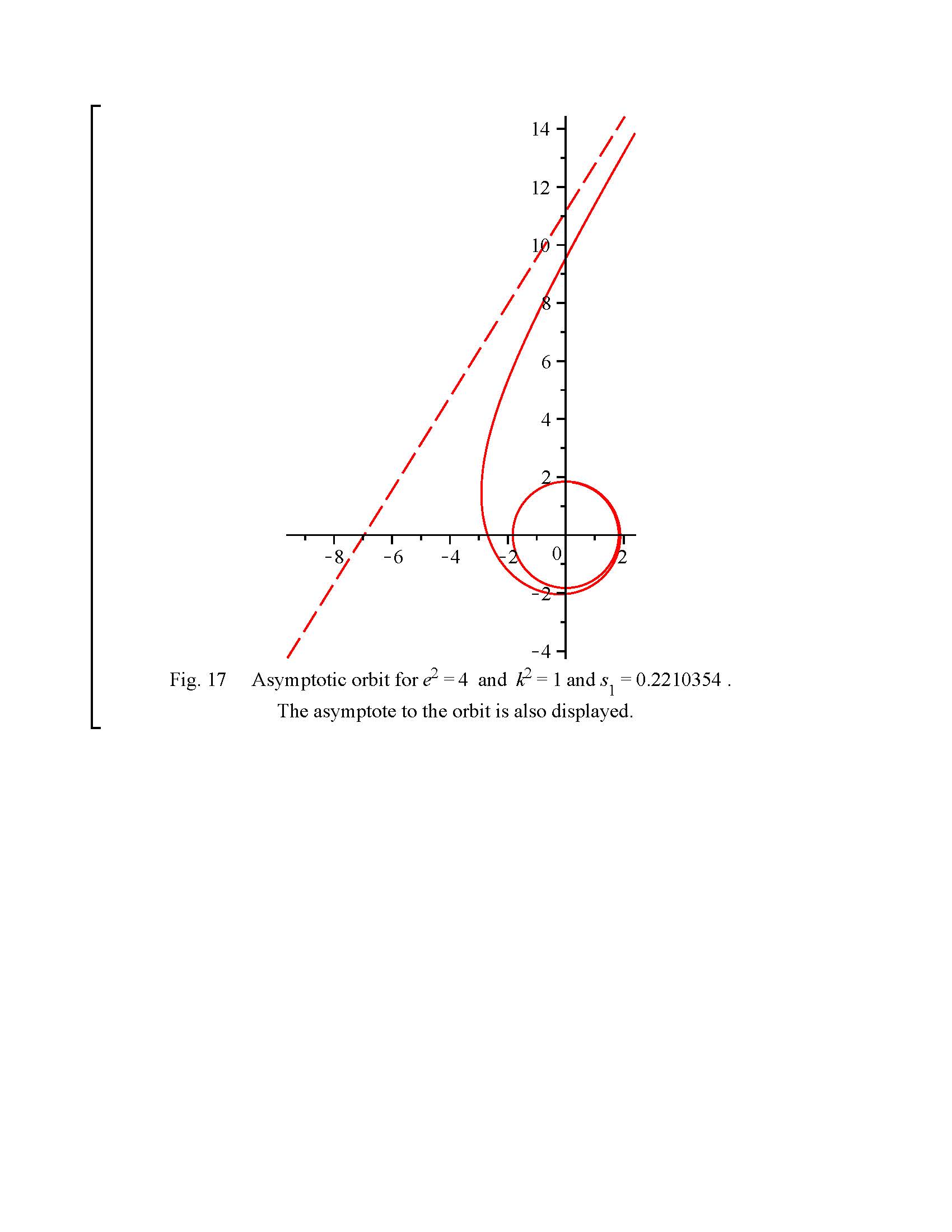}%
\caption{Asymptotic orbit for $e^{2}=4$ and $k^{2}=1$ and $s_{1}=0.2210354$.
The asymptote to the orbit is also displayed.}%
\label{Fig.17}%
\end{center}
\end{figure}

\begin{figure}
[ptb]
\begin{center}
\includegraphics[
natheight=11.000400in,
natwidth=8.500200in,
height=6.0001in,
width=5.553in
]%
{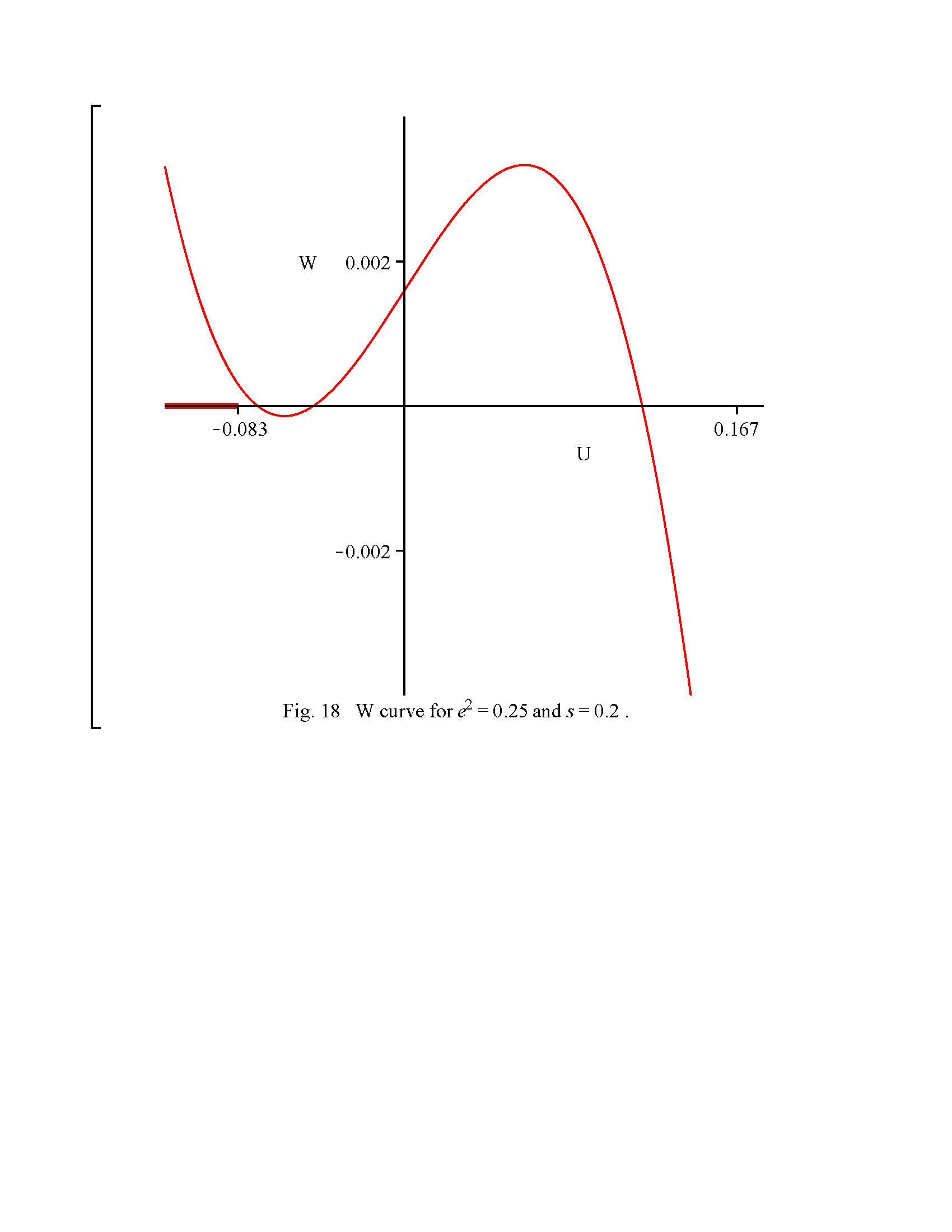}%
\caption{$W$ curve for $e^{2}=0.25$ and $s=0.2$.}%
\label{Fig.18}%
\end{center}
\end{figure}

\begin{figure}
[ptb]
\begin{center}
\includegraphics[
natheight=11.000400in,
natwidth=8.500200in,
height=6.0684in,
width=5.7199in
]%
{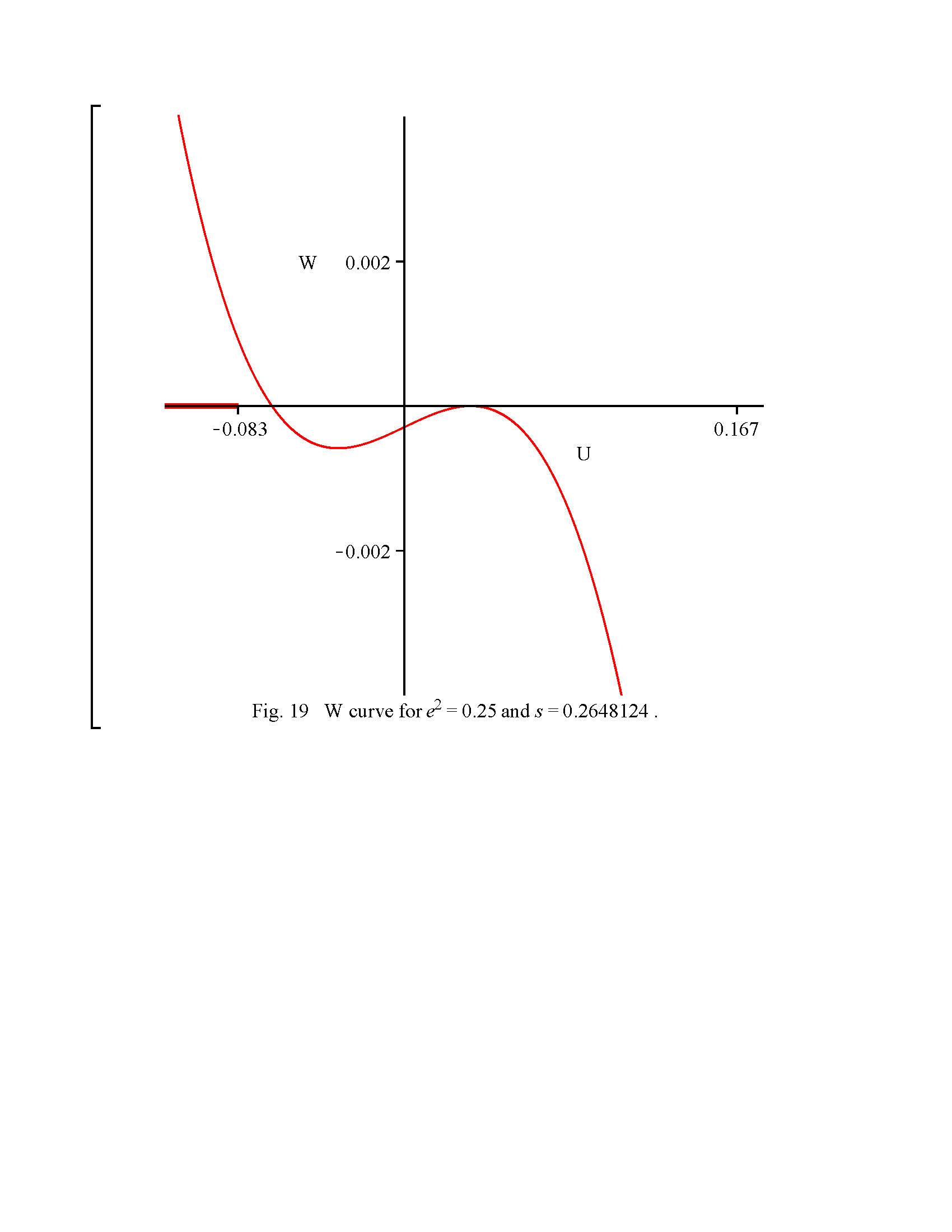}%
\caption{$W$ curve for $e^{2}=0.25$ and $s=0.2648124$.}%
\label{Fig.19}%
\end{center}
\end{figure}

\begin{figure}
[ptb]
\begin{center}
\includegraphics[
natheight=11.000400in,
natwidth=8.500200in,
height=5.8513in,
width=5.4699in
]%
{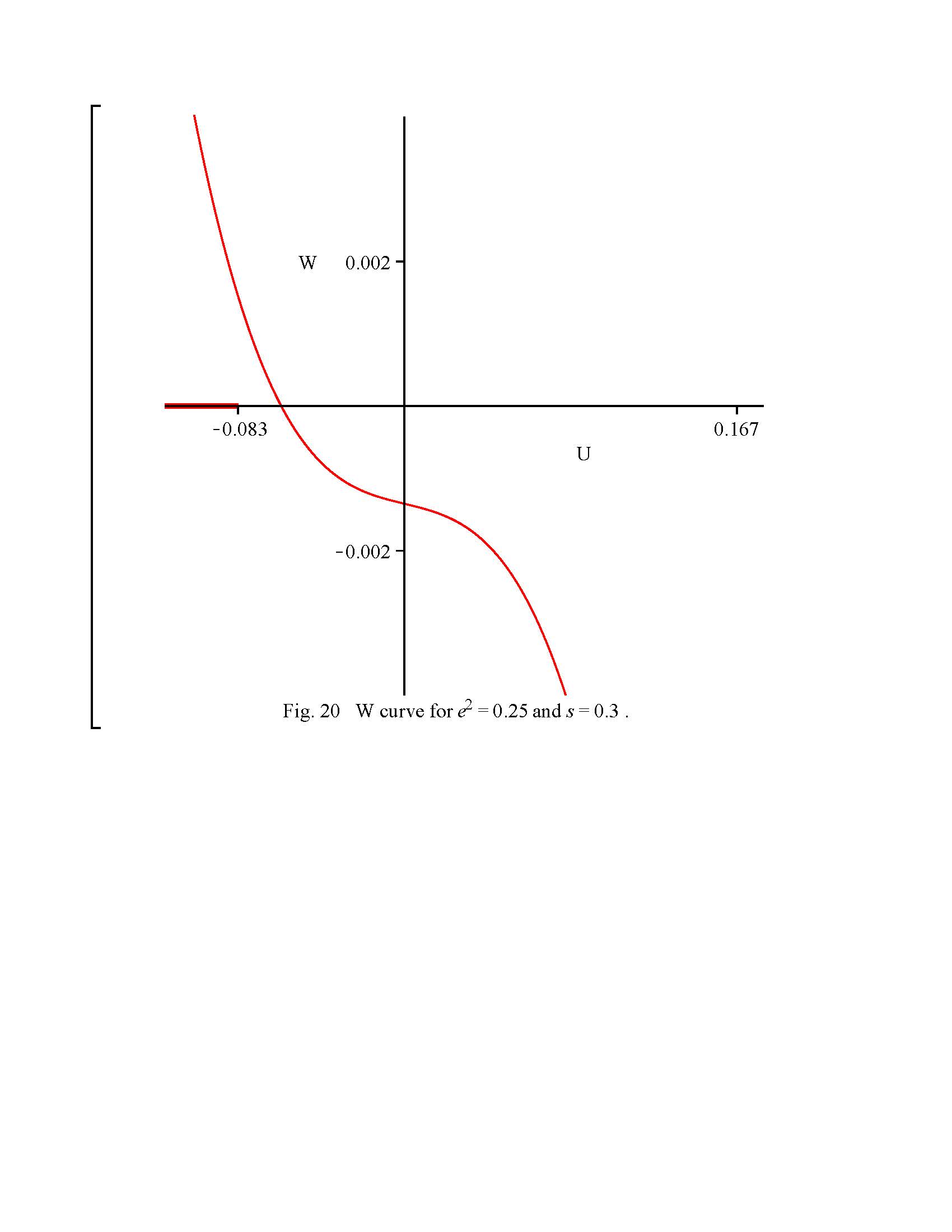}%
\caption{$W$ curve for $e^{2}=0.25$ and $s=0.3$.}%
\label{Fig.20}%
\end{center}
\end{figure}

\begin{figure}
[ptb]
\begin{center}
\includegraphics[
natheight=11.000400in,
natwidth=8.500200in,
height=6.5631in,
width=5.8652in
]%
{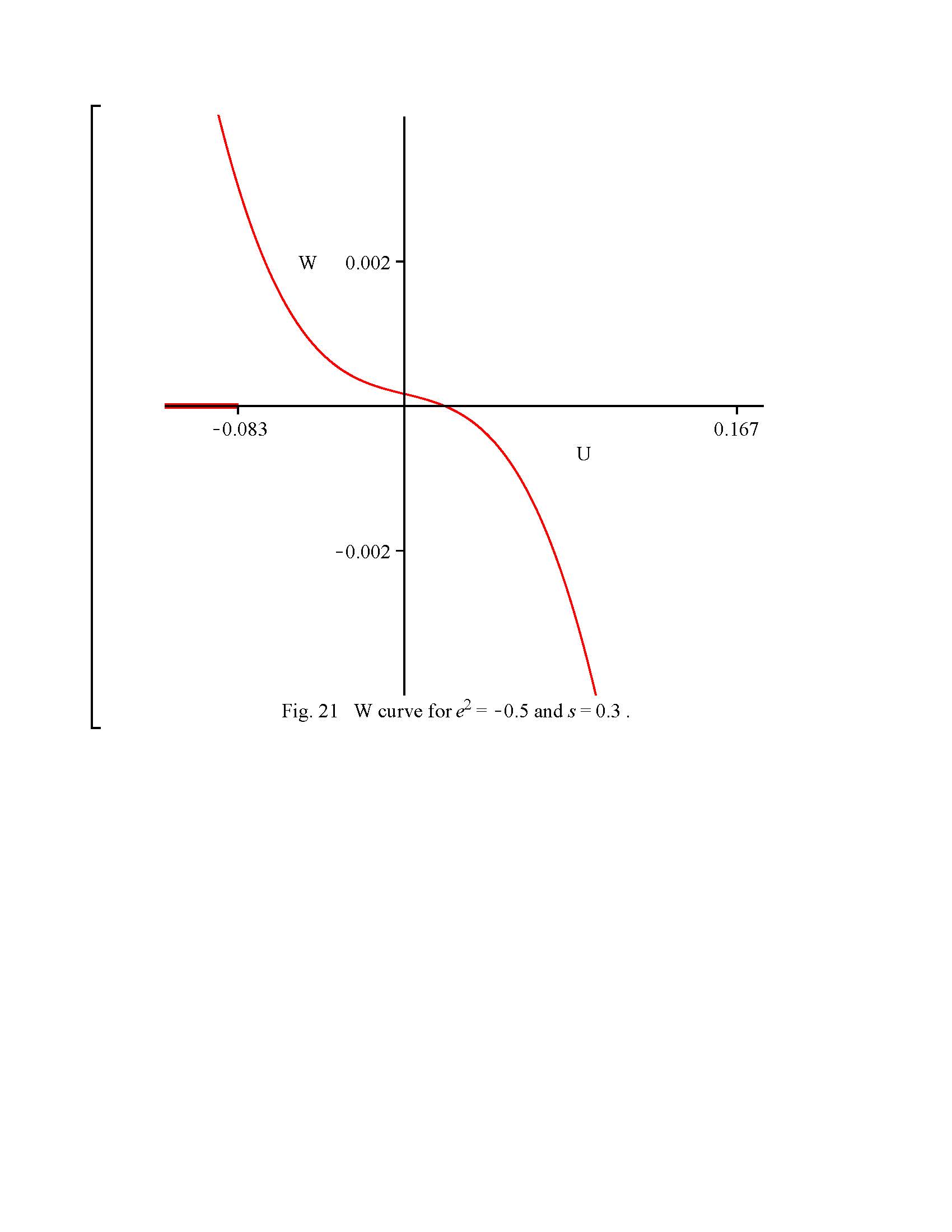}%
\caption{$W$ curve for $e^{2}=-0.5$ and $s=0.3$.}%
\label{Fig.21}%
\end{center}
\end{figure}

\begin{figure}
[ptb]
\begin{center}
\includegraphics[
natheight=11.000400in,
natwidth=8.500200in,
height=6.6054in,
width=5.6887in
]%
{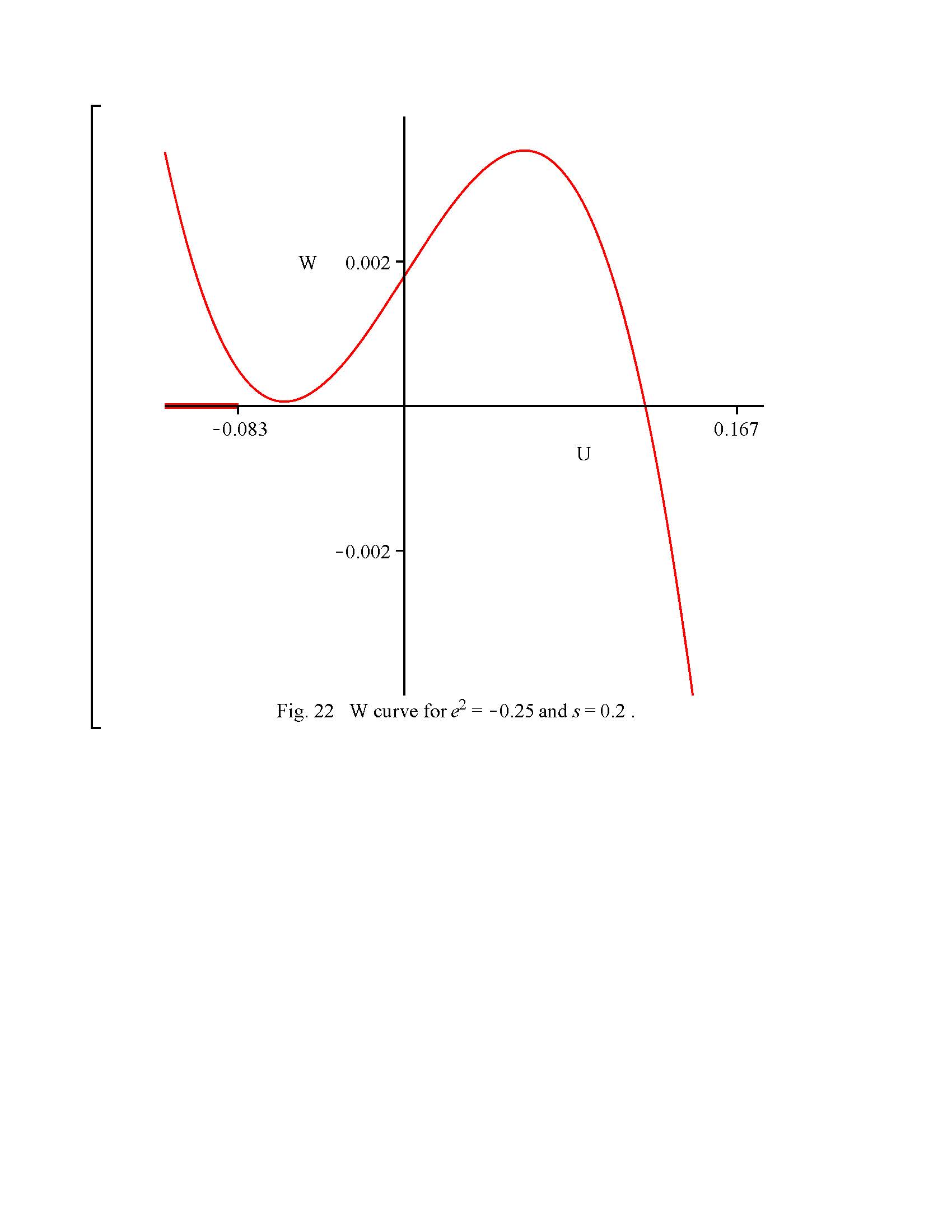}%
\caption{$W$ curve for $e^{2}=-0.25$ and $s=0.2$.}%
\label{Fig.22}%
\end{center}
\end{figure}

\begin{figure}
[ptb]
\begin{center}
\includegraphics[
natheight=11.000400in,
natwidth=8.500200in,
height=6.6271in,
width=5.9698in
]%
{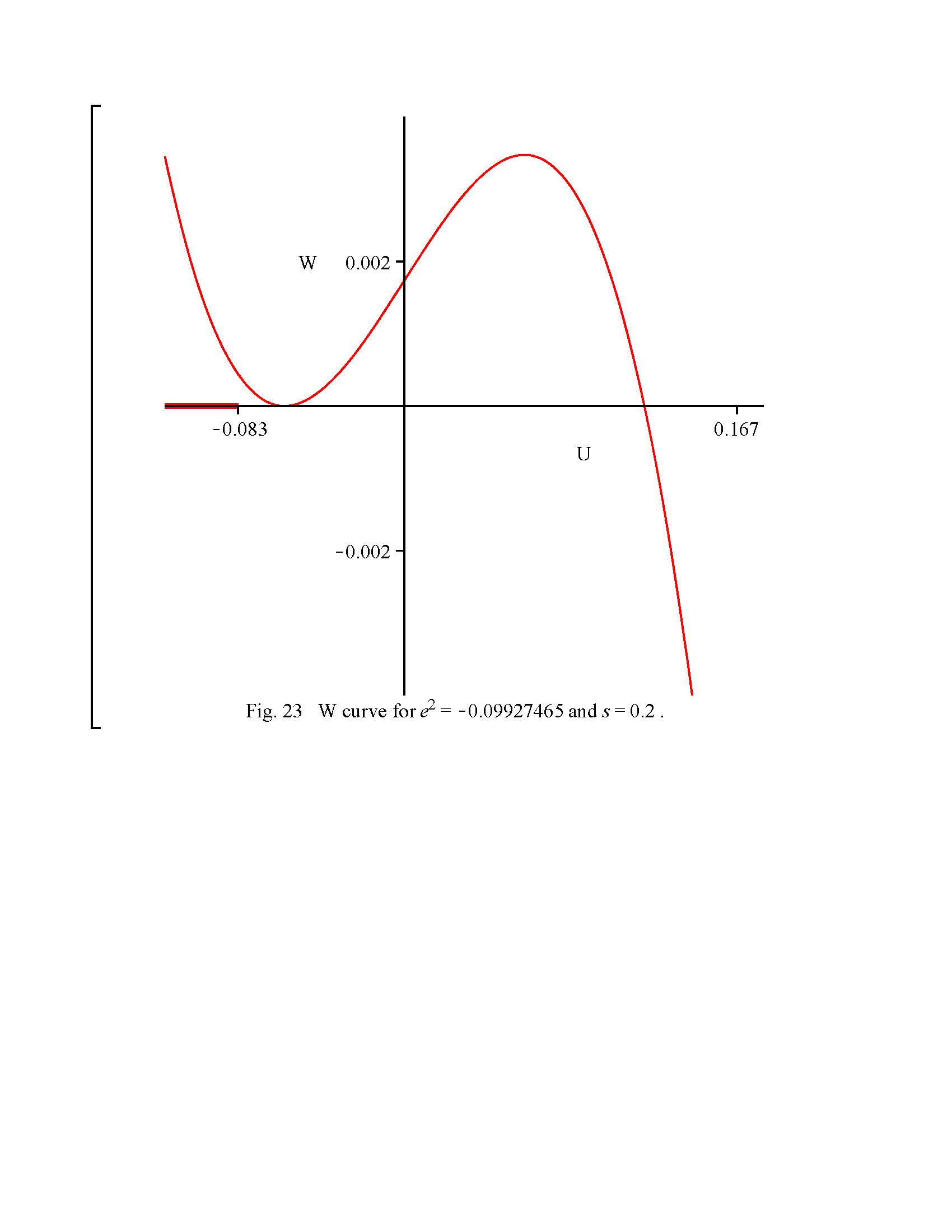}%
\caption{$W$ curve for $e^{2}=-0.09927465$ and $s=0.2$.}%
\label{Fig.23}%
\end{center}
\end{figure}

Two types of orbits occur on the boundary $s_{1}^{\prime}$: stable circular
orbits of radii $q_{c}$ given by eq.(57) and terminating orbits given by
eq.(67). Two types of orbits occur on the boundary $s_{1}$: \ unstable
circular orbits of radii $q_{u}$ given by eq.(69), and asymptotic orbits given
by eq.(62) with the particle starting from $q=q_{\max}$ for $-1/3\leq e^{2}<1$
(\ref{Fig.15}), or with the particle starting from infinity at the polar angle
$\phi=\Psi_{1}$ given by eq.(66) for $e^{2}\geq1$ (\ref{Fig.16} and
\ref{Fig.17}). The asymptotic orbits end up circling the star or black hole
with a radius that asymptotically approaches $q_{\min}$ given by eq.(65). As
the boundary $s_{1}$ is crossed from Region I to Region II, the
non-terminating orbit of Region I becomes an asymptotic orbit on $s_{1}$ and
then becomes the terminating orbit of Region II, while the terminating orbit
of Region I becomes an unstable circular orbit on $s_{1}$ and becomes part of
the terminating orbit in Region II. On the other hand, as the boundary
$s_{1}^{\prime}$ is crossed from Region II to Region I, the terminating orbit
of Region II becomes a terminating orbit on $s_{1}^{\prime}$ and then becomes
the terminating orbit of Region I, while a stable circular orbit becomes
possible on $s_{1}^{\prime}$ and then becomes an elliptical orbit in Region I.
These transitions are illustrated by pictures of the corresponding changes in
the "effective potential" curve $W$ (\ref{Fig.18}-\ref{Fig.23}). All these
observations could not have been so clearly presented without the help of the
two curves that clearly mark the boundaries $s_{1}^{\prime}$ and $s_{1}$
between Regions I and II on a universal map (\ref{Fig.1} and \ref{Fig.2}). We
consider the presentation of all these clearly identifiable transitions of
orbits from Region I to Region II to be one of the distinctive results of this paper.

The use of $e^{2}$ for characterizing the orbits is also related to the fact
that in the weak field limit and for $e^{2}\geq0$, $e$ becomes the Newtonian
eccentricity $e_{N}$ that describes the classical Newtonian orbit given by
eq.(110) or (121), as can be seen from eqs.(107)-(109) and (112). Thus the
usual division of the Newtonian orbit into elliptic for $0\leq e_{N}<1$,
parabolic for $e_{N}=1$, and hyperbolic for $e_{N}>1$ is seen to be a special
case of the division for the general relativistic non-terminating orbits in
Region I. We showed that in Region I close to $s_{1}^{\prime}$, there exist
elliptic orbits with a non-Newtonian eccentricity as shown in eqs. (127) and
(129). For the hyperbolic-type orbits in the Newtonian limit, an approximate
analytic expression [eq.(134)] has been presented that gives the difference
between the general relativistic and the Newtonian calculations for the angle
between the two asymptotes and this expression is different from the one given
by other authors [13].

The power of our analytic expressions for the orbits in the Schwarzschild
geometry is further illustrated by the numerous analytic simple expressions
for many quantities that are presented for various special cases in Section 7.
These examples include the cases of $k^{2}=0,$ $1/2,$ and $1,$ and of
$e^{2}=1$ and $-1/3$, and the expressions include those for $\varepsilon$, the
geometrical eccentricity defined by eq.(25), $q_{1}$ and $q_{2}$ discussed
above, and also $q_{2\max}$ in the region $e^{2}<0$. Every expression from
Eq.(73) to Eq.(106) has been used to check the extensive numerical data for
various orbits that were presented in refs.[2] and [3], and we consider these
simple relationships to be a bonus of our analytic formulation of the problem.

To better appreciate our presentation and characterization of all possible
orbits in the Schwarzschild geometry, we described in Section 9 the analysis
of Chandrasekhar [5] in terms of our notations for the purpose of comparisons.
Equations (135), (149) and (159) are, respectively, mathematically equivalent
to our eqs.(18), (35) and (43). Instead of using the parameters $e^{2}$ and
$s$ as we adopted, Chandrasekhar's formulation essentially used the parameters
$\varepsilon$ defined by eqs.(26) and (158), and $\mu$ defined by eq.(138),
for characterizing the orbits, but he often had to refer to the total energy
[$E$ of eq.(6)] and total angular momentum $L$ [which is a constant multiple
of $h$ of eq.(5)] of the particle and the effective potential energy curve
when discussing certain features of the orbits. Although the orbit equations
appear to involve only simple trigonometric functions, the "polar angles"
$\chi$ and $\xi$ are not the actual physical polar angle $\phi$ but are
complicated implicit functions of $\phi$ [see eqs.(137), (150), and the quote
after eq.(159)]. The two important physical quantities, $E$ and $h$, are not
only complicated functions of $\varepsilon$ and $\mu$, but the relations are
different for different types of orbits [see eqs.(143), (144), (160), and
(161)]. The parameter $\varepsilon$ can be real or purely imaginary, but it
has a physical meaning given by eq.(25) only in the range of real values
between $0 $ and $1$. The boundaries between different regions for different
types of orbits are expressed as certain equations that relate $\varepsilon$
and $\mu$. Chandrasekhar's formalism does not present all possible orbits on a
single universal map. The numerous equations in Section 9, however, provide
the connections of Chandrasekhar's formulation to ours that would be helpful
to many readers who are familiar with the former. Equation (163) or (34) also
corrects an erroneous expression (162) given by Chandrasekhar.

Acknowledgements

We are very grateful to Dr. J.L. Martin for his many comments and for drawing
our attention to his work on the circular orbits, and to Christopher Berry for
his helpful comments and interest.

\bigskip

\bigskip

References

*Electronic address: fhioe@sjfc.edu

[1] F.T. Hioe, Phys. Lett. A 373, 1506 (2009).

[2] F.T. Hioe and D. Kuebel, Phys. Rev. D 81, 084017 (2010).

[3] F.T. Hioe and D. Kuebel, arXiv:1008.1964 v1 (2010).

[4] F.T. Hioe and D. Kuebel, arXiv:1010.0996 v2 (2010).

[5] S. Chandrasekhar: The Mathematical Theory of Black Holes, Oxford
University Press, 1983, Chapter 3.

[6] C. Darwin, Proc. Roy. Soc. Lond. A249, 180 (1958), ibid. A263, 39 (1961).

[7] A.R. Forsyth, Proc. Roy. Soc. Lond. A97, 145 (1920).

[8] E.T. Whittaker: A Treatise on the Analytical Dynamics of Particles and
Rigid Bodies, 4th Edition, Dover, New York, 1944, Chapter XV.

[9] Y. Hagihara, Japanese Journal of Astronomy and Geophysics VIII, 67 (1931).

[10] M.P. Hobson, G. Efstathiou and A.N. Lasenby: General Relativity,
Cambridge University Press, 2006, Chapters 9 and 10.

[11] P.F. Byrd and M.D. Friedman: Handbook of Elliptic Integrals for Engineers
and Scientists, 2nd Edition, Springer-Verlag, New York, 1971.

[12] J.L. Martin: General Relativity, Revised Edition, Prentice Hall, New York
1996, Chapter 4, and private communication.

[13] J.M. Longuski, E. Fischbach, and D.J. Scheeres, Phys. Rev. Lett. 86, 2942
(2001), J.M. Longuski, E. Fischbach, D.J. Scheeres, G. Giampierri, and R.S.
Park, Phys. Rev. D 69, 042001 (2004).

\bigskip

\bigskip

\bigskip

\bigskip

\bigskip

\bigskip%
%TCIMACRO{\FRAME{ftbpFU}{5.303in}{5.2641in}{0pt}{\Qcb{The boundaries of Regions
%I, II and II' are displayed. The vertex point $V$ occurs at the intersection
%of the left boundary $s_{1}^{\prime}$ and the upper boundary $s_{1}$.}%
%}{\Qlb{Fig.1}}{parfig1.jpg}{\special{ language "Scientific Word";
%type "GRAPHIC";  display "USEDEF";  valid_file "F";  width 5.303in;
%height 5.2641in;  depth 0pt;  original-width 8.5002in;
%original-height 11.0004in;  cropleft "0";  croptop "1";  cropright "1";
%cropbottom "0";  filename 'parfig1.jpg';file-properties "XNPEU";}}}%
%BeginExpansion
%EndExpansion%
%TCIMACRO{\FRAME{ftbpFU}{4.7193in}{5.1188in}{0pt}{\Qcb{Curves of constant
%$k^{2}$. Within Region I only three values of $k^{2}$ are displayed. All
%curves intersect the point $V$. The dashed horizontal line through $V$ has
%$s^{2}=1/12$.}}{\Qlb{Fig.2}}{parfig2.jpg}%
%{\special{ language "Scientific Word";  type "GRAPHIC";  display "USEDEF";
%valid_file "F";  width 4.7193in;  height 5.1188in;  depth 0pt;
%original-width 8.5002in;  original-height 11.0004in;  cropleft "0";
%croptop "1";  cropright "1";  cropbottom "0";
%filename 'parfig2.JPG';file-properties "XNPEU";}}}%
%BeginExpansion
%EndExpansion%
%TCIMACRO{\FRAME{ftbpFU}{4.7297in}{5.3074in}{0pt}{\Qcb{Elliptic-type orbit in
%Region I for $e^{2}=0.25$ and $s=0.2$.}}{\Qlb{Fig.3}}{reg125q.jpg}%
%{\special{ language "Scientific Word";  type "GRAPHIC";  display "USEDEF";
%valid_file "F";  width 4.7297in;  height 5.3074in;  depth 0pt;
%original-width 8.5002in;  original-height 11.0004in;  cropleft "0";
%croptop "1";  cropright "1";  cropbottom "0";
%filename 'reg125q.jpg';file-properties "XNPEU";}}}%
%BeginExpansion
%EndExpansion%
%TCIMACRO{\FRAME{ftbpFU}{4.3033in}{4.785in}{0pt}{\Qcb{Parabolic-type orbit in
%Region I for $e^{2}=1$ and $s=0.2$.}}{\Qlb{Fig.4}}{reg11q.jpg}%
%{\special{ language "Scientific Word";  type "GRAPHIC";  display "USEDEF";
%valid_file "F";  width 4.3033in;  height 4.785in;  depth 0pt;
%original-width 8.5002in;  original-height 11.0004in;  cropleft "0";
%croptop "1";  cropright "1";  cropbottom "0";
%filename 'reg11q.jpg';file-properties "XNPEU";}}}%
%BeginExpansion
%EndExpansion%
%TCIMACRO{\FRAME{ftbpFU}{5.1984in}{6.0494in}{0pt}{\Qcb{Hyperbolic-type orbit in
%Region I for $e^{2}=4$ and $s=0.2$. The asymptotes to the orbit are also
%displayed.}}{\Qlb{Fig.5}}{reg14q.jpg}{\special{ language "Scientific Word";
%type "GRAPHIC";  display "USEDEF";  valid_file "F";  width 5.1984in;
%height 6.0494in;  depth 0pt;  original-width 8.5002in;
%original-height 11.0004in;  cropleft "0";  croptop "1";  cropright "1";
%cropbottom "0";  filename 'reg14q.jpg';file-properties "XNPEU";}}}%
%BeginExpansion
%EndExpansion%
%TCIMACRO{\FRAME{ftbpFU}{5.3964in}{6.1471in}{0pt}{\Qcb{Terminating orbit in
%Region I for $e^{2}=0.25$ and $s=0.2$.}}{\Qlb{Fig.6}}{treg125q.jpg}%
%{\special{ language "Scientific Word";  type "GRAPHIC";  display "USEDEF";
%valid_file "F";  width 5.3964in;  height 6.1471in;  depth 0pt;
%original-width 8.5002in;  original-height 11.0004in;  cropleft "0";
%croptop "1";  cropright "1";  cropbottom "0";
%filename 'treg125q.jpg';file-properties "XNPEU";}}}%
%BeginExpansion
%EndExpansion%
%TCIMACRO{\FRAME{ftbpFU}{5.2823in}{5.9785in}{0pt}{\Qcb{Terminating orbit in
%Region I for $e^{2}=1$ and $s=0.2$.}}{\Qlb{Fig.7}}{treg11q.jpg}%
%{\special{ language "Scientific Word";  type "GRAPHIC";  display "USEDEF";
%valid_file "F";  width 5.2823in;  height 5.9785in;  depth 0pt;
%original-width 8.5002in;  original-height 11.0004in;  cropleft "0";
%croptop "1";  cropright "1";  cropbottom "0";
%filename 'treg11q.jpg';file-properties "XNPEU";}}}%
%BeginExpansion
%EndExpansion%
%TCIMACRO{\FRAME{ftbpFU}{5.3342in}{6.026in}{0pt}{\Qcb{Terminating orbit in
%Region I for $e^{2}=4$ and $s=0.2$.}}{\Qlb{Fig.8}}{treg14q.jpg}%
%{\special{ language "Scientific Word";  type "GRAPHIC";  display "USEDEF";
%valid_file "F";  width 5.3342in;  height 6.026in;  depth 0pt;
%original-width 8.5002in;  original-height 11.0004in;  cropleft "0";
%croptop "1";  cropright "1";  cropbottom "0";
%filename 'treg14q.jpg';file-properties "XNPEU";}}}%
%BeginExpansion
%EndExpansion%
%TCIMACRO{\FRAME{ftbpFU}{5.188in}{5.8574in}{0pt}{\Qcb{Terminating orbit in
%Region II for $e^{2}=0.25$ and $s=0.3$.}}{\Qlb{Fig.9}}{treg225q.jpg}%
%{\special{ language "Scientific Word";  type "GRAPHIC";  display "USEDEF";
%valid_file "F";  width 5.188in;  height 5.8574in;  depth 0pt;
%original-width 8.5002in;  original-height 11.0004in;  cropleft "0";
%croptop "1";  cropright "1";  cropbottom "0";
%filename 'treg225q.jpg';file-properties "XNPEU";}}}%
%BeginExpansion
%EndExpansion%
%TCIMACRO{\FRAME{ftbpFU}{5.0427in}{5.6654in}{0pt}{\Qcb{Terminating orbit in
%Region II for $e^{2}=1$ and $s=0.3$.}}{\Qlb{Fig.10}}{treg21q.jpg}%
%{\special{ language "Scientific Word";  type "GRAPHIC";  display "USEDEF";
%valid_file "F";  width 5.0427in;  height 5.6654in;  depth 0pt;
%original-width 8.5002in;  original-height 11.0004in;  cropleft "0";
%croptop "1";  cropright "1";  cropbottom "0";
%filename 'treg21q.jpg';file-properties "XNPEU";}}}%
%BeginExpansion
%EndExpansion%
%TCIMACRO{\FRAME{ftbpFU}{5.1984in}{5.732in}{0pt}{\Qcb{Terminating orbit in
%Region II for $e^{2}=4$ and $s=0.3$. The asymptote to the orbit is also
%displayed.}}{\Qlb{Fig.11}}{treg24q.jpg}{\special{ language "Scientific Word";
%type "GRAPHIC";  display "USEDEF";  valid_file "F";  width 5.1984in;
%height 5.732in;  depth 0pt;  original-width 8.5002in;
%original-height 11.0004in;  cropleft "0";  croptop "1";  cropright "1";
%cropbottom "0";  filename 'treg24q.jpg';file-properties "XNPEU";}}}%
%BeginExpansion
%EndExpansion%
%TCIMACRO{\FRAME{ftbpFU}{5.3238in}{5.9975in}{0pt}{\Qcb{Terminating orbit in
%Region II for $e^{2}=-0.25$ and $s=0.2$.}}{\Qlb{Fig.12}}{treg2neg25q.jpg}%
%{\special{ language "Scientific Word";  type "GRAPHIC";  display "USEDEF";
%valid_file "F";  width 5.3238in;  height 5.9975in;  depth 0pt;
%original-width 8.5002in;  original-height 11.0004in;  cropleft "0";
%croptop "1";  cropright "1";  cropbottom "0";
%filename 'treg2neg25q.jpg';file-properties "XNPEU";}}}%
%BeginExpansion
%EndExpansion%
%TCIMACRO{\FRAME{ftbpFU}{5.4596in}{6.0347in}{0pt}{\Qcb{$(\kappa^{2},s^{2})$
%parameter space. The thick lines marked by $k^{2}=0$ and $k^{2}=1$ represent
%the boundaries of Region I and the dashed solid lines represent curves of
%constant $e^{2}$ with values of, from left to right, $-10,-5,-2,0,1,2,5,$ and
%$10$ respectively with the highlighted vertical dashed line representing
%$e^{2}=1$.}}{\Qlb{Fig.13}}{kappa2s2.jpg}%
%{\special{ language "Scientific Word";  type "GRAPHIC";  display "USEDEF";
%valid_file "F";  width 5.4596in;  height 6.0347in;  depth 0pt;
%original-width 8.5002in;  original-height 11.0004in;  cropleft "0";
%croptop "1";  cropright "1";  cropbottom "0";
%filename 'kappa2s2.jpg';file-properties "XNPEU";}}}%
%BeginExpansion
%EndExpansion%
%TCIMACRO{\FRAME{ftbpFU}{5.7303in}{6.2958in}{0pt}{\Qcb{Curves of constant
%$e^{2}$ using $g_{2}$ and $g_{3}$ as coordinates. The thick symmetric curves
%below the horizontal axis are the boundaries of Region I and represent
%$k^{2}=1$ to the right of the vertical axis and $k^{2}=0$ to the left of the
%vertical axis. The thick dashed line represents the boundary of Region II.}%
%}{\Qlb{Fig.14}}{g2g3.jpg}{\special{ language "Scientific Word";
%type "GRAPHIC";  display "USEDEF";  valid_file "F";  width 5.7303in;
%height 6.2958in;  depth 0pt;  original-width 8.5002in;
%original-height 11.0004in;  cropleft "0";  croptop "1";  cropright "1";
%cropbottom "0";  filename 'g2g3.jpg';file-properties "XNPEU";}}}%
%BeginExpansion
%EndExpansion%
%TCIMACRO{\FRAME{ftbpFU}{5.7821in}{6.4515in}{0pt}{\Qcb{Asymptotic orbit for
%$e^{2}=0.25$ and $k^{2}=1$ and $s_{1}=0.2648124$.}}{\Qlb{Fig.15}}%
%{asym25q.jpg}{\special{ language "Scientific Word";  type "GRAPHIC";
%display "USEDEF";  valid_file "F";  width 5.7821in;  height 6.4515in;
%depth 0pt;  original-width 8.5002in;  original-height 11.0004in;
%cropleft "0";  croptop "1";  cropright "1";  cropbottom "0";
%filename 'asym25q.jpg';file-properties "XNPEU";}}}%
%BeginExpansion
%EndExpansion%
%TCIMACRO{\FRAME{ftbpFU}{5.9387in}{6.6936in}{0pt}{\Qcb{Asymptotic orbit for
%$e^{2}=1$ and $k^{2}=1$ and $s_{1}=0.25$.}}{\Qlb{Fig.16}}{asym1q.jpg}%
%{\special{ language "Scientific Word";  type "GRAPHIC";  display "USEDEF";
%valid_file "F";  width 5.9387in;  height 6.6936in;  depth 0pt;
%original-width 8.5002in;  original-height 11.0004in;  cropleft "0";
%croptop "1";  cropright "1";  cropbottom "0";
%filename 'asym1q.jpg';file-properties "XNPEU";}}}%
%BeginExpansion
%EndExpansion%
%TCIMACRO{\FRAME{ftbpFU}{5.8237in}{6.5267in}{0pt}{\Qcb{Asymptotic orbit for
%$e^{2}=4$ and $k^{2}=1$ and $s_{1}=0.2210354$. The asymptote to the orbit is
%also displayed.}}{\Qlb{Fig.17}}{asym4q.jpg}%
%{\special{ language "Scientific Word";  type "GRAPHIC";  display "USEDEF";
%valid_file "F";  width 5.8237in;  height 6.5267in;  depth 0pt;
%original-width 8.5002in;  original-height 11.0004in;  cropleft "0";
%croptop "1";  cropright "1";  cropbottom "0";
%filename 'asym4q.jpg';file-properties "XNPEU";}}}%
%BeginExpansion
%EndExpansion
%

%TCIMACRO{\TeXButton{TeX}{\TeX{\clearpage}}}%
%BeginExpansion
\TeX{\clearpage}%
%EndExpansion

\bigskip%

%TCIMACRO{\FRAME{ftbpFU}{5.553in}{6.0001in}{0pt}{\Qcb{$W$ curve for
%$e^{2}=0.25$ and $s=0.2$.}}{\Qlb{Fig.18}}{plotrootsa.jpg}%
%{\special{ language "Scientific Word";  type "GRAPHIC";  display "USEDEF";
%valid_file "F";  width 5.553in;  height 6.0001in;  depth 0pt;
%original-width 8.5002in;  original-height 11.0004in;  cropleft "0";
%croptop "1";  cropright "1";  cropbottom "0";
%filename 'plotrootsa.jpg';file-properties "XNPEU";}}}%
%BeginExpansion
%EndExpansion%
%TCIMACRO{\FRAME{ftbpFU}{5.7199in}{6.0684in}{0pt}{\Qcb{$W$ curve for
%$e^{2}=0.25$ and $s=0.2648124$.}}{\Qlb{Fig.19}}{plotrootsb.jpg}%
%{\special{ language "Scientific Word";  type "GRAPHIC";  display "USEDEF";
%valid_file "F";  width 5.7199in;  height 6.0684in;  depth 0pt;
%original-width 8.5002in;  original-height 11.0004in;  cropleft "0";
%croptop "1";  cropright "1";  cropbottom "0";
%filename 'plotrootsb.jpg';file-properties "XNPEU";}}}%
%BeginExpansion
%EndExpansion%
%TCIMACRO{\FRAME{ftbpFU}{5.4699in}{5.8513in}{0pt}{\Qcb{$W$ curve for
%$e^{2}=0.25$ and $s=0.3$.}}{\Qlb{Fig.20}}{plotrootsc.jpg}%
%{\special{ language "Scientific Word";  type "GRAPHIC";  display "USEDEF";
%valid_file "F";  width 5.4699in;  height 5.8513in;  depth 0pt;
%original-width 8.5002in;  original-height 11.0004in;  cropleft "0";
%croptop "1";  cropright "1";  cropbottom "0";
%filename 'plotrootsc.jpg';file-properties "XNPEU";}}}%
%BeginExpansion
%EndExpansion%
%TCIMACRO{\FRAME{ftbpFU}{5.8652in}{6.5631in}{0pt}{\Qcb{$W$ curve for
%$e^{2}=-0.5$ and $s=0.3$.}}{\Qlb{Fig.21}}{plotrootsd.jpg}%
%{\special{ language "Scientific Word";  type "GRAPHIC";  display "USEDEF";
%valid_file "F";  width 5.8652in;  height 6.5631in;  depth 0pt;
%original-width 8.5002in;  original-height 11.0004in;  cropleft "0";
%croptop "1";  cropright "1";  cropbottom "0";
%filename 'plotrootsd.jpg';file-properties "XNPEU";}}}%
%BeginExpansion
%EndExpansion%
%TCIMACRO{\FRAME{ftbpFU}{5.6887in}{6.6054in}{0pt}{\Qcb{$W$ curve for
%$e^{2}=-0.25$ and $s=0.2$.}}{\Qlb{Fig.22}}{plotrootse.jpg}%
%{\special{ language "Scientific Word";  type "GRAPHIC";  display "USEDEF";
%valid_file "F";  width 5.6887in;  height 6.6054in;  depth 0pt;
%original-width 8.5002in;  original-height 11.0004in;  cropleft "0";
%croptop "1";  cropright "1";  cropbottom "0";
%filename 'plotrootse.jpg';file-properties "XNPEU";}}}%
%BeginExpansion
%EndExpansion%
%TCIMACRO{\FRAME{ftbpFU}{5.9698in}{6.6271in}{0pt}{\Qcb{$W$ curve for
%$e^{2}=-0.09927465$ and $s=0.2$.}}{\Qlb{Fig.23}}{plotrootsf.jpg}%
%{\special{ language "Scientific Word";  type "GRAPHIC";  display "USEDEF";
%valid_file "F";  width 5.9698in;  height 6.6271in;  depth 0pt;
%original-width 8.5002in;  original-height 11.0004in;  cropleft "0";
%croptop "1";  cropright "1";  cropbottom "0";
%filename 'plotrootsf.jpG';file-properties "XNPEU";}}}%
%BeginExpansion
%EndExpansion

\end{document}